%% file: AC_Josephson_CdAs_combined_2.tex
\documentclass[%
reprint,
superscriptaddress,
amsmath,amssymb,
aps,
prb,
floatfix,
]{revtex4-2}

\usepackage[pdftex]{graphicx}
\usepackage{epstopdf}
\usepackage{dcolumn}% Align table columns on decimal point
\usepackage{bm}% bold math
\usepackage{amsmath}
\usepackage{xr-hyper}
\usepackage{hyperref}% add hypertext capabilities

\newcommand{\vb}{\textbf}

\begin{document}

\preprint{APS/123-QED}

\title{AC Josephson effect in a gate-tunable Cd$_3$As$_2$ nanowire superconducting weak link}
% Force line breaks with \\
%%%%%
\author{R.~Haller}
 \email{roy.haller@unibas.ch}
\affiliation{
Department of Physics, University of Basel, Klingelbergstrasse 82 CH-4056, Switzerland
}
%%%%%
\author{M.~Osterwalder}
\affiliation{
Department of Physics, University of Basel, Klingelbergstrasse 82 CH-4056, Switzerland
}
%%%%%
\author{G.~Fülöp}
\affiliation{
Department of Physics, University of Basel, Klingelbergstrasse 82 CH-4056, Switzerland
}
\affiliation{%
Department of Physics, Institute of Physics,  Budapest University of Technology and Economics and MTA-BME Nanoelectronics Momentum Research Group, Műegyetem rkp. 3., H-1111 Budapest, Hungary
}
%%%%%
\author{J.~Ridderbos}
\affiliation{
MESA+ Institute for Nanotechnology, University of Twente, P.O. Box 217, 7500 AE Enschede, Netherlands
}
%%%%%
\author{M.~Jung}
\affiliation{
DGIST Research Institute, DGIST, Daegu 42988, Republic of Korea
}
%%%%%
\author{C.~Schönenberger}
\homepage{http://www.nanoelectronics.unibas.ch/}
\affiliation{
Department of Physics, University of Basel, Klingelbergstrasse 82 CH-4056, Switzerland
}
\affiliation{
Swiss Nanoscience Institute, University of Basel, Klingelbergstrasse 82 CH-4056, Switzerland
}

\date{\today}% It is always \today, today,
% Abstract
%=======================================================================================================================================

\begin{abstract}

Three-dimensional topological Dirac semimetals have recently gained significant attention, since they possess exotic quantum states.
When constructing Josephson junctions utilizing these materials as the weak link, the fractional ac Josephson effect emerges in the presence of a topological supercurrent contribution.
We investigate the ac Josephson effect in a Dirac semimetal Cd$_3$As$_2$ nanowire using two complementary methods: by probing the radiation spectrum and by measuring Shapiro patterns.
With both techniques, we find that conventional supercurrent dominates at all investigated doping levels and that any potentially present topological contribution falls below our detection threshold.
The inclusion of thermal noise in a resistively and capacitively shunted junction (RCSJ) model allows us to reproduce the microwave characteristics of the junction.
With this refinement, we explain how weak superconducting features can be masked and provide a framework to account for elevated electronic temperatures present in realistic experimental scenarios.

\end{abstract}

\maketitle

%\tableofcontents

%=======================================================================================================================================
% Introduction
%=======================================================================================================================================

\section{\label{sec:intro}Introduction}
Topological superconductivity is a hot topic in condensed matter physics, since the quasi-particle excitations that govern its attributes, so-called Majorana fermions, follow non-Abelian exchange statistics -- a property that could revolutionize quantum computation, as it allows the implementation of fault-tolerant operations~\cite{Kitaev2001, KITAEV20032, Ivanov2001, Nayak2008,Wilczek2009, Stern2013, Karzig2017}.
While theoretical concepts on those exotic particles have already been developed in the past decades~\cite{Fu2009,Lutchyn2010,Sau2010,Alicea2010,Qi2011a}, platforms to synthesize Majoranas are still being established and clear experimental evidence of their existence is lacking~\cite{FrolovT2020}.

We employ Dirac semimetal Cd$_3$As$_2$ nanowire Josephson junctions as a platform for the search of topological superconductivity.
Dirac semimetals possess exotic electronic properties emerging from three dimensional Dirac cones - crossings between valence and conduction bands with linear dispersion relations~\cite{Young2012,Wang2012,Wang2013a,Burkov2016,Armitage2018,Crassee2018}. 
They host topological surface states~\cite{Jeon2014a} and develop Majorana flat bands in Josephson junctions~\cite{Chen2017a}.
Recently, the superconducting proximity effect has been observed in Cd$_3$As$_2$ nanowires and nanoplates, including Josephson supercurrent carried by surface states~\cite{Li2018b}. In this context, $\pi$- and $4\pi$-periodic supercurrent states, and topological supercurrent oscillations haven been reported~\cite{Yu2018,Wang2018c}. 

In our Cd$_3$As$_2$ nanowire junctions, we hunt for signatures of the fractional ac Josephson effect arising from $4\pi$-periodic contributions in the current-phase relation (CPR) that are a direct consequence of topologically protected Andreev bound states, i.e., Majorana modes~\cite{Fu2009,Kwon2004b,Feng2020}.
To this end, we investigate the ac Josephson effect by two complementary methods: (i) via the detection of Josephson radiation and (ii) by measuring Shapiro steps. 

In the latter experiment, missing odd Shapiro steps are the fingerprint of topological supercurrent admixtures~\cite{Dominguez2012, Dominguez2017, Pico2017}.
This effect has been observed for different material platforms:
InSb nanowire junctions exposed to in-plane magnetic fields~\cite{Rokhinson2012}, Bi$_{\mathrm{x}}$Se$_{\mathrm{y}}$-based~\cite{Calvez2019, Schuffelgen2019,Yano2020} and HgTe-based~\cite{Wiedenmann2016a,Bocquillon2017} topological insulator junctions, Bi$_{\mathrm{x}}$Sb$_{\mathrm{y}}$ Dirac semimetal nanoplate junctions~\cite{LiDS2018}, (Bi$_{1-{\mathrm{x}}}$Sb$_{\mathrm{x}}$)$_2$Te$_3$-based topological insulator nanowire junctions~\cite{Bai2022} and includes Cd$_3$As$_2$ nanowire junctions~\cite{Wang2018c}.

When measuring Josephson radiation, the signature of a topologically non-trivial supercurrent is the halving of the fundamental Josephson frequency $f_J=q^*V/h$, with $q^*$ being the effective charge, $V$ the dc voltage bias and $h$ the Planck constant. 
This occurs since the topological supercurrent is carried by single electrons $\left(q^*=e\right)$ instead of Cooper-pairs in the case of conventional junctions $\left(q^*=2e\right)$~\cite{Kwon2004b}. 
Evidence for topological emission has only been claimed from a gate-tunable Al-HgTe-Al topological insulator junction~\cite{Deaconb} and on an Al-InAs-Al nanowire junction exposed to in-plane magnetic fields~\cite{Laroche2019}.

The (fractional) ac Josephson effect principally distinguishes conventional from topological junctions, but many experimental aspects pose a challenge to unambiguously prove the emergence of Majorana states such as: marginal $4\pi$-supercurrent contributions, elevated electronic temperatures, environmental circuit effects~\cite{Kamata2018, Mudi2021}, Landau-Zener transitions~\cite{Billangeon2007, Dartiailh2021,Rosen2021,Elfeky2023}, hardly tunable junctions parameters~\cite{Houzet2013,Park2021,Jang2021}, lifetime broadening and a finite detection bandwidth~\cite{San-Jose2012, Houzet2013}. 
The approach to combine Shapiro measurements with radiation measurements on a single Josephson junction, as in Ref.\,\cite{Takeshige2020}, should greatly reduce ambiguity on its topological nature.
Furthermore, the fractional radiation signal can be detected independent of the magnitude of the trivial supercurrent. 
This contrasts Shapiro step measurements where mainly the ratio of the topological to trivial supercurrent determines its resolvability~\cite{Wang2018c}.

In this work, we investigate a gate-tunable Cd$_3$As$_2$ nanowire Josephson junction by assessing the dc and the ac Josephson effect.
By including thermal fluctuations in the resistively and capacitively shunted junction (RCSJ) model, we successfully reproduce the obtained $IV$-curves, Josephson emission spectra and Shapiro patterns using only a conventional, sinusoidal $2\pi$-periodic CPR. 
The comparison enables us to determine the effective electron temperature of the system which, in turn, imposes limitations to the sensitivity and may serve as a possible explanation for the absence of fractional signatures in the ac Josephson effect.

%=======================================================================================================================================
% Device & Measurement setup
%=======================================================================================================================================

\section{\label{sec:deviceandsetup}Device \& Measurement setup }

\begin{figure}[t!]
\includegraphics[width=8.6cm]{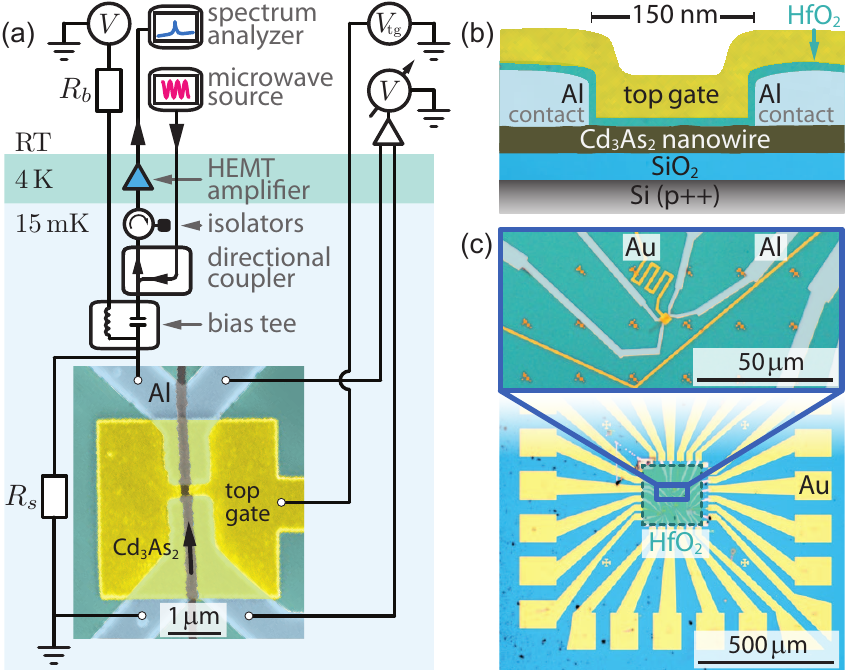} 
\caption{\label{fig:cdas_device}Device and measurement setup. 
(a) Cd$_3$As$_2$ nanowire Josephson junction embedded in a radiation (irradiation) setup using a spectrum analyzer (microwave source). 
(Inset) False-colored SEM image of the junction defined by superconducting Al leads (blue) in the Cd$_3$As$_2$ nanowire (center axis along black arrow). 
The top gate electrode (yellow) is isolated from the junction by a 20\,nm thick HfO$_2$ layer. 
The voltage $V_{\mathrm{tg}}$ applied on the Au gate allows tuning the charge carrier density. 
The junction is shunted by a resistor $R_s=10\,\Omega$ that connects the rf line to the fridge ground on the PCB.
(b) Cross-sectional illustration of the junction along the black arrow in (a). 
(c) Optical image showing the electrodes extending from the junction (top) towards the Au bonding pads shown on the overview of the chip (bottom). 
}
\end{figure}

Fig.\,\ref{fig:cdas_device}(a) shows the Cd$_3$As$_2$ nanowire Josephson junction device connected to the schematic of the measurement setup. 
The nanowire has a diameter of $50$\,nm, and the junction has a length of $150$\,nm that is determined by the spacing of the superconducting Al leads. 
A locally deposited layer of HfO$_2$ (20\,nm) covering the whole inner part of the chip (see Fig.\,\ref{fig:cdas_device}(c)) serves as gate-dielectric for the Au top gate that allows tuning the charge carrier density inside the wire by applying a gate voltage $V_{\mathrm{tg}}$. 
A cross-sectional cut through the device structure is illustrated in Fig.\,\ref{fig:cdas_device}(b). 

The measurement setup shown in Fig.\,\ref{fig:cdas_device}(a) allows probing the ac Josephson effect in two ways. 
First, the Josephson radiation emitted from the junction under finite dc bias can be detected directly  with a spectrum analyzer. 
Second, the reverse experiment can be done: irradiating the device with an ac field emitted by a microwave source while measuring the dc transport response of the junction, resulting in Shapiro steps~\cite{Shapiro1963}. 
In this case, a directional coupler feeds the incoming microwave tone to the sample, whereas in the Josephson radiation experiment, it directs the signal emitted by the junction to the spectrum analyzer.
The out-coming signal passes through isolator stages, limiting the detection bandwidth to $2.5-3.8$\,GHz, after which the signal gets amplified by a high-electron-mobility-transistor (HEMT).

Our dc setup sources a current using a bias resistor $R_b=1$\,M$\Omega$ in series with a dc voltage source with a small ac component with frequency $f=177$\,Hz supplied by a lock-in amplifier. 
This current is applied via a bias tee to the microwave line that directly connects to the source of the sample. 
The other side of the sample is galvanically connected to ground, closing both the low- and high-frequency circuits. 
When the sample is in the non-superconducting regime, the signal is converted to a stable voltage bias by shunting the device with a resistor $R_s=10$\,$\Omega$. 
This shunt resistor is directly placed between the central conductor of the transmission line and the galvanic ground on the sample holder. 
We measure the differential resistance of the shunted device using a voltage amplifier and lock-in techniques. 
All measurements are performed in a dilution refrigerator with a base temperature of $\sim15$\,mK. 

Details about the fabrication, the measurement setup and the device integration on the PCB are given in the supplementary material (SM) Sec.\,\ref{sec:methodes} and Sec.\,\ref{sec:setup}~\footnote[52]{See Supplemental Material for details about the RCSJ model\,(\ref{sec:modeling}), the device fabrication\,(\ref{sec:methodes}), the measurement setup\,(\ref{sec:setup}); valuations of the Josephson emission power\,(\ref{sec:transfer}), the $I_{\mathrm{c}}R_N$ product\,(\ref{sec:icrn}); the evaluation of Shapiro patterns\,(\ref{sec:shapiro_eva}).}.

%=======================================================================================================================================
% Measurements
%=======================================================================================================================================

\section{\label{sec:measurement}Measurement results}

%=======================================================================================================================================
% Gate dependency
%=======================================================================================================================================

\begin{figure}[t!]
\includegraphics[width=8.6cm]{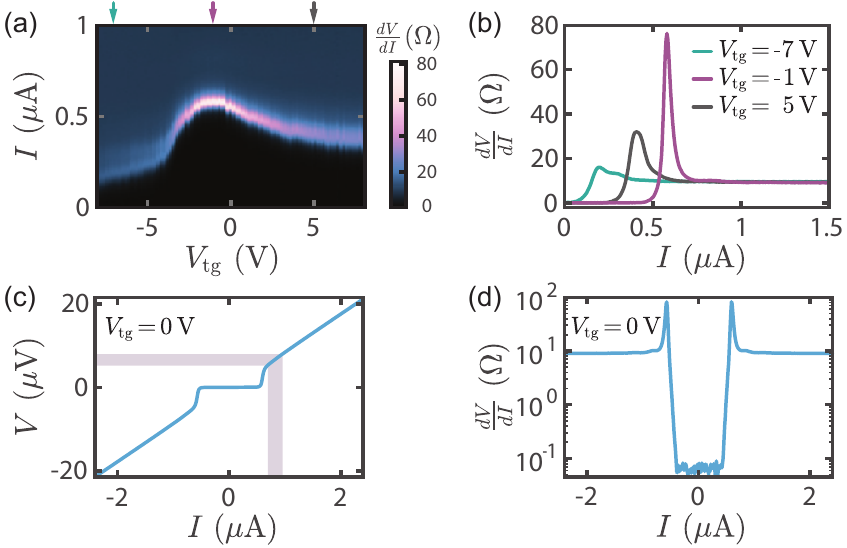}
\caption{\label{fig:cdas_gate} Gate dependent supercurrent. 
(a) Differential resistance $dV/dI$ as a function of current bias $I$ and top gate voltage $V_{\mathrm{tg}}$. 
(b) $dV/dI$ as a function of $I$ at different gate voltages taken from line cut in (a) at the positions indicated by the arrows. The $dV/dI$-peak position is identified as the critical current $I_\mathrm{c}$. 
(c) $IV$-characteristic at $V_{\mathrm{tg}}=0$\,V obtained by integrating the measured $dV/dI$ shown in logarithmic scale in (d). 
The violet shaded area indicates the current and voltage range that produces conventional Josephson radiation in a $2.5-3.8$\,GHz bandwidth.
}
\end{figure}

We first probe the gate dependent dc response of the junction, which is presented in Fig.~\ref{fig:cdas_gate}. 
The differential resistance as a function of current $I$ bias and top gate voltage $V_\mathrm{tg}$ in Fig.\,\ref{fig:cdas_gate}(a), reveals a clear switching behavior from a zero resistance superconducting state (black) to a normal resistive state (blue). 
We attribute the peaks in differential resistance to the critical current $I_{\mathrm{c}}$, which become sharper and more pronounced for higher $I_\mathrm{c}$ as seen in  Fig.\,\ref{fig:cdas_gate}(b). Under sufficiently large current bias the differential resistance approaches $R_s$.
As a function of $V_{\mathrm{tg}}$, we observe an anomalous evolution of $I_{\mathrm{c}}$; when sweeping the gate from negative to positive voltages, $I_{\mathrm{c}}$ first increases and then decreases again in an asymmetric way with a maximum of $I_{\mathrm{c}}=580$\,nA at $V_{\mathrm{tg}}=-1$\,V. 
This behavior contrasts ordinary semiconductor JJs, which generally exhibit a steady increase of the critical current as a function of larger electric-field induced doping. 
The unusual gate response has been previously reported in long Cd$_3$As$_2$ nanowire JJs and has been proposed to originate from scattering mechanisms between surface and bulk states that give rise to dephasing~\cite{Li2018b}. 
In this case, one assumes the majority of the supercurrent is carried by surface states and increasing the electron density enhances scattering with bulk modes which, in turn, results in a suppression of the coherent Cooper pair transport, and hence leads to a reduction of $I_{\mathrm{c}}$. 
This hypothesis is further supported by the $I_{\mathrm{c}}R_N$-product discussed in the SM Sec.\,\ref{sec:icrn}. 

In Fig.\,\ref{fig:cdas_gate}(c), the $IV$-curve at \mbox{$V_{\mathrm{tg}}=0$\,V}, obtained by integrating the measured $dV/dI$ curve presented in Fig.\,\ref{fig:cdas_gate}(d), shows a clear voltage plateau which we assume is zero. 
In the normal state, the junction shows ohmic behavior down to $5\,\mu{}V$, which allows steady voltage biasing in the regime of conventional Josephson radiation for the given detection bandwidth (violet shading).

%=======================================================================================================================================
% Josephson emission
%=======================================================================================================================================

\begin{figure}[t!]
\includegraphics[width=8.6cm]{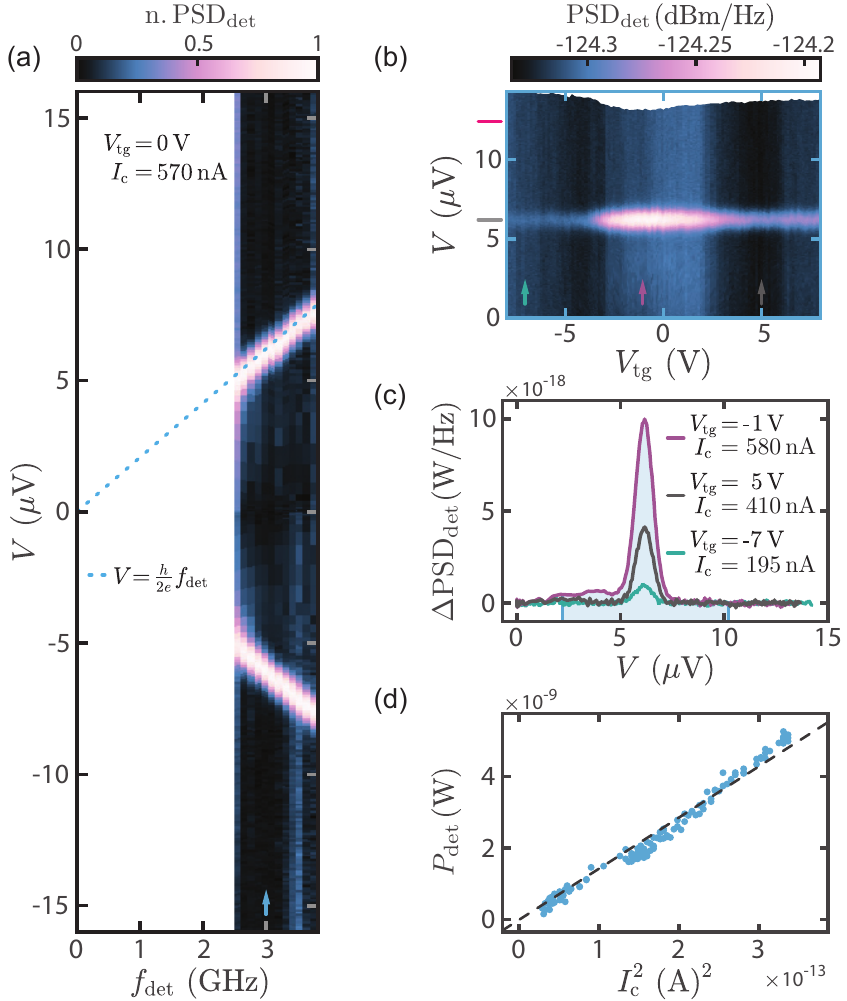}
\caption{\label{fig:cdas_radiation} Emission spectrum.
(a) Normalized power spectral density (n.\,PSD$_{\mathrm{det}}$) as a function of the voltage drop $V$ across the junction and detection frequencies $f_{\mathrm{det}}$ at $V_{\mathrm{tg}}=0$\,V. The radiation spectrum is overlaid with the expected peak position of conventional Josephson emission. 
(b) Power spectral density (PSD$_{\mathrm{det}}$) as a function of $V$ and $V_{\mathrm{tg}}$, measured at a fixed detection frequency $f_{\mathrm{det}}=3$\,GHz indicated by the blue arrow in (a). The gray mark on the voltage axis indicates the expected conventional emission peak position, whereas the pink mark is placed at the position of a potential topologically non-trivial emission peak.
(c) PSD$_{\mathrm{det}}$ in linear scale with subtracted background ($\Delta$PSD$_{\mathrm{det}}$) as a function of $V$ at three gate voltages obtained from line cuts in (b) at the positions indicated by the arrows. 
(d) Emission strength $P_{\mathrm{det}}$\,(W) plotted against the corresponding critical current squared ($I_{\mathrm{c}}^2$). 
}
\end{figure}

In the next step, we investigate the Josephson radiation spectrum at $V_{\mathrm{tg}}=0$\,V, for which $I_{\mathrm{c}}=570$\,nA. 
In Fig.\,\ref{fig:cdas_radiation}(a), the normalized power spectral density collected for a $20$\,MHz bandwidth is plotted as a function of voltage bias $V$ and detection frequency $f_\mathrm{det}$. 
The normalization for each $f_\mathrm{det}$ allows us to compensate for the frequency dependent background. 
The detected radiation features are symmetric in voltage and solely determined by conventional Josephson emission arising due to the inelastic transfer of Cooper pairs. 
The signal follows the fundamental ac Josephson relation $V=hf_J/\left(2e\right)$, illustrated by the dashed line. 
When the detection frequency of the spectrum analyzer is aligned with the Josephson frequency $\left(f_J=f_\mathrm{det}\right)$ the power spectral density peaks.
Neither signatures from topological supercurrent contributions that would evolve as $V=hf_J/e$, nor clear higher-order emission features are observed.
The latter is expected for a Josephson junction possessing a CPR consisting of $\sin(n\varphi)$-harmonics with $n=2,\,3,\,4,\dots$, i.e., a skewed CPR where multiple Cooper pairs counted by $n$ are involved in one effective transfer event~\cite{Heikkila,Chauvin2006, Dayem1966, Bretheau2013, Basset2019}, resulting in higher-order emission peaks following the relation $V=hf_J/\left(n2e\right)$.
Therefore, we conclude that our junction dynamics are dominated by a sinusoidal $2\pi$-periodic CPR.

Next, we look for the emergence of a topological phase as a function of doping level in Fig.\,\ref{fig:cdas_radiation}(b), where we measured the power spectral density with fixed $f_{\mathrm{det}}=3$\,GHz, while sweeping $V$ across the junction for different $V_\mathrm{tg}$.
We note the voltage position for topologically non-trivial emission $\left(V=hf_J/e\right)$ in pink, at which no additional radiation peaks appear.
The conventional emission peak modulates in intensity but remains visible throughout the whole gate range and aligns to the conventional voltage position $\left(V=hf_J/2e\right)$ marked in gray.

To estimate a lower bound of sensitivity in terms of $I_\mathrm{c}$, we relate the emission strength to the critical current. 
We first convert the raw data from dBm/Hz-scale to Watt/Hz-scale and subtract for each gate voltage a linear background such that regions far away from the radiation peaks have vanishing small contributions. 
The hereby obtained relative emission density ($\Delta$PSD$_{\mathrm{det}}$) as a function of $V$ is shown in Fig.\,\ref{fig:cdas_radiation}(c) for different $V_\mathrm{tg}$, corresponding to different $I_{\mathrm{c}}$ that are extracted from Fig.\,\ref{fig:cdas_gate}(a).   
In Fig.\,\ref{fig:cdas_radiation}(d) we correlate the emission strength $P_{\mathrm{det}}$ to the square of the critical current $I_{\mathrm{c}}^2$. 
$P_{\mathrm{det}}$ is obtained by integrating $\Delta$PSD$_{\mathrm{det}}$ over the voltage interval ($2.2-10.2\,\mu$V) that is indicated with the light blue shading in Fig.\,\ref{fig:cdas_radiation}(c) as $P_{\mathrm{det}}(\mathrm{W})=\frac{2e}{h}\int_{2.2\,\mu{\mathrm{V}}}^{10.2\,\mu{\mathrm{V}}} \Delta{\mathrm{PSD_{\mathrm{det}}\,[W/Hz]}}\,dV$.
We recognize a clear quadratic dependence between $P_{\mathrm{det}}$ and $I_{\mathrm{c}}$. 
Considering the junction as a sinusoidal ac current source with amplitude $I_{\mathrm{c}}$, indeed a power $P_{\mathrm{det}}=\mathcal{R}I_{\mathrm{c}}^2/2$ is dissipated in the detector, where $\mathcal{R}$ describes the power transfer ratio, detailed in SM Sec.\,\ref{sec:transfer}.
From the power to critical current relation, we estimate for the given experimental configuration, a lower detection limit of $I_{\mathrm{c}}\approx55$\,nA. 
This is a potential explanation for the lack of topological signatures in the radiation spectrum since it is known that the associated non-trivial supercurrent in Cd$_3$As$_2$ nanowires may only be a small fraction $\left(\leq10\%\right)$ of the overall critical current~\cite{Wang2018c}.

%=======================================================================================================================================
% Shapiro steps
%=======================================================================================================================================

\begin{figure}[t!]
\includegraphics[width=8.6cm]{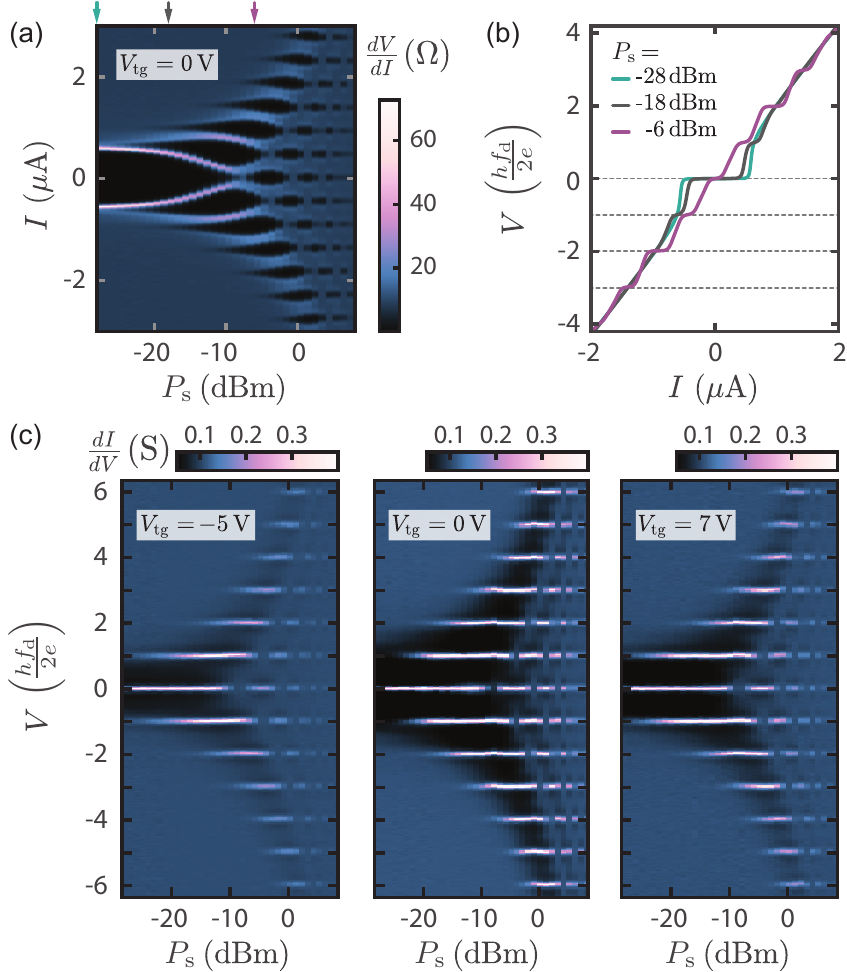}
\caption{\label{fig:cdas_shapiro} Shapiro step measurement with drive frequency $f_{\mathrm{d}}=2$\,GHz.
(a) Differential resistance $dV/dI$ as a function of drive power $P_{\mathrm{s}}$ at the microwave source and current bias $I$. 
(b) $IV$-curves obtained by integrating $dV/dI$ for irradiation powers indicated by the arrows in (a), with the voltage axis normalized to the Shapiro step voltage $hf_{\mathrm{d}}/2e$.
(c) Differential conductance $dI/dV$ as a function of $P_{\mathrm{s}}$ and $V$ in units of the Shapiro step voltage $hf_{\mathrm{d}}/2e$, measured on the hole side ($V_\mathrm{tg}=-5$\,V), the electron side ($V_\mathrm{tg}=7$\,V), and close to the Dirac point ($V_\mathrm{tg}=0$\,V) . Data is numerically converted from current-biased data; the center map constitutes the same measurement as (a).
}
\end{figure}

We now expand the investigation of the ac Josephson effect by measuring the Shapiro step pattern.
Here, the junction is irradiated by a microwave tone of fixed frequency $f_{\mathrm{d}}=2$\,GHz at variable output power $P_{\mathrm{s}}$ of the signal generator. 
We probe the differential resistance $dV/dI$ as a function of current bias $I$ and $P_{\mathrm{s}}$ for a constant gate voltage $V_{\mathrm{tg}}=0$\,V. 
The resulting map shown in Fig.\,\ref{fig:cdas_shapiro}(a), reveals the characteristic voltage plateaus in regions of vanishing $dV/dI$ (dark teardrop shapes). 
With increasing $P_{\mathrm{s}}$ , the $0^{\mathrm{th}}$-plateau $\left(V=0\right)$ diminishes, while higher-order Shapiro steps progressively emerge.
In Fig.\,\ref{fig:cdas_shapiro}(b) we present the $IV$-curve obtained by integrating $dV/dI$ for different drive powers, and the voltage axis in units of the Shapiro voltage matches with the measured voltage steps.
By numerically differentiating the interpolated $IV$-curves we can extract the differential conductance $dI/dV$, which is presented in Fig.\,\ref{fig:cdas_shapiro}(c) as a function of $V$ and $P_{\mathrm{s}}$ for different top gate voltages. 
With this visualization technique, no binning is required and the resolution is maintained~\cite{Ridderbos2019}. 
The voltage plateaus appear as lines in the color map, whereas their intensities reflect the width of the Shapiro steps.
In agreement with the lack of topological signatures in the radiation measurement, no modulation in the width of subsequent steps is observed in the experimentally accessed parameter space, as evaluated in the SM Sec.\,\ref{sec:shapiro_eva}.

%=======================================================================================================================================
% RCSJ model
%=======================================================================================================================================

\section{\label{sec:rcsj}RCSJ simulation with thermal fluctuations}

\begin{figure}[t!]
\includegraphics[width=8.6cm]{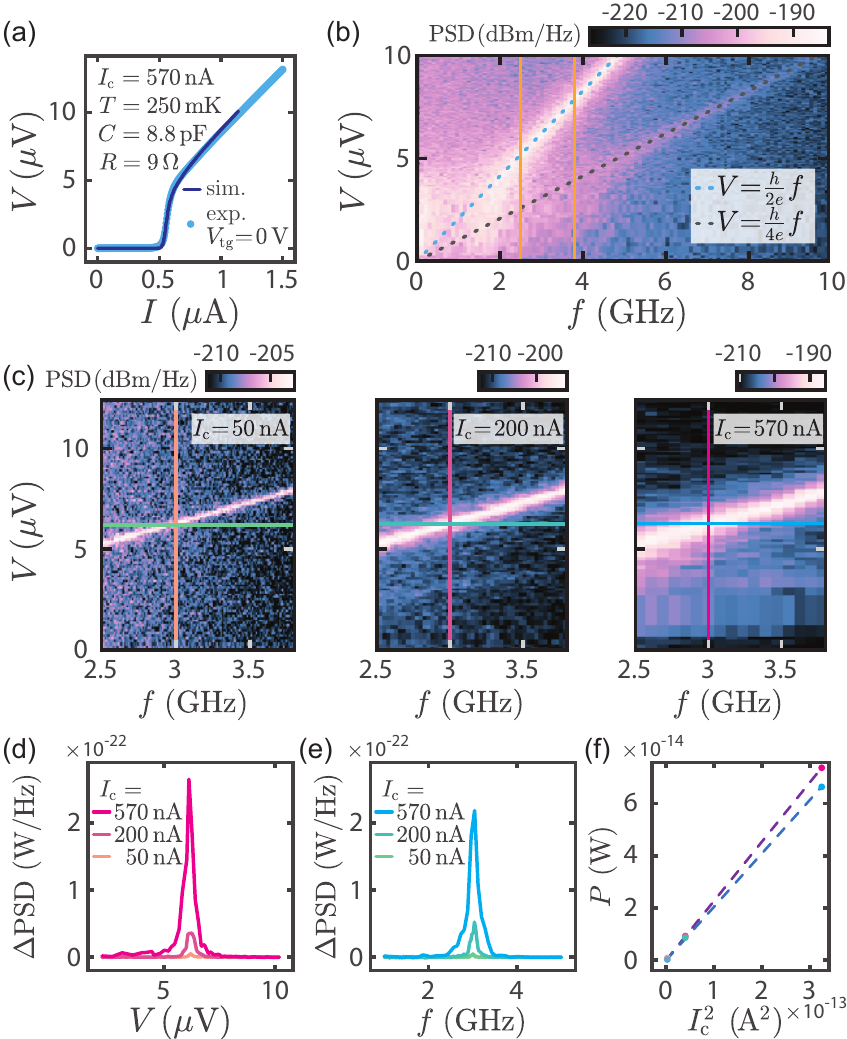}
\caption{\label{fig:cdas_sim} 
Results of RCSJ modeling with thermal fluctuations.
(a) Simulated $IV$-characteristics (dark blue) fitted to the experimental curve at $V_{\mathrm{tg}}=0$\,V (light blue). 
The fit parameters: critical current $I_{\mathrm{c}}$, effective temperature $T$, shunt capacitance $C$, and shunt resistance $R$ are listed. 
$T$,\,$C$,\,$R$ are kept constant for the subsequent plots.
(b) Simulated power spectral density (PSD) as a function of mean voltage $V$ and frequency $f$ deduced from the voltage evolution in time for $I_{\mathrm{c}}=570$\,nA. 
The map is overlaid with the fundamental (blue) voltage-to-frequency conversion and the first higher-harmonic (gray).
The orange vertical lines indicate the frequency range ($2.5-3.8$\,GHz) of interest for the comparison with the experiment.
(c) Simulated emission spectrum for different $I_{\mathrm{c}}$.
(d-e) PSD in linear scale with subtracted background ($\Delta$PSD) for different $I_{\mathrm{c}}$ deduced from the maps presented in (c) for a given frequency and voltage, respectively. In (d) $f$=3\,GHz and in (e) its voltage equivalent $V=6.2\,\mu$V that relate to vertical and horizontal cuts presented in (c).
(f) Emission strength $P$ plotted against the corresponding critical current squared ($I_{\mathrm{c}}^2$) by integrating over the voltage ($P_V$, red) and by integrating over the frequency ($P_f$, blue).
}
\end{figure}

To gain a deeper understanding of the junction dynamics, we aim to reproduce our experimental results in the framework of the resistively and capacitively shunted junction (RCSJ) model. 
To be able to reproduce our experimental data, we include a noise term to account for thermal fluctuations~\cite{Kautz}.
Consequently, the model~\footnote[60]{Python code of the RCSJ model is available on\\\href{https://github.com/donfuge/rcsj_sde}{https://github.com/donfuge/rcsj\_sde}}, which is detailed in the SM Sec.\,\ref{sec:modeling} contains the following parameters: the critical current $I_{\mathrm{c}}$, the shunt resistance $R$, the shunt capacitance $C$ and the effective temperature $T$.
Furthermore, the shape of the CPR matters, which in the following is set to a sinusoidal, $2\pi$-periodic function $I_{\mathrm{s}}=I_{\mathrm{c}}\sin\left(\varphi\right)$, where $\varphi$ represents the phase drop and $I_{\mathrm{s}}$ the supercurrent across the junction. 
The discussion is extended to admixtures of $2\pi$- and $4\pi$-periodic CPRs in the SM Sec.\,\ref{sec:modeling}.

The output of the model is the time evolution of the junction phase $\varphi(t)$, from which the junction voltage is calculated using the ac Josephson relation, ${V(t) = \frac{\hbar}{2e}\frac{d\varphi(t)}{dt}}$ with $\hbar=h/2\pi$. First, we consider the time-averaged results of the modified RCSJ model to fit the experimentally obtained $IV$-curve at $V_{\mathrm{tg}}=0$\,V as presented in Fig.\,\ref{fig:cdas_sim}(a). 
With the hereby obtained model parameters that are listed in the plot, a junction quality factor $Q=\sqrt{2eI_{\mathrm{c}}/\left(\hbar{}C\right)}\cdot{}RC\approx1$ is obtained, which classifies the junction into the intermediately damped regime~\cite{Stewart}, which is further supported by the smooth switching behavior accompanied with a negligible hysteresis between up/down current bias sweeps (not shown).
The elevated effective temperature ($T=250$\,mK) with respect to the base temperature ($T_b=15$\,mK) originates from external noise, likely generated by the HEMT configured at a sub-optimal operating point, an issue that was only later identified.

In the next step, we employ the extracted parameters and focus on the time-dependent $V(t)$ solutions of the RCSJ model to reproduce the emission spectrum. 
By taking the squared magnitude of the Fourier transform of $V(t)$, we calculate the voltage noise spectral density as a function of frequency $f$, and convert it to power spectral density (PSD). 
Here, we do not include any amplification of the signal; the comparison of absolute signal levels between the experiment and simulation is discussed in the SM Sec.\,\ref{sec:transfer}. 
In Fig.\,\ref{fig:cdas_sim}(b) we plot the simulated PSD as a function of $f$ and averaged voltage $V$. 
Within the experimentally accessible range, indicated by the orange lines, the simulation qualitatively reproduces the detected emission spectrum previously presented in Fig.\,\ref{fig:cdas_radiation}(a). 
Interestingly, the simulation reveals, in addition to the conventional voltage to frequency relation (blue dashed), an emission peak at double the frequency for a given voltage (gray dashed). 
We attribute this feature to the current bias in our model, leading to higher order modes of frequency $f=n2eV/h$~\cite{Bretheau2013}.
In this manner, higher-order features can be generated from a purely sinusoidal CPR, as seen in the simulations for different model parameters presented in the SM Sec.\,\ref{sec:modeling}.

We now direct our attention to the modeled emission strength as a function of $I_{\mathrm{c}}$. Fig.\,\ref{fig:cdas_sim}(c) shows simulated radiation maps for increasing $I_{\mathrm{c}}$, with $f$ corresponding to our experimental range.
For each $I_{\mathrm{c}}$ we generate voltage linecuts for fixed $f=3$\,GHz (orange, pink, magenta lines), and extract their corresponding frequency linecuts for fixed $V=6.2$\,$\mu$V (green, aqua, blue). In Fig.\,\ref{fig:cdas_sim}(d) and (e) we evaluate $\Delta$PSD by subtracting a linear background for the voltage-cuts and a $1/f$ background for the frequency-cuts, from which we obtain the emission strength of the junction:
(i)\,By integrating $\Delta$PSD over the voltage interval ($2.2-10.2\,\mu$V) as 
$P_V(\mathrm{W})=\frac{2e}{h}\int_{2.2\,\mu{\mathrm{V}}}^{10.2\,\mu{\mathrm{V}}} \Delta{\mathrm{PSD\,[W/Hz]}}\,dV$ and
(ii)\,by integrating $\Delta$PSD over the voltage corresponding frequency interval ($1-5$\,GHz) as $P_f(\mathrm{W})=\int_{1\,{\mathrm{GHz}}}^{5\,{\mathrm{GHz}}} \Delta{\mathrm{PSD\,[W/Hz]}}\,df$.
The hereby obtained emission strength is linearly correlated to $I_{\mathrm{c}}^2$ as illustrated in Fig.\,\ref{fig:cdas_sim}(f), where the procedure via integration over the voltage (frequency) is shown in red (blue) color. 
We attributed the slightly enhanced emission strength obtained from the voltage integration procedure to higher-order contributions appearing at integer fractions of the main emission peak.

\begin{figure}[t!]
\includegraphics[width=8.6cm]{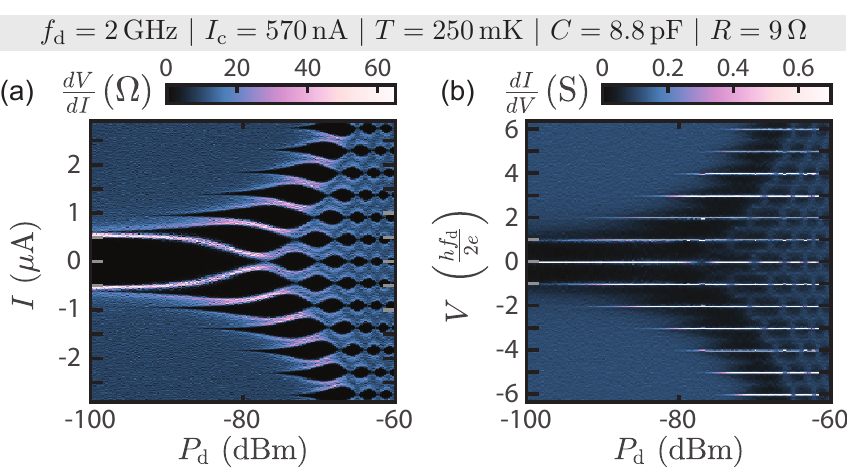}
\caption{\label{fig:cdas_sim_shapiro} 
Simulated Shapiro pattern reproducing the experimentally obtained features at $V_{\mathrm{tg}}=0$\,V presented in Fig.\,\ref{fig:cdas_shapiro}. 
The parameters used in the RCSJ model with thermal fluctuations are listed. 
(a) Differential resistance $dV/dI$ as a function of drive power $P_{\mathrm{d}}$ at the device and current bias $I$.
(b) Differential conductance $dI/dV$ as a function of $P_{\mathrm{d}}$ and $V$ in units of the Shapiro step voltage $hf_{\mathrm{d}}/2e$.
}
\end{figure}

By adding an ac current bias excitation in the RCSJ model expressed in drive power $P_{\mathrm{d}}$ at the device, it is also capable of generating Shapiro step patterns. 
Employing the same model parameters as before, we find good agreement in the Shapiro pattern between the simulation presented in Fig.\,\ref{fig:cdas_sim_shapiro} and the experiment shown in Fig.\,\ref{fig:cdas_shapiro}, again by using a $2\pi$-periodic sinusoidal CPR.
In the SM Sec.\,\ref{sec:shapiro_eva} we analyze simulated Shapiro step patterns with CPRs containing an admixture of topologically non-trivial $\sin(\varphi/2)$ CPR contributions.
In contrast to the evaluation of the radiation experiment, where an absolute detection limit can be defined, with the step width analysis a relative detection threshold can be established. 
For the specific parameters used, we find that below a topological to conventional CPR ratio of $\sim10\,\%$, the topological features are masked by the dominating $2\pi$-periodic supercurrent.

%=======================================================================================================================================
% Conclusion
%=======================================================================================================================================

\section{\label{sec:conclusion}Conclusion}

We investigated the dc and ac Josephson effect in a single Cd$_3$As$_2$ nanowire Josephson junction.
Although we observe an anomalous evolution of the critical current as a function of gate voltage that points to the presence of supercurrent carrying surface states, no direct evidence of topological superconductivity is found in both the Josephson radiation measurement and Shapiro pattern evaluation.
Importantly, we reproduce our experimental findings using a refined RCSJ model that includes thermal fluctuations, and we obtain an effective electronic temperature $T\approx250$\,mK which masks emission signals from supercurrents $<55$\,nA, highlighting the importance of reducing thermal noise.

The work on (Bi$_{1-{\mathrm{x}}}$Sb$_{\mathrm{x}}$)$_2$Te$_3$~\cite{Takeshige2020} similarly finds no evidence of the fractional Josephson effect, which they attribute to a large shunt capacitance that reduces the power reaching their detector. 
We face similar detection limit challenges which could be overcome by improving the filtering, minimizing the ground noise, matching the junction and detector impedance, optimizing the detector settings, and adding a Josephson parametric amplifier to the amplification chain. 
The resulting gain in the signal-to-noise ratio hopefully allows us to resolve much lower supercurrents and reveal the topological nature of these materials.

%=======================================================================================================================================
% Acknowledgments
%=======================================================================================================================================

\vspace{-1em}

\begin{acknowledgments}

We are grateful for scientific discussions with Chuan~Li.
This research was supported by the Swiss National Science Foundation through a) grants No 172638 and 192027, b) the National Centre of Competence in Research Quantum Science and Technology (QSIT), and c) the QuantEra project SuperTop; the János Bolyai Research Scholarship of the Hungarian
Academy of Sciences, the National Research Development and Innovation Office (NKFIH) through the OTKA Grants FK 132146 and NN127903 (FlagERA Topograph), and the National Research, Development and Innovation Fund of Hungary within the Quantum Technology National Excellence Program (Project
Nr. 2017-1.2.1-NKP-2017-00001), the Quantum Information National Laboratory of Hungary and the ÚNKP-22-5 New National Excellence Program. We further acknowledge funding from the European Union’s Horizon 2020 research and innovation program, specifically a) from the European Research Council (ERC) grant agreement No 787414, ERC-Adv TopSupra, b) grant agreement No. 862046, FET TOPSQUAD and c) grant agreement No 828948, FET-open project AndQC. M.J. is supported by the midcareer research program (NRF-2023R1A2C1004832) through the National Research Foundation of Korea (NRF) grant.

All raw- and metadata in this publication are available in numerical form together with the processing codes at DOI:~\href{https://doi.org/10.5281/zenodo.7961884}{10.5281/zenodo.7961884}.
\end{acknowledgments}

%=======================================================================================================================================
% BIBLIOGRAPHY
%=======================================================================================================================================
\input{AC_Josephson_CdAs_main_2.bbl}% Produces the bibliography via BibTeX.

%=======================================================================================================================================
% SUPPLEMENT
%=======================================================================================================================================

\pagebreak
\clearpage
\widetext

\setcounter{equation}{0}
\setcounter{figure}{0}
\setcounter{table}{0}
\setcounter{page}{1}
\setcounter{section}{0}
\setcounter{footnote}{0}

\renewcommand{\thefigure}{S\arabic{figure}}
\renewcommand{\theequation}{S\arabic{equation}}
\renewcommand{\thesection}{S\Roman{section}}
\renewcommand{\bibnumfmt}[1]{[S#1]}
\renewcommand{\citenumfont}[1]{S#1}

\textbf{\centering\large Supplementary Material:\\}
\textbf{\centering\large AC Josephson effect in a gate-tunable Cd$_3$As$_2$ nanowire superconducting weak link\\}

\vspace{1em}

%=======================================================================================================================================
% Modeling
%=======================================================================================================================================

\section{\label{sec:modeling}Modeling}

In the RCSJ model~\footnote{Python code of the RCSJ model is available on\\\href{https://github.com/donfuge/rcsj_sde}{https://github.com/donfuge/rcsj\_sde}} the junction phase $\varphi$ is governed by the following differential equation \cite{S_Kautz,S_AppliedS}: 
\begin{equation}
    \frac{\hbar}{2e} C \frac{d^2 \varphi}{dt^2} + \frac{\hbar}{2e} \frac{1}{R} \frac{d\varphi}{dt} + I_{\mathrm{c}} f(\varphi) + I_{F}(t) = I(t), 
\end{equation}
where $C$ is the junction capacitance, $R$ is the parallel resistance, $f(\varphi)$ is the dimensionless shape of the current-phase relation, $I_{\mathrm{c}}$ is the critical current, $I_F(t)$ is the fluctuating noise term, and $I(t)=I_{\mathrm{dc}} + I_{\mathrm{ac}}\sin(\omega t)$ is the bias current. 

To be able to accommodate a mixture of a trivial and a topological supercurrent contribution in the current-phase relation, we set 
\begin{equation}
    f(\varphi) = a \sin \left( \varphi \right) + b \sin \left( \varphi/2 \right),
\end{equation}
where we encode the weight of the two components in the dimensionless parameters $a$ and $b$. We remark that if both $a$ and $b$ are non-zero, the maximal supercurrent can be larger than the $I_{\mathrm{c}}$ parameter. 
We assume thermal noise for $I_F(t)$, which is characterized by 
\begin{align}
\left\langle I_F (t) \right \rangle & = 0 \\
\left\langle I_F(t) I_F(t')\right\rangle & = \frac{2 k_B T}{R} \delta(t-t'),
\end{align}
where $\delta(x)$ is the Dirac function.
Let us introduce the normalized units $i=I/I_{\mathrm{c}}$, $i_F = I_F /I_{\mathrm{c}}$ and $\tau = {t}/{t_c}$, with
\begin{equation}
    t_c = \left( \frac{2e}{\hbar} I_{\mathrm{c}} R \right) ^{-1}.
\end{equation}
Using these normalized units, the differential equation takes the form of
\begin{equation}
    \label{eq:sde_norm}
    \frac{1}{\beta} \frac{d^2 \varphi}{d\tau^2} + \frac{d\varphi}{d\tau} +  f(\varphi) - i(\tau) + i_F(\tau) = 0,
\end{equation}
where $\beta$ is the Stewart-McCumber parameter,
\begin{equation}
    \beta = \frac{2e}{\hbar} I_{\mathrm{c}} R^2 C.
\end{equation}
The noise term in normalized units satisfies 
\begin{align}
\left\langle i_F (\tau) \right \rangle & = 0 \\
\left\langle i_F(\tau) i_F(\tau')\right\rangle & = \frac{2e}{\hbar}\frac{2 k_B T}{I_{\mathrm{c}}} \delta(\tau-\tau') = \sigma^2 \delta(\tau-\tau').
\end{align}
The second-order differential equation in \ref{eq:sde_norm} can be written as two coupled first-order equations. By introducing $\vb{y} = [\varphi, \dot{\varphi}]^T$, where $\dot{\varphi} = d\varphi/d\tau$, we write \ref{eq:sde_norm} as
\begin{align}
\dot y_0 & =  y_1 \\
\dot y_1 & = \frac{1}{\beta} \left(i(\tau) - f(y_0) -  y_1 + i_F(\tau) \right)
\end{align}
Let us write the equation set in a matrix form, and separate the deterministic and random terms,
\begin{equation}
d\vb{y} = \vb{F}(\vb{y}, \tau) d\tau + \vb{G}(\vb{y}, \tau) d\vb{W},     
\end{equation}
where $d\vb{W}$ is the random noise originating from the current fluctuations, and
\begin{align}
     \vb{F}(\vb{y}, \tau)  &= \begin{pmatrix} y_1 \\ \left(i(\tau) - f(y_0) -  y_1 \right)/\beta \end{pmatrix} \\
     \vb{G}(\vb{y}, \tau) & \equiv \vb{G} =  \begin{pmatrix} 0 \\ \sigma/\beta \end{pmatrix}.
\end{align}
For the simulations, we calculate the time evolution of $\vb{y}(\tau)$ numerically, using Heun's method \cite{S_Rumelin}. We set the initial conditions, $\vb{y}[0]=\vb{y}_0$, and calculate $\vb{y}[n+1]$ from $\vb{y}[n]$ using the following update rules: 
\begin{align}
\vb{f}_1 & = \vb{F}(\vb{y}[n]) \cdot h \\
\vb{k} & = \vb{y}[n] + \vb{f}_1 \\
\vb{f}_2 & = \vb{F}(\vb{k}) \cdot h \\
\vb{y}[n+1] & =  \vb{y}[n] + (\vb{f}_1 + \vb{f}_2)/2  + \vb{G} \delta W, 
\end{align}
where $h$ is the size of the time step, and $\delta W$ is a random number sampled from a Gaussian distribution with zero mean and variance $h$, independently for each time step. 
For the $IV$-curve, we calculate the DC voltage on the junction with Josephson's equation using the time-averaged $\varphi(t)$,
\begin{equation}
    V = \frac{\hbar}{2e} \left\langle \frac{d\varphi(t)}{dt} \right\rangle
      = \frac{\hbar}{2e} \frac{1}{t_c} \left\langle y_1(\tau) \right\rangle.
\end{equation}
We use the \texttt{scipy.signal.welch} implementation of the Welch method \cite{S_1161901} in the scipy Python package to calculate the power spectral density (PSD) of the junction voltage $V(t) = ({\hbar}/{2e})\cdot d\varphi(t)/dt $. Then, we convert the PSD from V$^2$/Hz to dBm/Hz using the formula
\begin{equation}
  \mathrm{PSD}\, [\mathrm{dBm/Hz}] = 10\cdot\log_{10} \left(\frac{\mathrm{PSD}\,[\mathrm{V}^2/\mathrm{Hz}]}{50\, \Omega} \right ) + 30.
\end{equation}

\begin{figure}%[t!]
\centering
\includegraphics[width=15cm]{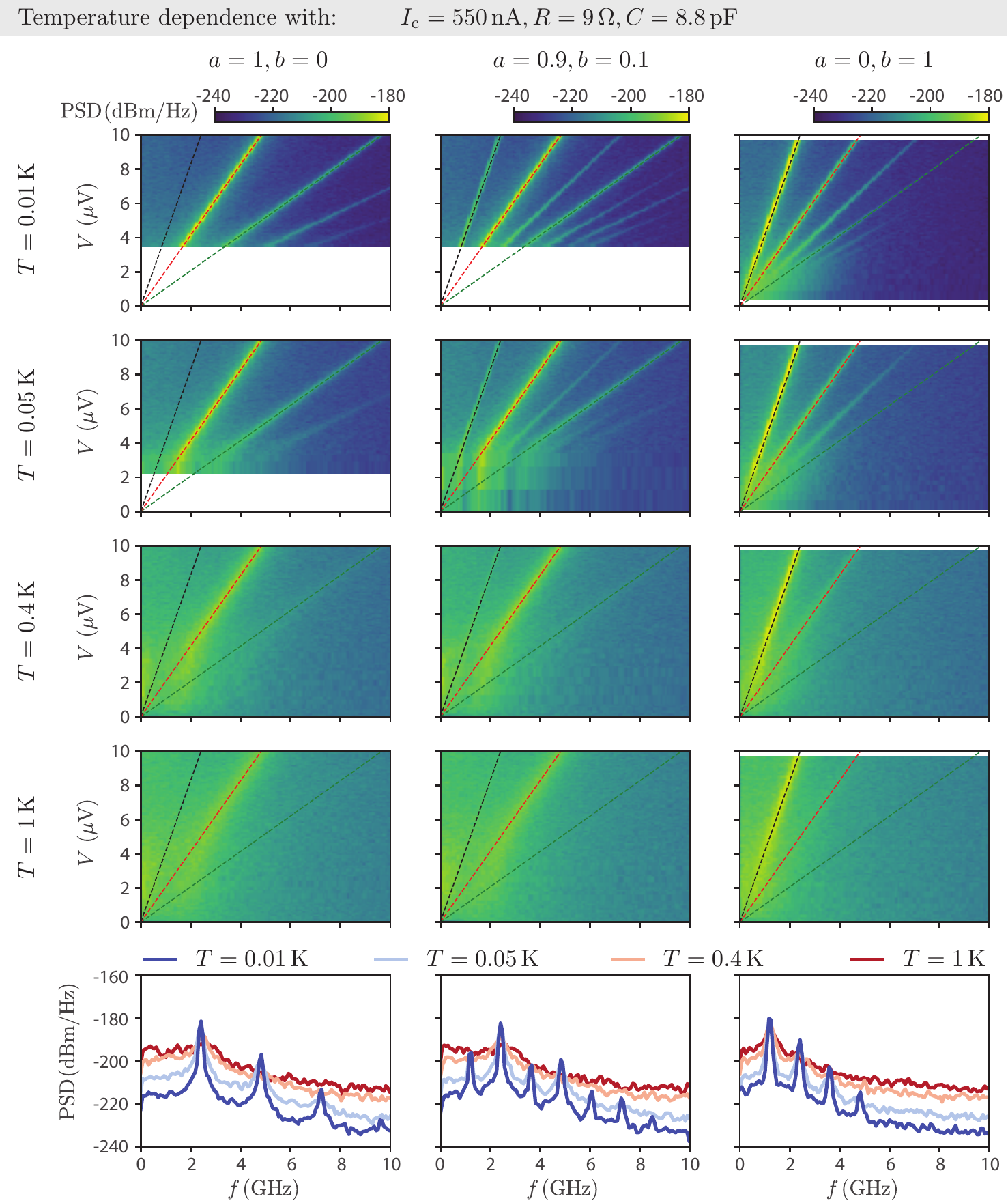}
\caption{\label{fig:cdas_Tdep}
Simulated Josephson radiation as a function of effective electronic temperature $T$ for different combinations of trivial ($a$) and topological ($b$) supercurrent contributions in the current-phase relation for the circuit parameters given in the header.
Radiation maps: Power spectral density (PSD) as a function of mean voltage $V$ and frequency $f$.
The bottom row shows cuts through the different maps at $V=5\,\mu$V.
Guides for the eye: topological feature $V=hf/e$ (black dashed), trivial feature $V=hf/(2e)$ (red dashed), higher-order feature $V=hf/(4e)$ (green dashed).
}
\end{figure}

\begin{figure}%[t!]
\centering
\includegraphics[width=15cm]{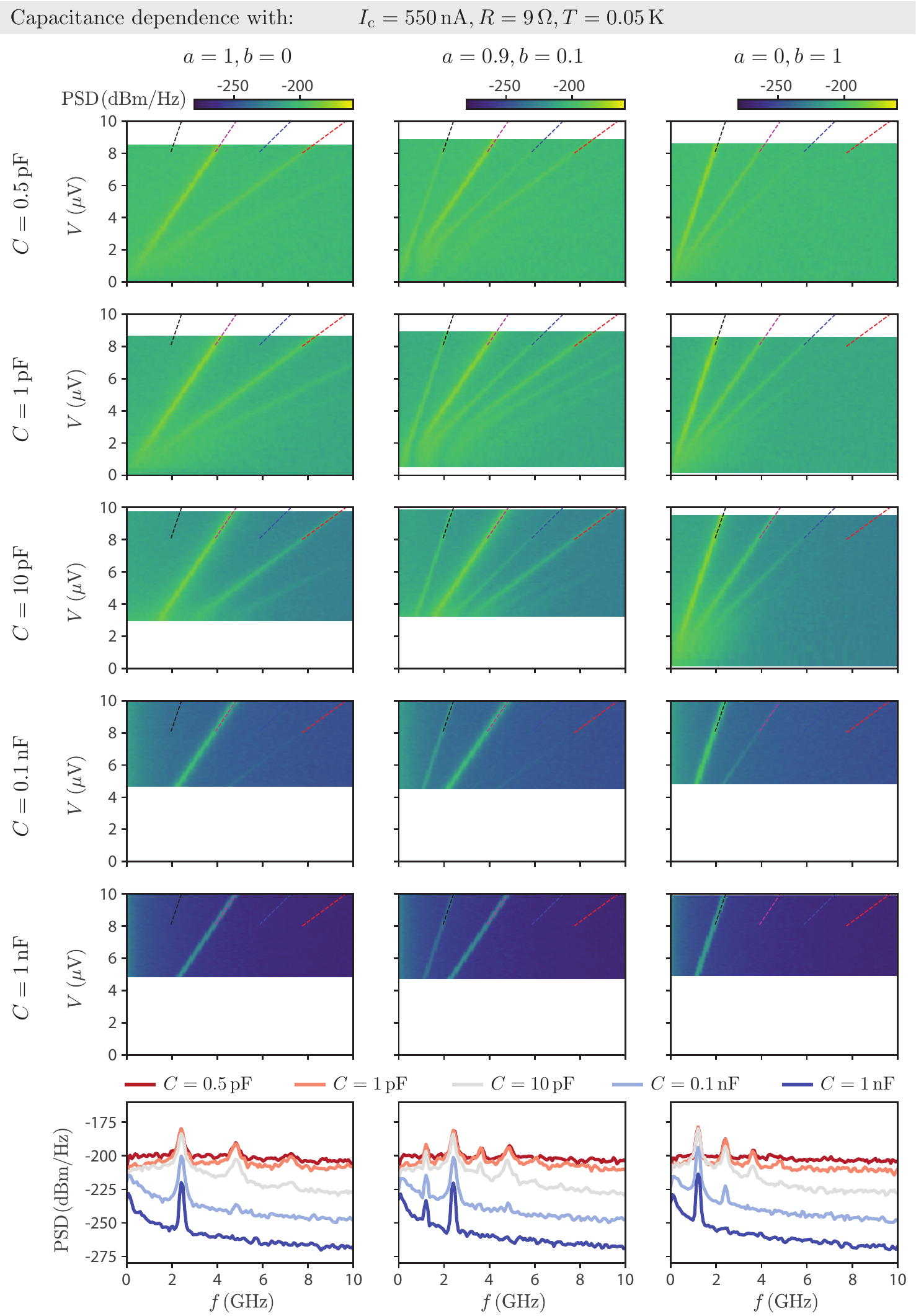}
\caption{\label{fig:cdas_Cdep}
Simulated Josephson radiation as a function of shunt capacitance $C$ for different combinations of trivial ($a$) and topological ($b$) supercurrent contributions in the current-phase relation for the circuit parameters given in the header.
Radiation maps: Power spectral density (PSD) as a function of mean voltage $V$ and frequency $f$.
The bottom row shows cuts through the different maps at $V=5\,\mu$V.
Guides for the eye: topological feature $V=hf/e$ (black dashed), trivial feature $V=hf/(2e)$ (violet dashed), higher-order topological feature $V=hf/(3e)$ (blue dashed), higher-order feature $V=hf/(4e)$ (red dashed).
}
\end{figure}

Having the model established we now explore parts of the parameter space, namely, we consider different compositions of the current-phase relation determined by the normalized amplitudes $a$ and $b$ describing a conventional $a\sin(\varphi)$ component added to a topologically non-trivial $b\sin(\varphi/2)$ component. 
In Fig.\,\ref{fig:cdas_Tdep} and Fig.\,\ref{fig:cdas_Cdep} we display the radiation spectra resulting from the simulation for fixed $I_{\mathrm{c}}$ and $R$. 
The blank or blurry regions at low voltages are due to a sudden jump or transition in the $IV$-characteristics for the specific parameter. 
In Fig.\,\ref{fig:cdas_Tdep} we investigate the Josephson emission as a function of effective electronic temperature $T$ at fixed $C$, and in Fig.\,\ref{fig:cdas_Cdep} we investigate the dependence on the shunt capacitance $C$ at fixed $T$.
In the purely conventional case ($a=1$ and $b=0$), the lines of strong radiation follow the Josephson relation $V=hf/(n\cdot 2e)$, where $n=1,2,3,\dots$, whereas in the purely topological case ($a=0$ and $b=1$), we clearly observe fractions in the radiation pattern following the relation $V=hf/(n\cdot e)$. 
In both cases, the $n=1$ line is the most dominant, and higher-order lines weaken with increasing $n$. 
In the mixed case ($a\neq0$ and $b\neq0$), both sets of lines appear, with weights defined by the parameters $a$ and $b$.

In the temperature dependence (Fig.\,\ref{fig:cdas_Tdep}), we note an increasing noise floor for increasing temperatures. 
Eventually, small radiative features are hidden in the noise floor, which explains the transition from a rich emission spectrum at low temperature to a spectrum in which only the single, most dominant line is discernible at elevated temperature. 

In the capacitance dependence (Fig.\,\ref{fig:cdas_Cdep}), we observe a suppression of high-order signatures for a larger shunt capacitance. 
In addition, we note a reduction of the noise floor and the overall signal. 
Hence, although the radiation spectrum becomes cleaner for larger $C$, the emitted signal eventually becomes too small to be detected in a standard amplification chain, which could be a motivation to employ quantum-limited parametric amplifiers in a Josephson radiation setup. 

\clearpage
%%%%%%%%%%%%%%%%%%%%%%%%%%

\section{\label{sec:methodes}Fabrication}
Cd$_3$As$_2$ nanowires were grown by K.\,Park with the vapor transport method~\cite{S_Minkyung2018,S_Jeon2014a,S_Kidong2020,S_Li2018b,S_Li2015giant,S_Zhang2015}.
The nanowires are transferred from the growth batch with a cleanroom wipe snippet onto a highly $p$-doped Si/SiO$_2$ (500\,$\mu$m/305\,nm) wafer with pre-patterned Au markers and bonding pads.
After removing the native oxide of the nanowires with Ar-milling, the wire is in-situ contacted by Ti/Al (3\,nm/200\,nm) electrodes in a quasi-four-probe configuration as seen in Fig.\,\ref{fig:cdas_device}(a), in the main text.
The wire has a diameter of 50\,nm and the junction length is 150\,nm. 
The Al leads extend from the nanowire up to the inner border of the base structures. 
We cover this inner region, indicated by the green box in Fig.\,\ref{fig:cdas_device}(c) in the main text, with a 20\,nm thick locally deposited HfO$_2$ gate-dielectric. 
A subsequently deposited Au top gate allows tuning of the charge carrier density.

\section{\label{sec:setup}Measurement setup}
Both Josephson radiation and Shapiro step measurements are performed in a Triton 200 dilution refrigerator, of which the experimental setup is similar to Ref.\,\cite{S_Deaconb}.

\begin{figure}[h!]
\centering
\includegraphics[width=10cm]{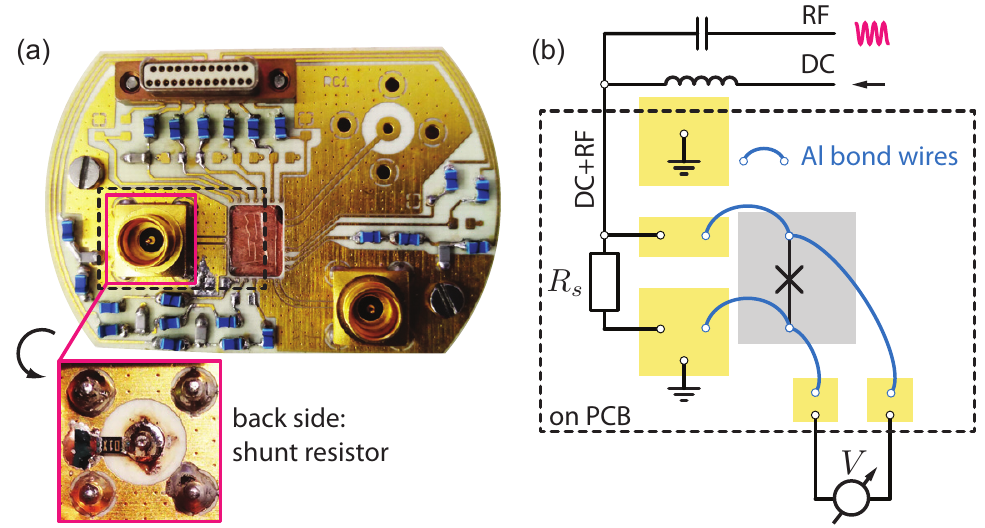}
\caption{\label{fig:cdas_PCB}
PCB for Josephson radiation and Shapiro step measurements. 
(a) Optical image of the PCB with a length of 4\,cm and width of 2.5\,cm. 
The metal film shunt resistor ($R_s=10$\,$\Omega$ from Welwyn: PCD0603R-10RBT1) is soldered between the central pin and the ground plane on the back side of the rf SMP connector. 
(b) Representation of the bonding scheme to link the Josephson junction device to the measurement setup.  
}
\end{figure}

We start our description with the PCB shown in Fig.\,\ref{fig:cdas_PCB}(a) that hosts co-planar transmission lines with SMP connectors and dc lines connecting to a nano-D adapter.
The in-house designed double-sided Ni/Au plated Rogers$^\copyright$ 4350 PCB is mounted onto a copper plate.
A $10\,\Omega$ metal film resistor is soldered on the back side of the PCB in between the ground plane and the central conductor of the SMP connector.
The hole in the PCB exposes the bare copper backplate to which we mount the sample close to the co-planar transmission line.
The SMP central conductor is bonded to one side of the junction, whereas the other side is bonded to the PCB ground resulting in a resistively shunted junction configuration.
Both wire bonds are made as short as possible. 
Additional bond wires connect to the differential voltage measure and gate voltage supplies (not shown).

\begin{figure}[h!]
\centering
\includegraphics[width=10cm]{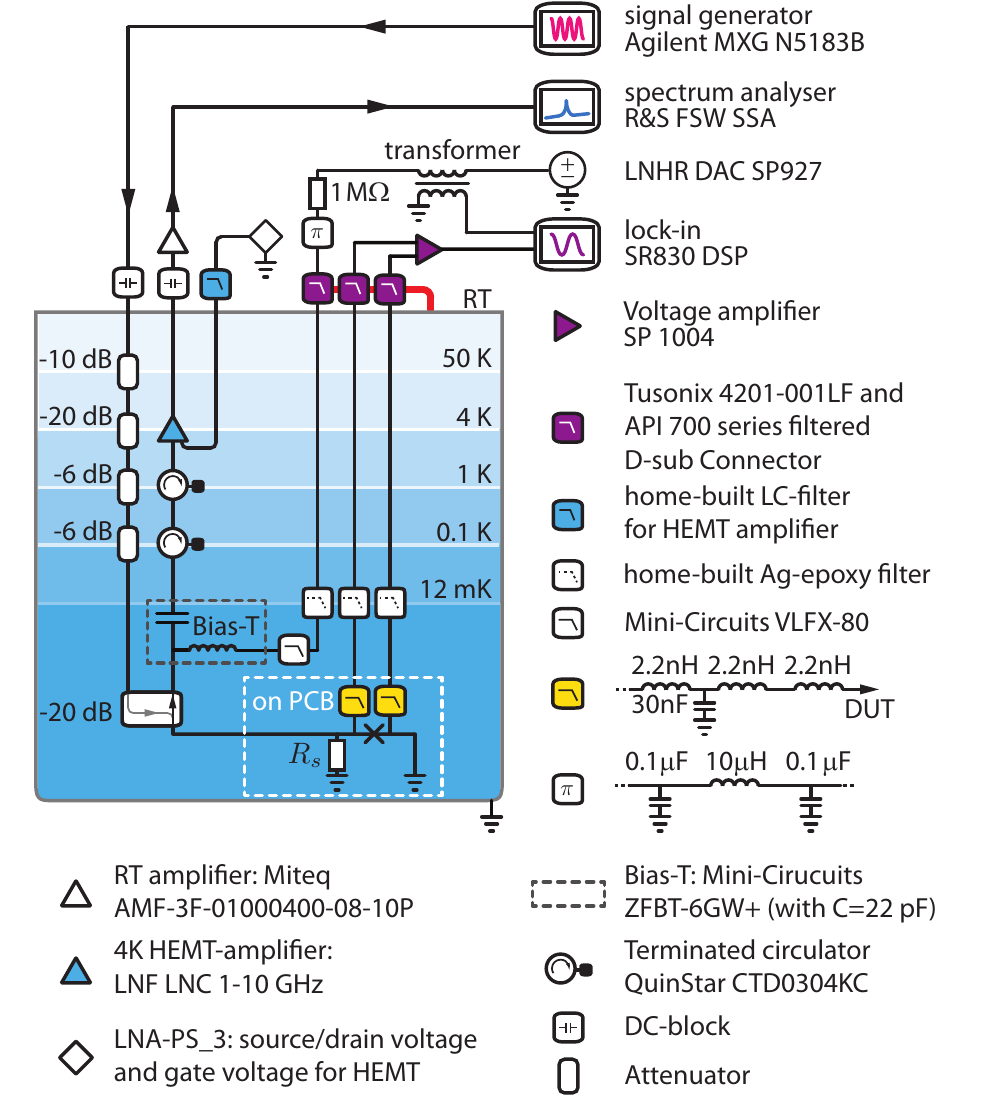}
\caption{\label{fig:cdas_setup}
Detailed overview of a hybrid Josephson radiation and Shapiro step measurement setup.}
\end{figure}

After inserting the PCB, the device connects to the full measurement setup presented in Fig.\,\ref{fig:cdas_setup}.
The dc supply lines are filtered at room temperature, on the PCB, and are thermalized to the mixing chamber plate via silver-epoxy filters that provide a cut-off of $\sim6$\,MHz~\cite{S_hasler2016microwave,S_scheller2014silver}. 
We use home-built $LC$-$\pi$-filters ($L=10\,\mu$H, $C=100$\,nF) to minimize the noise on the dc supply for the HEMT amplifier~\cite{S_haller2021probing}.
The Fisher cable connecting the break-out box to the cryostat looms and the cable connecting the 4K HEMT amplifier with its power supply are surrounded by several ferrites. 
The break-out box shield is linked to the cryostat ground with a massive copper mesh cable, which is kept as short as possible (indicated with the red line).
A current bias is generated by a $1$\,M$\Omega$ resistor in series with a voltage source. 
The current couples via a bias-tee to the rf line that connects through the device to ground. 
The differential voltage drop across the junction is measured with a voltage amplifier and lock-in techniques. 
In the superconducting regime, most of the current flows through the junction, and once the junction switches to the normal state, most of the current will flow through the shunt resistor where the exact current division depends on the ratio between the junction normal state resistance $R_N$ and the shunt resistance $R_s$. 
Since $R_N \gg R_s$ as discussed in Sec.\,\ref{sec:icrn}, the $1$\,M$\Omega$ source and the $10\,\Omega$ shunt resistor act as a voltage divider and provide a stable voltage drop across the junction.

The radiation generated by the junction is amplified using a cryogenic HEMT amplifier at the $4$~K stage, a room temperature amplifier, and finally measured with a spectrum analyzer.
Two circulators at the HEMT input limit back-action noise, but limit the detection bandwidth to $2.5-3.8$\,GHz.
The following parameters on the spectrum analyzer were used: detection bandwidth $20$\,MHz, span $24$\,MHz, radio bandwidth $10$\,MHz, video bandwidth $5$\,MHz and 1001 points resulting in a sweep time of $2$\,s over the detection frequency range. 
In addition to the sensing line, a drive line links to the device via a directional coupler used for irradiating the junction with a signal generator in Shapiro step measurements.

\clearpage

\section{\label{sec:transfer}Josephson emission power}
\noindent\textbf{Comparison of power levels in the experiment and simulation}

\begin{figure}[h!]
\centering
\includegraphics[width=11.6cm]{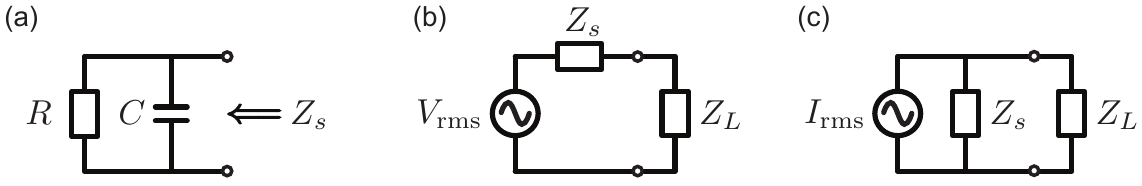}
\caption{\label{fig:cdas_power}
(a) Impedance $Z_s$ of a parallel $RC$-circuit modeling the source impedance of the Josephson junction.
(b) AC voltage source with root-mean-square voltage $V_{\mathrm{rms}}$ and impedance $Z_s$ connected to a load impedance $Z_L$.
(c) AC current source with root-mean-square current $I_{\mathrm{rms}}$ and impedance $Z_s$ connected to a load impedance $Z_L$.
}
\end{figure}

In this section, we discuss the difference in absolute power level between experiment and simulation. For this purpose, we calculate the total radiation power $P$ by integrating the spectral density of the conventional radiation peak defined with the relation $V=(h/2e)f$, for different values of the critical current.
The experimental relation $P_{\mathrm{det}}(I_{\mathrm{c}}^2)$ is fitted with a line going through the origin, and results in a slope of $s_{\mathrm{exp}}^V=1.425\times10^{4}\,\Omega$ as shown in gray dashed in Fig.\,\ref{fig:cdas_radiation} in the main text.
The simulation allows us to carry out the integration of the peak power over frequency and voltage as well, and the subsequent linear fit results in $s_{\mathrm{sim}}^V=0.227\,\Omega$ for the  integration over voltage (red dashed), and $s_{\mathrm{sim}}^f=0.205\,\Omega$ for the integration over frequency (blue dashed) as shown in Fig.\,\ref{fig:cdas_sim} in the main text. 
To be consistent with the evaluation of the experimental data, we calculate in the following with $s_{\mathrm{sim}}^V$. 
In the main text, we evaluated the modeled radiation power with $P={V_{\mathrm{rms}}^2}/{\left|Z_L\right|}$, by assuming the junction to be an ideal voltage source. However, taking into account a finite output impedance $Z_s$ as illustrated in the circuit shown in Fig.\,\ref{fig:cdas_power}(b), the power dissipated in the load $Z_L$ is $P_L=V_{\mathrm{rms}}^2 {\left|Z_L\right|}/{\left|Z_s+Z_L\right|^2}$. This yields a correction factor of $t_P = P_L/P = \left|Z_L/(Z_L + Z_s)\right|^2$. We model the junction source impedance with the shunt resistor $R$ in parallel with the shunt capacitor $C$ as shown in Fig.\,\ref{fig:cdas_power}(a), resulting in $Z_s = R Z_c/(R+Z_c)$, where $Z_c=1/(i 2 \pi f C)$ . The load impedance is set to $Z_L=50\,\Omega$ according to the input impedance of the spectrum analyzer. At $f=3$\,GHz, with $C=8.8$\,pF and $R=9\,\Omega$, we get $t_P \approx 0.83$.
The ratio $s_{\mathrm{exp}}^V/(t_P s_{\mathrm{sim}}^V)$ evaluates the power translation between the experiment and the simulation, which reads in dB-scale: 
\begin{equation}
\label{eq:rad:gain}
g_{\mathrm{exp}}=10\cdot\log_{10}\left(s_{\mathrm{exp}}^V/(t_P s_{\mathrm{sim}}^V)\right)=48.8\,{\mathrm{dB}}.
\end{equation}
Next, we compare this amplification to the gain of the amplification chain present in the experimental setup (see Fig.\ref{fig:cdas_setup}). By considering the gain of the HEMT amplifier ($+27$\,dB) and the room-temperature amplifier ($+39$\,dB), as well as the attenuation of the measurement line ($-13$\,dB), we estimate a total effective gain of
\begin{equation}
\label{eq:rad:gain}
g_{\mathrm{eff}}\approx 51\,{\mathrm{dB}},
\end{equation}
which gives us a reasonable agreement between the power level in the experiment and the simulation.
After the experiment has been conducted, it turned out that the supply lines to the HEMT-amplifier were high-ohmic, such that its ideal working point with the highest gain (nominally $+40$\,dB) and lowest thermal noise could not be reached.
The small discrepancy between $g_{\mathrm{eff}}$ and $g_{\mathrm{exp}}$ can be attributed to uncertainties in the experimental setup. 

\vspace{0.5cm}
\noindent\textbf{Interpretation of the $P(I_{\mathrm{c}}^2)$-relation}

We explain the quadratic dependence between $P_\mathrm{det}$ and $I_{\mathrm{c}}$ as follows. We model the junction as a sinusoidal current source, with a current amplitude $I_{\mathrm{c}}$ and source impedance $Z_s$ as illustrated in Fig.\,\ref{fig:cdas_power}(c). Here, $Z_s$ is as well $R$ in parallel with $C$. Since $I_\mathrm{rms}=I_{\mathrm{c}}/\sqrt{2}$, the power transferred to the load $Z_L$ is 
\begin{equation}
\label{eq:rad:mismatch}
P_L=\frac{1}{2}\left|\frac{Z_s}{Z_s+Z_L}\right|^2 \left|Z_L\right| \cdot I_{\mathrm{c}}^2. 
\end{equation}
With $C=8.8$\,pF, $R=9\,\Omega$ and $Z_L=50\,\Omega$, we obtain at $f=3$\,GHz, $P_L\approx0.224\,\Omega\cdot I_{\mathrm{c}}^2$, which resembles the $P(I_{\mathrm{c}}^2)$-relation extracted from the model.
In the experimental configuration, the transferred power is amplified with a gain of $g$, and the detected power is $P_\mathrm{det} = g P_L = \mathcal{R} I_{\mathrm{c}}^2/2$. We note that formula \ref{eq:rad:mismatch} should be considered when designing experiments. For a high power transfer, $Z_s \gg Z_L$ is preferred, however, for stable biasing in the finite detection window, the junction should not be in the hysteretic regime. The trade-off between the sufficient damping of the junction and the optimization of the power transfer could be balanced with an in-situ variable shunt resistance.

\section{\label{sec:icrn}$I_{\mathrm{c}}R_N$ product}
\begin{figure}[h!]
\centering
\includegraphics[width=10cm]{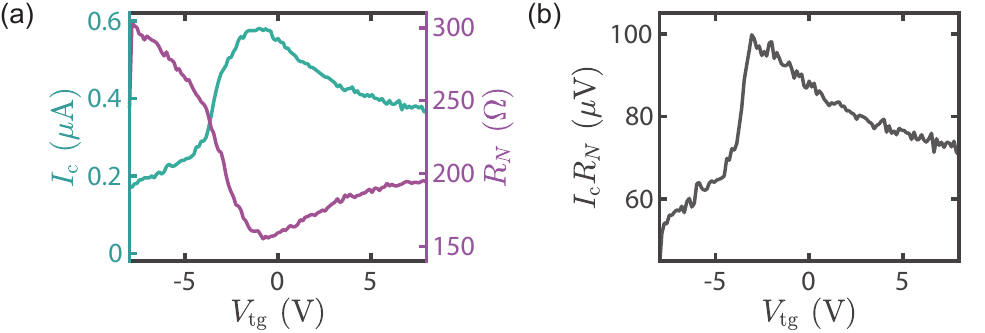}
\caption{\label{fig:cdas_IcRn}
Gate dependent critical current $I_{\mathrm{c}}$ and normal state resistance $R_N$. 
(a) Left axis (green) $I_{\mathrm{c}}$ and right axis (violet) $R_N$ as a function of gate voltage $V_{\mathrm{tg}}$. 
(b) $I_{\mathrm{c}}R_N$-product as a function of $V_{\mathrm{tg}}$.
}
\end{figure}

We obtained the normal state resistance $R_N$ from measurements carried out in a magnetic field of 20\,mT, above the critical field of the junction. We use a 100 nA ac excitation on the lock-in amplifier and correct for the parallel shunt resistor $R_s$. In Fig.\,\ref{fig:cdas_IcRn}(a) we initially observe the expected inverse correlation between $I_{\mathrm{c}}$ and $R_N$. However, when looking at the $I_{\mathrm{c}}R_N$ product in Fig.\,\ref{fig:cdas_IcRn}(b), we observe an $I_{\mathrm{c}}R_N\sim50\,\mu$V in the far negatively doped regime up to $\sim100\,\mu$V where $I_{\mathrm{c}}$ is at its maximum value. For more positive doping $I_{\mathrm{c}}R_N$ again decreases with increasing $R_N$. 
Both the peak in $I_{\mathrm{c}}$ and $I_{\mathrm{c}}R_N$ close to $V_\mathrm{tg}=-2$\,V (i.e., the Dirac point), hint towards supercurrent carried by the surface states, and an increase of bulk-surface interaction when the doping level is increased away from the Dirac point~\cite{S_Li2018b}.

\section{\label{sec:shapiro_eva}Shapiro steps evaluation}
In this section, we compare simulated Shapiro step patterns with measurement data, and use parameters that describe the modification of the pattern due to the presence of a topological supercurrent.
In a conventional Josephson junction with a $2\pi$-periodic CPR, steps at the Shapiro voltages $V_k=k\frac{h}{2e}f_{\mathrm{d}}$ appear. 
Here, $f_{\mathrm{d}}$ is the drive frequency and $k$ is an integer that represents the number of $2\pi$ phase evolutions that the particle undergoes in the tilted washboard potential. 
For topological supercurrents, a $4\pi$-periodic CPR results in a $4\pi$ phase evolution and thus results in Shapiro voltages of height $V_k=k\frac{h}{e}f_{\mathrm{d}}$, double compared to the trivial Shapiro voltage.
In Shapiro step patterns, the presence of topological supercurrent therefore effectively suppresses the relative current magnitude associated with odd steps with $2\pi$ supercurrent periodicity, in favor of the even current steps which entail both $2\pi$ and $4\pi$ contributions~\cite{S_Park2021}.

\begin{figure}[b!]
\centering
\includegraphics[width=17.2cm]{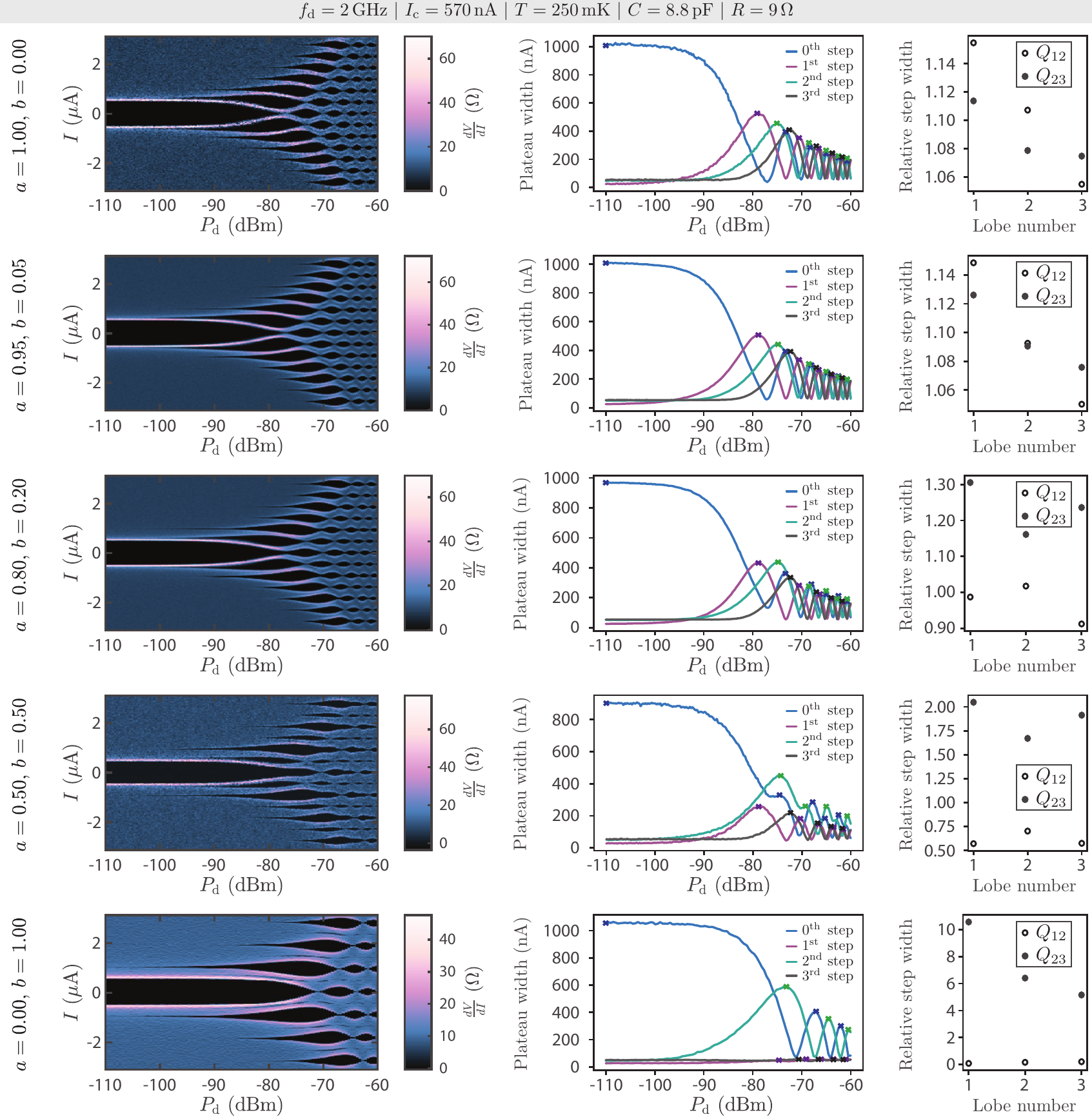}
\caption{\label{fig:cdas_shaprioana_sim}
Simulated Shapiro patterns and current plateau widths for various trivial ($a$) and topological ($b$) supercurrent contributions. Circuit parameters used for the circuit parameters are given in the header.
Left column: Shapiro maps showing differential resistance $dV/dI$ as a function of drive power $P_{\mathrm{d}}$ and current bias $I$.
Middle column: Evolution of the plateau widths as a function of drive power $P_{\mathrm{d}}$, corresponding to the left column for the first four steps. The maxima of the plateaus are indicated with crosses. 
Right column: Relative step widths for the first three lobes. $Q_{nm}$ describes the ratio between the maximal current plateau widths of step $n$ divided by the ones of $m=n+1$.
}
\end{figure}

Fig.\,\ref{fig:cdas_shaprioana_sim}, left column, shows simulated data of Shapiro steps as a function of the microwave power $P_{\mathrm{d}}$ and current bias $I$, while the middle column shows the accompanying dc current plateau widths for which the constant Shapiro voltages appear. 
Starting from a pure conventional current ($a=1-b=1$, $b=0$), an increased ratio of conventional to topological current (see $a$ and $b$ values) is plotted when moving down vertically. The maximum step width is quantified by the lobe maxima (colored crosses) and we show a reduction of the step width for odd $k$ for increasing $b$, which is a direct consequence of the different admixture of $2\pi$ and $4\pi$ CPR components. 
The odd steps ($n=1,3$) almost vanish when only topological current is present ($a=0$, $b=1$), but a small step due to higher order modes remains. 
We quantify the change in the even to odd ratios by introducing the relative plateau width parameter $Q_{nm}=I_{\mathrm{max,}n}/I_{\mathrm{max,}m}$, defined as the ratio between the maximal current plateau width $I_{\mathrm{max,}n}$ of step $n$, divided by the current plateau with of $I_{\mathrm{max,}m}$ of $m=n+1$. 
These ratios are plotted for the first three lobes.

The $Q_{12}$ and $Q_{23}$ ratios for the conventional case ($a=1$, $b=0$), are monotonically decreasing for each successive lobe, while $Q_{12}$ is always larger than $Q_{23}$. 
Comparing this case $a=0.95$ and $b=0.05$, the small fraction of topological current moves $Q_{12}$ and $Q_{23}$ value closer together (the second lobe values almost overlap), indicating that the odd $n=1$ step decreases in relative magnitude. 
For $b=0.2$, $Q_{12}$ is smaller than $Q_{23}$ for all lobes, a trend that continues until for $b=1$, the $n=1$ step almost completely disappears, reflected by the high value of $Q_{23}$. 

\begin{figure}%[H!]
\centering
\includegraphics[width=17.2cm]{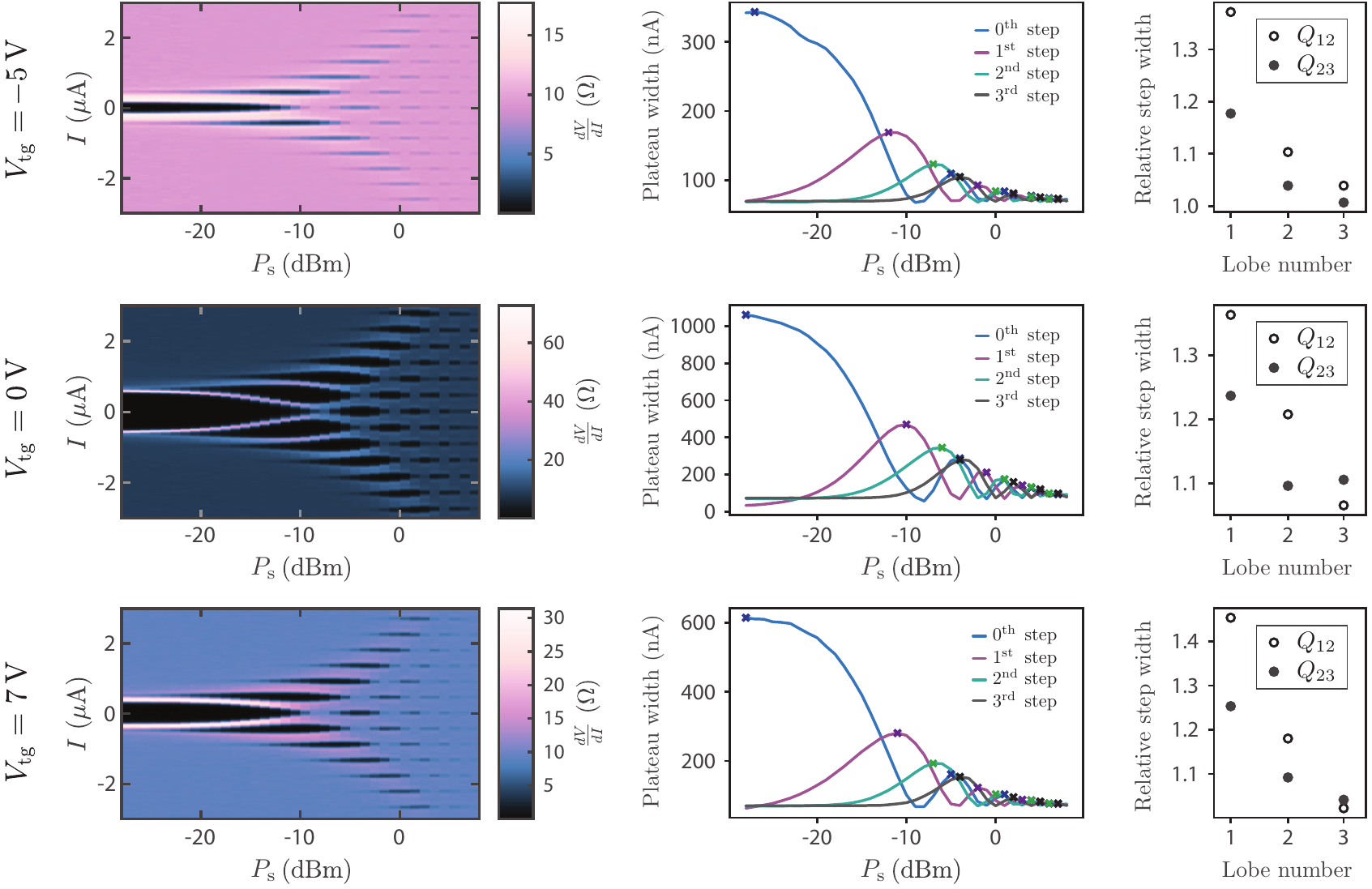}
\caption{\label{fig:cdas_shaprioana}
Measured Shapiro patterns with drive frequency $f_{\mathrm{d}}=2$\,GHz and current plateau widths for three gate voltages.
Left column: Shapiro maps showing differential resistance $dV/dI$ as a function of drive power $P_{\mathrm{s}}$ at the microwave source and current bias $I$.
Middle column:  Evolution of the plateau widths as a function of drive power $P_{\mathrm{s}}$, corresponding to the left column for the first four steps. The maxima of the plateaus are indicated with crosses.
Right column: Relative step widths for the first three lobes. $Q_{nm}$ describes the ratio between the maximal current plateau widths of step $n$ divided by the ones of $m=n+1$.
}
\end{figure}

We now compare our results from the simulation with experimental results shown in Fig.\,\ref{fig:cdas_shaprioana}. 
Again, the left column shows Shapiro steps as a function of $P_{\mathrm{s}}$ and $I$, the middle column shows the extracted step heights in $I$, while the right column shows the associated $Q_{12}$ and $Q_{23}$ values. 
From top to bottom, we move from hole doping at $V_\mathrm{tg}=-5$\,V, to close to the Dirac point at $V_\mathrm{tg}=0$\,V, and finally to electron doping at $V_\mathrm{tg}=7$\,V. 
Although small changes in the absolute and relative $Q_{12}$ and $Q_{23}$ can be distinguished, no clear evidence of suppression of the first $n=1$ step can be observed. 
An important note is that the junction parameters used in the simulations in Fig.\,\ref{fig:cdas_shaprioana} are fitted to match the measurement data for $V_\mathrm{bg}=0$\,V. 
The lower $I_\mathrm{c}$ for $V_\mathrm{bg}=-5$\,V and $V_\mathrm{bg}=7$\,V changes the dynamics of the junction which, in turn, modifies the plateau widths and the $Q_{12}$ and $Q_{23}$ values, on top of any potential modification due to a topological supercurrent contribution. 
To establish signatures of topological supercurrent, one therefore desires to measure Shapiro patterns over a wide frequency range, while keeping the junction parameters constant~\cite{S_Park2021,S_Jang2021}. 

\clearpage

\begin{figure}%[H!]
\centering
\includegraphics[width=11cm]{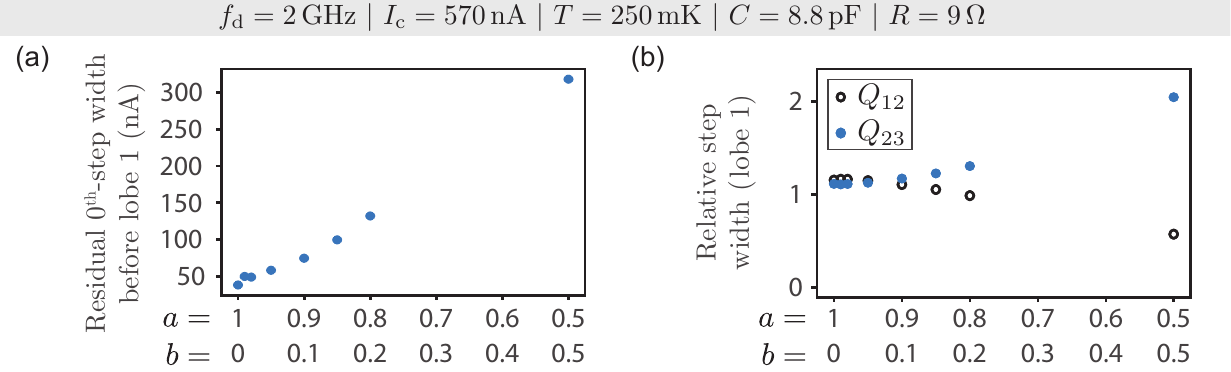}
\caption{\label{fig:cdas_shaprioana_sim_summary}
Simulated topological indicators when sweeping the $a$ to $b$ ratio for the circuit parameters given in the header.
(a) Minimum of the current plateau width between the first and second lobe for the $0^{\mathrm{th}}$-step (blue).
(b) Relative step widths for the first lobes. $Q_{nm}$ describes the ratio between the maximal current plateau widths of step $n$ divided by the ones of $m=n+1$.
}
\end{figure} 

\vspace*{-2em}

We additionally perform a simulated parameter sweep of $Q_{12}$ and $Q_{23}$ for the first lobe as a function of $a$ and $b$ in Fig.\,\ref{fig:cdas_shaprioana_sim_summary}(b). 
The same trend is observed: for an increasing topological current contribution, $Q_{12}$ increases while $Q_{23}$ decreases. 
Although this trend is visible even for small topological current fractions of $b\sim0.05$, the crossing point where $Q_{12}=Q_{23}$ at $b\sim0.1$ can be used as a clear indicator in an experiment for a topological supercurrent contribution.
Finally, we look at another indicator for a topological current contribution shown in Fig.\,\ref{fig:cdas_shaprioana_sim_summary}(a), namely the minimum, e.g., depression of the current plateau width between the first and second lobe of the $0^\mathrm{th}$ step~\cite{S_Wang2018c}. 
This supplementary indicator is a stronger function for increasing $b$ than the $Q$ parameters, which might give indications for even smaller topological supercurrents. However, more in-depth research is required to investigate this indicator as a function of the junction parameters.
The scripts used to perform the Shapiro step width analysis and the generation of the plots can be found in Ref.\,\footnote{See the Python script in folder '\texttt{shapiro\_evaluation}' in the Zenodo repository~\href{https://doi.org/10.5281/zenodo.7961884}{10.5281/zenodo.7961884} to extract the current plateau widths from Shapiro maps.}.

\clearpage
%=======================================================================================================================================
% BIBLIOGRAPHY
%=======================================================================================================================================
\input{AC_Josephson_CdAs_SM_2.bbl}% Produces the bibliography via BibTeX.

\end{document}

%% file: AC_Josephson_CdAs_main_2.bbl

%% file: AC_Josephson_CdAs_SM_2.bbl

%% file: AC_Josephson_CdAs_combined_2.bbl
\begin{thebibliography}{61}%
\makeatletter
\providecommand \@ifxundefined [1]{%
 \@ifx{#1\undefined}
}%
\providecommand \@ifnum [1]{%
 \ifnum #1\expandafter \@firstoftwo
 \else \expandafter \@secondoftwo
 \fi
}%
\providecommand \@ifx [1]{%
 \ifx #1\expandafter \@firstoftwo
 \else \expandafter \@secondoftwo
 \fi
}%
\providecommand \natexlab [1]{#1}%
\providecommand \enquote  [1]{``#1''}%
\providecommand \bibnamefont  [1]{#1}%
\providecommand \bibfnamefont [1]{#1}%
\providecommand \citenamefont [1]{#1}%
\providecommand \href@noop [0]{\@secondoftwo}%
\providecommand \href [0]{\begingroup \@sanitize@url \@href}%
\providecommand \@href[1]{\@@startlink{#1}\@@href}%
\providecommand \@@href[1]{\endgroup#1\@@endlink}%
\providecommand \@sanitize@url [0]{\catcode `\\12\catcode `\$12\catcode
  `\&12\catcode `\#12\catcode `\^12\catcode `\_12\catcode `\%12\relax}%
\providecommand \@@startlink[1]{}%
\providecommand \@@endlink[0]{}%
\providecommand \url  [0]{\begingroup\@sanitize@url \@url }%
\providecommand \@url [1]{\endgroup\@href {#1}{\urlprefix }}%
\providecommand \urlprefix  [0]{URL }%
\providecommand \Eprint [0]{\href }%
\providecommand \doibase [0]{https://doi.org/}%
\providecommand \selectlanguage [0]{\@gobble}%
\providecommand \bibinfo  [0]{\@secondoftwo}%
\providecommand \bibfield  [0]{\@secondoftwo}%
\providecommand \translation [1]{[#1]}%
\providecommand \BibitemOpen [0]{}%
\providecommand \bibitemStop [0]{}%
\providecommand \bibitemNoStop [0]{.\EOS\space}%
\providecommand \EOS [0]{\spacefactor3000\relax}%
\providecommand \BibitemShut  [1]{\csname bibitem#1\endcsname}%
\let\auto@bib@innerbib\@empty
%</preamble>
\bibitem [{\citenamefont {Kitaev}(2001)}]{Kitaev2001}%
  \BibitemOpen
  \bibfield  {author} {\bibinfo {author} {\bibfnamefont {A.~Y.}\ \bibnamefont
  {Kitaev}},\ }\bibfield  {title} {\bibinfo {title} {{Unpaired Majorana
  fermions in quantum wires}},\ }\href
  {https://doi.org/10.1070/1063-7869/44/10s/s29} {\bibfield  {journal}
  {\bibinfo  {journal} {Physics-Uspekhi}\ }\textbf {\bibinfo {volume} {44}},\
  \bibinfo {pages} {131} (\bibinfo {year} {2001})}\BibitemShut {NoStop}%
\bibitem [{\citenamefont {Kitaev}(2003)}]{KITAEV20032}%
  \BibitemOpen
  \bibfield  {author} {\bibinfo {author} {\bibfnamefont {A.}~\bibnamefont
  {Kitaev}},\ }\bibfield  {title} {\bibinfo {title} {{Fault-tolerant quantum
  computation by anyons}},\ }\href
  {https://doi.org/https://doi.org/10.1016/S0003-4916(02)00018-0} {\bibfield
  {journal} {\bibinfo  {journal} {Annals of Physics}\ }\textbf {\bibinfo
  {volume} {303}},\ \bibinfo {pages} {2} (\bibinfo {year} {2003})}\BibitemShut
  {NoStop}%
\bibitem [{\citenamefont {Ivanov}(2001)}]{Ivanov2001}%
  \BibitemOpen
  \bibfield  {author} {\bibinfo {author} {\bibfnamefont {D.~A.}\ \bibnamefont
  {Ivanov}},\ }\bibfield  {title} {\bibinfo {title} {{Non-Abelian Statistics of
  Half-Quantum Vortices in $\mathit{p}$-Wave Superconductors}},\ }\href
  {https://doi.org/10.1103/PhysRevLett.86.268} {\bibfield  {journal} {\bibinfo
  {journal} {Phys. Rev. Lett.}\ }\textbf {\bibinfo {volume} {86}},\ \bibinfo
  {pages} {268} (\bibinfo {year} {2001})}\BibitemShut {NoStop}%
\bibitem [{\citenamefont {Nayak}\ \emph {et~al.}(2008)\citenamefont {Nayak},
  \citenamefont {Simon}, \citenamefont {Stern}, \citenamefont {Freedman},\ and\
  \citenamefont {Das~Sarma}}]{Nayak2008}%
  \BibitemOpen
  \bibfield  {author} {\bibinfo {author} {\bibfnamefont {C.}~\bibnamefont
  {Nayak}}, \bibinfo {author} {\bibfnamefont {S.~H.}\ \bibnamefont {Simon}},
  \bibinfo {author} {\bibfnamefont {A.}~\bibnamefont {Stern}}, \bibinfo
  {author} {\bibfnamefont {M.}~\bibnamefont {Freedman}},\ and\ \bibinfo
  {author} {\bibfnamefont {S.}~\bibnamefont {Das~Sarma}},\ }\bibfield  {title}
  {\bibinfo {title} {{Non-Abelian anyons and topological quantum
  computation}},\ }\href {https://doi.org/10.1103/RevModPhys.80.1083}
  {\bibfield  {journal} {\bibinfo  {journal} {Rev. Mod. Phys.}\ }\textbf
  {\bibinfo {volume} {80}},\ \bibinfo {pages} {1083} (\bibinfo {year}
  {2008})}\BibitemShut {NoStop}%
\bibitem [{\citenamefont {Wilczek}(2009)}]{Wilczek2009}%
  \BibitemOpen
  \bibfield  {author} {\bibinfo {author} {\bibfnamefont {F.}~\bibnamefont
  {Wilczek}},\ }\bibfield  {title} {\bibinfo {title} {{Majorana returns}},\
  }\href {https://doi.org/10.1038/nphys1380} {\bibfield  {journal} {\bibinfo
  {journal} {Nat. Phys.}\ }\textbf {\bibinfo {volume} {5}},\ \bibinfo {pages}
  {614} (\bibinfo {year} {2009})}\BibitemShut {NoStop}%
\bibitem [{\citenamefont {Stern}\ and\ \citenamefont
  {Lindner}(2013)}]{Stern2013}%
  \BibitemOpen
  \bibfield  {author} {\bibinfo {author} {\bibfnamefont {A.}~\bibnamefont
  {Stern}}\ and\ \bibinfo {author} {\bibfnamefont {N.~H.}\ \bibnamefont
  {Lindner}},\ }\bibfield  {title} {\bibinfo {title} {{Topological Quantum
  Computation -- From Basic Concepts to First Experiments}},\ }\href
  {https://doi.org/10.1126/science.1231473} {\bibfield  {journal} {\bibinfo
  {journal} {Science}\ }\textbf {\bibinfo {volume} {339}},\ \bibinfo {pages}
  {1179} (\bibinfo {year} {2013})}\BibitemShut {NoStop}%
\bibitem [{\citenamefont {Karzig}\ \emph {et~al.}(2017)\citenamefont {Karzig},
  \citenamefont {Knapp}, \citenamefont {Lutchyn}, \citenamefont {Bonderson},
  \citenamefont {Hastings}, \citenamefont {Nayak}, \citenamefont {Alicea},
  \citenamefont {Flensberg}, \citenamefont {Plugge}, \citenamefont {Oreg},
  \citenamefont {Marcus},\ and\ \citenamefont {Freedman}}]{Karzig2017}%
  \BibitemOpen
  \bibfield  {author} {\bibinfo {author} {\bibfnamefont {T.}~\bibnamefont
  {Karzig}}, \bibinfo {author} {\bibfnamefont {C.}~\bibnamefont {Knapp}},
  \bibinfo {author} {\bibfnamefont {R.~M.}\ \bibnamefont {Lutchyn}}, \bibinfo
  {author} {\bibfnamefont {P.}~\bibnamefont {Bonderson}}, \bibinfo {author}
  {\bibfnamefont {M.~B.}\ \bibnamefont {Hastings}}, \bibinfo {author}
  {\bibfnamefont {C.}~\bibnamefont {Nayak}}, \bibinfo {author} {\bibfnamefont
  {J.}~\bibnamefont {Alicea}}, \bibinfo {author} {\bibfnamefont
  {K.}~\bibnamefont {Flensberg}}, \bibinfo {author} {\bibfnamefont
  {S.}~\bibnamefont {Plugge}}, \bibinfo {author} {\bibfnamefont
  {Y.}~\bibnamefont {Oreg}}, \bibinfo {author} {\bibfnamefont {C.~M.}\
  \bibnamefont {Marcus}},\ and\ \bibinfo {author} {\bibfnamefont {M.~H.}\
  \bibnamefont {Freedman}},\ }\bibfield  {title} {\bibinfo {title} {{Scalable
  designs for quasiparticle-poisoning-protected topological quantum computation
  with Majorana zero modes}},\ }\href
  {https://doi.org/10.1103/PhysRevB.95.235305} {\bibfield  {journal} {\bibinfo
  {journal} {Phys. Rev. B}\ }\textbf {\bibinfo {volume} {95}},\ \bibinfo
  {pages} {235305} (\bibinfo {year} {2017})}\BibitemShut {NoStop}%
\bibitem [{\citenamefont {Fu}\ and\ \citenamefont {Kane}(2009)}]{Fu2009}%
  \BibitemOpen
  \bibfield  {author} {\bibinfo {author} {\bibfnamefont {L.}~\bibnamefont
  {Fu}}\ and\ \bibinfo {author} {\bibfnamefont {C.~L.}\ \bibnamefont {Kane}},\
  }\bibfield  {title} {\bibinfo {title} {Josephson current and noise at a
  superconductor/quantum-spin-{Hall}-insulator/super\-conductor junction},\
  }\href {https://doi.org/10.1103/PhysRevB.79.161408} {\bibfield  {journal}
  {\bibinfo  {journal} {Phys. Rev. B}\ }\textbf {\bibinfo {volume} {79}},\
  \bibinfo {pages} {161408(R)} (\bibinfo {year} {2009})}\BibitemShut {NoStop}%
\bibitem [{\citenamefont {Lutchyn}\ \emph {et~al.}(2010)\citenamefont
  {Lutchyn}, \citenamefont {Sau},\ and\ \citenamefont
  {Das~Sarma}}]{Lutchyn2010}%
  \BibitemOpen
  \bibfield  {author} {\bibinfo {author} {\bibfnamefont {R.~M.}\ \bibnamefont
  {Lutchyn}}, \bibinfo {author} {\bibfnamefont {J.~D.}\ \bibnamefont {Sau}},\
  and\ \bibinfo {author} {\bibfnamefont {S.}~\bibnamefont {Das~Sarma}},\
  }\bibfield  {title} {\bibinfo {title} {{Majorana Fermions and a Topological
  Phase Transition in Semiconductor-Superconductor Heterostructures}},\ }\href
  {https://doi.org/10.1103/PhysRevLett.105.077001} {\bibfield  {journal}
  {\bibinfo  {journal} {Phys. Rev. Lett.}\ }\textbf {\bibinfo {volume} {105}},\
  \bibinfo {pages} {077001} (\bibinfo {year} {2010})}\BibitemShut {NoStop}%
\bibitem [{\citenamefont {Sau}\ \emph {et~al.}(2010)\citenamefont {Sau},
  \citenamefont {Lutchyn}, \citenamefont {Tewari},\ and\ \citenamefont
  {Das~Sarma}}]{Sau2010}%
  \BibitemOpen
  \bibfield  {author} {\bibinfo {author} {\bibfnamefont {J.~D.}\ \bibnamefont
  {Sau}}, \bibinfo {author} {\bibfnamefont {R.~M.}\ \bibnamefont {Lutchyn}},
  \bibinfo {author} {\bibfnamefont {S.}~\bibnamefont {Tewari}},\ and\ \bibinfo
  {author} {\bibfnamefont {S.}~\bibnamefont {Das~Sarma}},\ }\bibfield  {title}
  {\bibinfo {title} {{Generic New Platform for Topological Quantum Computation
  Using Semiconductor Heterostructures}},\ }\href
  {https://doi.org/10.1103/PhysRevLett.104.040502} {\bibfield  {journal}
  {\bibinfo  {journal} {Phys. Rev. Lett.}\ }\textbf {\bibinfo {volume} {104}},\
  \bibinfo {pages} {040502} (\bibinfo {year} {2010})}\BibitemShut {NoStop}%
\bibitem [{\citenamefont {Alicea}(2010)}]{Alicea2010}%
  \BibitemOpen
  \bibfield  {author} {\bibinfo {author} {\bibfnamefont {J.}~\bibnamefont
  {Alicea}},\ }\bibfield  {title} {\bibinfo {title} {{Majorana fermions in a
  tunable semiconductor device}},\ }\href
  {https://doi.org/10.1103/PhysRevB.81.125318} {\bibfield  {journal} {\bibinfo
  {journal} {Phys. Rev. B}\ }\textbf {\bibinfo {volume} {81}},\ \bibinfo
  {pages} {125318} (\bibinfo {year} {2010})}\BibitemShut {NoStop}%
\bibitem [{\citenamefont {Qi}\ and\ \citenamefont {Zhang}(2011)}]{Qi2011a}%
  \BibitemOpen
  \bibfield  {author} {\bibinfo {author} {\bibfnamefont {X.~L.}\ \bibnamefont
  {Qi}}\ and\ \bibinfo {author} {\bibfnamefont {S.~C.}\ \bibnamefont {Zhang}},\
  }\bibfield  {title} {\bibinfo {title} {{Topological insulators and
  superconductors}},\ }\href {https://doi.org/10.1103/RevModPhys.83.1057}
  {\bibfield  {journal} {\bibinfo  {journal} {Rev. Mod. Phys.}\ }\textbf
  {\bibinfo {volume} {83}},\ \bibinfo {pages} {1057} (\bibinfo {year}
  {2011})}\BibitemShut {NoStop}%
\bibitem [{\citenamefont {Frolov}\ \emph {et~al.}(2020)\citenamefont {Frolov},
  \citenamefont {Manfra},\ and\ \citenamefont {Sau}}]{FrolovT2020}%
  \BibitemOpen
  \bibfield  {author} {\bibinfo {author} {\bibfnamefont {S.~M.}\ \bibnamefont
  {Frolov}}, \bibinfo {author} {\bibfnamefont {M.~J.}\ \bibnamefont {Manfra}},\
  and\ \bibinfo {author} {\bibfnamefont {J.~D.}\ \bibnamefont {Sau}},\
  }\bibfield  {title} {\bibinfo {title} {{Topological superconductivity in
  hybrid devices}},\ }\href {https://doi.org/10.1038/s41567-020-0925-6}
  {\bibfield  {journal} {\bibinfo  {journal} {Nature Physics}\ }\textbf
  {\bibinfo {volume} {16}},\ \bibinfo {pages} {718} (\bibinfo {year}
  {2020})}\BibitemShut {NoStop}%
\bibitem [{\citenamefont {Young}\ \emph {et~al.}(2012)\citenamefont {Young},
  \citenamefont {Zaheer}, \citenamefont {Teo}, \citenamefont {Kane},
  \citenamefont {Mele},\ and\ \citenamefont {Rappe}}]{Young2012}%
  \BibitemOpen
  \bibfield  {author} {\bibinfo {author} {\bibfnamefont {S.~M.}\ \bibnamefont
  {Young}}, \bibinfo {author} {\bibfnamefont {S.}~\bibnamefont {Zaheer}},
  \bibinfo {author} {\bibfnamefont {J.~C.~Y.}\ \bibnamefont {Teo}}, \bibinfo
  {author} {\bibfnamefont {C.~L.}\ \bibnamefont {Kane}}, \bibinfo {author}
  {\bibfnamefont {E.~J.}\ \bibnamefont {Mele}},\ and\ \bibinfo {author}
  {\bibfnamefont {A.~M.}\ \bibnamefont {Rappe}},\ }\bibfield  {title} {\bibinfo
  {title} {{Dirac Semimetal in Three Dimensions}},\ }\href
  {https://doi.org/10.1103/PhysRevLett.108.140405} {\bibfield  {journal}
  {\bibinfo  {journal} {Phys. Rev. Lett.}\ }\textbf {\bibinfo {volume} {108}},\
  \bibinfo {pages} {140405} (\bibinfo {year} {2012})}\BibitemShut {NoStop}%
\bibitem [{\citenamefont {Wang}\ \emph {et~al.}(2012)\citenamefont {Wang},
  \citenamefont {Sun}, \citenamefont {Chen}, \citenamefont {Franchini},
  \citenamefont {Xu}, \citenamefont {Weng}, \citenamefont {Dai},\ and\
  \citenamefont {Fang}}]{Wang2012}%
  \BibitemOpen
  \bibfield  {author} {\bibinfo {author} {\bibfnamefont {Z.}~\bibnamefont
  {Wang}}, \bibinfo {author} {\bibfnamefont {Y.}~\bibnamefont {Sun}}, \bibinfo
  {author} {\bibfnamefont {X.-Q.}\ \bibnamefont {Chen}}, \bibinfo {author}
  {\bibfnamefont {C.}~\bibnamefont {Franchini}}, \bibinfo {author}
  {\bibfnamefont {G.}~\bibnamefont {Xu}}, \bibinfo {author} {\bibfnamefont
  {H.}~\bibnamefont {Weng}}, \bibinfo {author} {\bibfnamefont {X.}~\bibnamefont
  {Dai}},\ and\ \bibinfo {author} {\bibfnamefont {Z.}~\bibnamefont {Fang}},\
  }\bibfield  {title} {\bibinfo {title} {{Dirac semimetal and topological phase
  transitions in ${A}_{3}$Bi ($A=\text{Na}$, K, Rb)}},\ }\href
  {https://doi.org/10.1103/PhysRevB.85.195320} {\bibfield  {journal} {\bibinfo
  {journal} {Phys. Rev. B}\ }\textbf {\bibinfo {volume} {85}},\ \bibinfo
  {pages} {195320} (\bibinfo {year} {2012})}\BibitemShut {NoStop}%
\bibitem [{\citenamefont {Wang}\ \emph {et~al.}(2013)\citenamefont {Wang},
  \citenamefont {Weng}, \citenamefont {Wu}, \citenamefont {Dai},\ and\
  \citenamefont {Fang}}]{Wang2013a}%
  \BibitemOpen
  \bibfield  {author} {\bibinfo {author} {\bibfnamefont {Z.}~\bibnamefont
  {Wang}}, \bibinfo {author} {\bibfnamefont {H.}~\bibnamefont {Weng}}, \bibinfo
  {author} {\bibfnamefont {Q.}~\bibnamefont {Wu}}, \bibinfo {author}
  {\bibfnamefont {X.}~\bibnamefont {Dai}},\ and\ \bibinfo {author}
  {\bibfnamefont {Z.}~\bibnamefont {Fang}},\ }\bibfield  {title} {\bibinfo
  {title} {{Three-dimensional Dirac semimetal and quantum transport in
  Cd${}_{3}$As${}_{2}$}},\ }\href {https://doi.org/10.1103/PhysRevB.88.125427}
  {\bibfield  {journal} {\bibinfo  {journal} {Phys. Rev. B}\ }\textbf {\bibinfo
  {volume} {88}},\ \bibinfo {pages} {125427} (\bibinfo {year}
  {2013})}\BibitemShut {NoStop}%
\bibitem [{\citenamefont {Burkov}(2016)}]{Burkov2016}%
  \BibitemOpen
  \bibfield  {author} {\bibinfo {author} {\bibfnamefont {A.}~\bibnamefont
  {Burkov}},\ }\bibfield  {title} {\bibinfo {title} {{Topological
  semimetals}},\ }\href@noop {} {\bibfield  {journal} {\bibinfo  {journal}
  {Nature materials}\ }\textbf {\bibinfo {volume} {15}},\ \bibinfo {pages}
  {1145} (\bibinfo {year} {2016})}\BibitemShut {NoStop}%
\bibitem [{\citenamefont {Armitage}\ \emph {et~al.}(2018)\citenamefont
  {Armitage}, \citenamefont {Mele},\ and\ \citenamefont
  {Vishwanath}}]{Armitage2018}%
  \BibitemOpen
  \bibfield  {author} {\bibinfo {author} {\bibfnamefont {N.~P.}\ \bibnamefont
  {Armitage}}, \bibinfo {author} {\bibfnamefont {E.~J.}\ \bibnamefont {Mele}},\
  and\ \bibinfo {author} {\bibfnamefont {A.}~\bibnamefont {Vishwanath}},\
  }\bibfield  {title} {\bibinfo {title} {{Weyl and Dirac semimetals in
  three-dimensional solids}},\ }\href
  {https://doi.org/10.1103/RevModPhys.90.015001} {\bibfield  {journal}
  {\bibinfo  {journal} {Rev. Mod. Phys.}\ }\textbf {\bibinfo {volume} {90}},\
  \bibinfo {pages} {015001} (\bibinfo {year} {2018})}\BibitemShut {NoStop}%
\bibitem [{\citenamefont {Crassee}\ \emph {et~al.}(2018)\citenamefont
  {Crassee}, \citenamefont {Sankar}, \citenamefont {Lee}, \citenamefont
  {Akrap},\ and\ \citenamefont {Orlita}}]{Crassee2018}%
  \BibitemOpen
  \bibfield  {author} {\bibinfo {author} {\bibfnamefont {I.}~\bibnamefont
  {Crassee}}, \bibinfo {author} {\bibfnamefont {R.}~\bibnamefont {Sankar}},
  \bibinfo {author} {\bibfnamefont {W.-L.}\ \bibnamefont {Lee}}, \bibinfo
  {author} {\bibfnamefont {A.}~\bibnamefont {Akrap}},\ and\ \bibinfo {author}
  {\bibfnamefont {M.}~\bibnamefont {Orlita}},\ }\bibfield  {title} {\bibinfo
  {title} {{3D Dirac semimetal ${\mathrm{Cd}}_{3}{\mathrm{As}}_{2}$: A review
  of material properties}},\ }\href
  {https://doi.org/10.1103/PhysRevMaterials.2.120302} {\bibfield  {journal}
  {\bibinfo  {journal} {Phys. Rev. Mater.}\ }\textbf {\bibinfo {volume} {2}},\
  \bibinfo {pages} {120302} (\bibinfo {year} {2018})}\BibitemShut {NoStop}%
\bibitem [{\citenamefont {Jeon}\ \emph {et~al.}(2014)\citenamefont {Jeon},
  \citenamefont {Zhou}, \citenamefont {Gyenis}, \citenamefont {Feldman},
  \citenamefont {Kimchi}, \citenamefont {Potter}, \citenamefont {Gibson},
  \citenamefont {Cava}, \citenamefont {Vishwanath},\ and\ \citenamefont
  {Yazdani}}]{Jeon2014a}%
  \BibitemOpen
  \bibfield  {author} {\bibinfo {author} {\bibfnamefont {S.}~\bibnamefont
  {Jeon}}, \bibinfo {author} {\bibfnamefont {B.~B.}\ \bibnamefont {Zhou}},
  \bibinfo {author} {\bibfnamefont {A.}~\bibnamefont {Gyenis}}, \bibinfo
  {author} {\bibfnamefont {B.~E.}\ \bibnamefont {Feldman}}, \bibinfo {author}
  {\bibfnamefont {I.}~\bibnamefont {Kimchi}}, \bibinfo {author} {\bibfnamefont
  {A.~C.}\ \bibnamefont {Potter}}, \bibinfo {author} {\bibfnamefont {Q.~D.}\
  \bibnamefont {Gibson}}, \bibinfo {author} {\bibfnamefont {R.~J.}\
  \bibnamefont {Cava}}, \bibinfo {author} {\bibfnamefont {A.}~\bibnamefont
  {Vishwanath}},\ and\ \bibinfo {author} {\bibfnamefont {A.}~\bibnamefont
  {Yazdani}},\ }\bibfield  {title} {\bibinfo {title} {{Landau quantization and
  quasiparticle interference in the three-dimensional Dirac semimetal
  Cd$_3$As$_2$}},\ }\href {https://doi.org/10.1038/nmat4023} {\bibfield
  {journal} {\bibinfo  {journal} {Nat. Mater.}\ }\textbf {\bibinfo {volume}
  {13}},\ \bibinfo {pages} {851} (\bibinfo {year} {2014})}\BibitemShut
  {NoStop}%
\bibitem [{\citenamefont {Chen}\ \emph {et~al.}(2017)\citenamefont {Chen},
  \citenamefont {Pikulin},\ and\ \citenamefont {Franz}}]{Chen2017a}%
  \BibitemOpen
  \bibfield  {author} {\bibinfo {author} {\bibfnamefont {A.}~\bibnamefont
  {Chen}}, \bibinfo {author} {\bibfnamefont {D.~I.}\ \bibnamefont {Pikulin}},\
  and\ \bibinfo {author} {\bibfnamefont {M.}~\bibnamefont {Franz}},\ }\bibfield
   {title} {\bibinfo {title} {{Josephson current signatures of Majorana flat
  bands on the surface of time-reversal-invariant Weyl and Dirac semimetals}},\
  }\href {https://doi.org/10.1103/PhysRevB.95.174505} {\bibfield  {journal}
  {\bibinfo  {journal} {Phys. Rev. B}\ }\textbf {\bibinfo {volume} {95}},\
  \bibinfo {pages} {174505} (\bibinfo {year} {2017})}\BibitemShut {NoStop}%
\bibitem [{\citenamefont {Li}\ \emph {et~al.}(2018{\natexlab{a}})\citenamefont
  {Li}, \citenamefont {Li}, \citenamefont {Wang}, \citenamefont {Wang},
  \citenamefont {Liao}, \citenamefont {Brinkman},\ and\ \citenamefont
  {Yu}}]{Li2018b}%
  \BibitemOpen
  \bibfield  {author} {\bibinfo {author} {\bibfnamefont {C.~Z.}\ \bibnamefont
  {Li}}, \bibinfo {author} {\bibfnamefont {C.}~\bibnamefont {Li}}, \bibinfo
  {author} {\bibfnamefont {L.~X.}\ \bibnamefont {Wang}}, \bibinfo {author}
  {\bibfnamefont {S.}~\bibnamefont {Wang}}, \bibinfo {author} {\bibfnamefont
  {Z.~M.}\ \bibnamefont {Liao}}, \bibinfo {author} {\bibfnamefont
  {A.}~\bibnamefont {Brinkman}},\ and\ \bibinfo {author} {\bibfnamefont
  {D.~P.}\ \bibnamefont {Yu}},\ }\bibfield  {title} {\bibinfo {title} {{Bulk
  and surface states carried supercurrent in ballistic Nb-Dirac semimetal
  Cd$_3$As$_2$ nanowire-Nb junctions}},\ }\href
  {https://doi.org/10.1103/PhysRevB.97.115446} {\bibfield  {journal} {\bibinfo
  {journal} {Phys. Rev. B}\ }\textbf {\bibinfo {volume} {97}},\ \bibinfo
  {pages} {115446} (\bibinfo {year} {2018}{\natexlab{a}})}\BibitemShut
  {NoStop}%
\bibitem [{\citenamefont {Yu}\ \emph {et~al.}(2018)\citenamefont {Yu},
  \citenamefont {Pan}, \citenamefont {Medlin}, \citenamefont {Rodriguez},
  \citenamefont {Lee}, \citenamefont {Bao},\ and\ \citenamefont
  {Zhang}}]{Yu2018}%
  \BibitemOpen
  \bibfield  {author} {\bibinfo {author} {\bibfnamefont {W.}~\bibnamefont
  {Yu}}, \bibinfo {author} {\bibfnamefont {W.}~\bibnamefont {Pan}}, \bibinfo
  {author} {\bibfnamefont {D.~L.}\ \bibnamefont {Medlin}}, \bibinfo {author}
  {\bibfnamefont {M.~A.}\ \bibnamefont {Rodriguez}}, \bibinfo {author}
  {\bibfnamefont {S.~R.}\ \bibnamefont {Lee}}, \bibinfo {author} {\bibfnamefont
  {Z.-q.}\ \bibnamefont {Bao}},\ and\ \bibinfo {author} {\bibfnamefont
  {F.}~\bibnamefont {Zhang}},\ }\bibfield  {title} {\bibinfo {title}
  {{$\ensuremath{\pi}$ and $4\ensuremath{\pi}$ Josephson Effects Mediated by a
  Dirac Semimetal}},\ }\href {https://doi.org/10.1103/PhysRevLett.120.177704}
  {\bibfield  {journal} {\bibinfo  {journal} {Phys. Rev. Lett.}\ }\textbf
  {\bibinfo {volume} {120}},\ \bibinfo {pages} {177704} (\bibinfo {year}
  {2018})}\BibitemShut {NoStop}%
\bibitem [{\citenamefont {Wang}\ \emph {et~al.}(2018)\citenamefont {Wang},
  \citenamefont {Li}, \citenamefont {Li}, \citenamefont {Liao}, \citenamefont
  {Brinkman},\ and\ \citenamefont {Yu}}]{Wang2018c}%
  \BibitemOpen
  \bibfield  {author} {\bibinfo {author} {\bibfnamefont {A.-Q.}\ \bibnamefont
  {Wang}}, \bibinfo {author} {\bibfnamefont {C.-Z.}\ \bibnamefont {Li}},
  \bibinfo {author} {\bibfnamefont {C.}~\bibnamefont {Li}}, \bibinfo {author}
  {\bibfnamefont {Z.-M.}\ \bibnamefont {Liao}}, \bibinfo {author}
  {\bibfnamefont {A.}~\bibnamefont {Brinkman}},\ and\ \bibinfo {author}
  {\bibfnamefont {D.-P.}\ \bibnamefont {Yu}},\ }\bibfield  {title} {\bibinfo
  {title} {{$4\ensuremath{\pi}$-Periodic Supercurrent from Surface States in
  ${\mathrm{Cd}}_{3}{\mathrm{As}}_{2}$ Nanowire-Based Josephson Junctions}},\
  }\href {https://doi.org/10.1103/PhysRevLett.121.237701} {\bibfield  {journal}
  {\bibinfo  {journal} {Phys. Rev. Lett.}\ }\textbf {\bibinfo {volume} {121}},\
  \bibinfo {pages} {237701} (\bibinfo {year} {2018})}\BibitemShut {NoStop}%
\bibitem [{\citenamefont {Kwon}\ \emph {et~al.}(2004)\citenamefont {Kwon},
  \citenamefont {Sengupta},\ and\ \citenamefont {Yakovenko}}]{Kwon2004b}%
  \BibitemOpen
  \bibfield  {author} {\bibinfo {author} {\bibfnamefont {H.-J.}\ \bibnamefont
  {Kwon}}, \bibinfo {author} {\bibfnamefont {K.}~\bibnamefont {Sengupta}},\
  and\ \bibinfo {author} {\bibfnamefont {V.~M.}\ \bibnamefont {Yakovenko}},\
  }\bibfield  {title} {\bibinfo {title} {{Fractional ac Josephson effect in
  $p$-and $d$-wave superconductors}},\ }\href
  {https://doi.org/10.1140/epjb/e2004-00066-4} {\bibfield  {journal} {\bibinfo
  {journal} {Eur. Phys. J. B}\ }\textbf {\bibinfo {volume} {37}},\ \bibinfo
  {pages} {349} (\bibinfo {year} {2004})}\BibitemShut {NoStop}%
\bibitem [{\citenamefont {Feng}\ \emph {et~al.}(2020)\citenamefont {Feng},
  \citenamefont {Huang}, \citenamefont {Wang},\ and\ \citenamefont
  {Niu}}]{Feng2020}%
  \BibitemOpen
  \bibfield  {author} {\bibinfo {author} {\bibfnamefont {J.-J.}\ \bibnamefont
  {Feng}}, \bibinfo {author} {\bibfnamefont {Z.}~\bibnamefont {Huang}},
  \bibinfo {author} {\bibfnamefont {Z.}~\bibnamefont {Wang}},\ and\ \bibinfo
  {author} {\bibfnamefont {Q.}~\bibnamefont {Niu}},\ }\bibfield  {title}
  {\bibinfo {title} {{Josephson radiation from nonlinear dynamics of Majorana
  zero modes}},\ }\href {https://doi.org/10.1103/PhysRevB.101.180504}
  {\bibfield  {journal} {\bibinfo  {journal} {Phys. Rev. B}\ }\textbf {\bibinfo
  {volume} {101}},\ \bibinfo {pages} {180504(R)} (\bibinfo {year}
  {2020})}\BibitemShut {NoStop}%
\bibitem [{\citenamefont {Dom\'{\i}nguez}\ \emph {et~al.}(2012)\citenamefont
  {Dom\'{\i}nguez}, \citenamefont {Hassler},\ and\ \citenamefont
  {Platero}}]{Dominguez2012}%
  \BibitemOpen
  \bibfield  {author} {\bibinfo {author} {\bibfnamefont {F.}~\bibnamefont
  {Dom\'{\i}nguez}}, \bibinfo {author} {\bibfnamefont {F.}~\bibnamefont
  {Hassler}},\ and\ \bibinfo {author} {\bibfnamefont {G.}~\bibnamefont
  {Platero}},\ }\bibfield  {title} {\bibinfo {title} {{Dynamical detection of
  Majorana fermions in current-biased nanowires}},\ }\href
  {https://doi.org/10.1103/PhysRevB.86.140503} {\bibfield  {journal} {\bibinfo
  {journal} {Phys. Rev. B}\ }\textbf {\bibinfo {volume} {86}},\ \bibinfo
  {pages} {140503(R)} (\bibinfo {year} {2012})}\BibitemShut {NoStop}%
\bibitem [{\citenamefont {Dom\'{\i}nguez}\ \emph {et~al.}(2017)\citenamefont
  {Dom\'{\i}nguez}, \citenamefont {Kashuba}, \citenamefont {Bocquillon},
  \citenamefont {Wiedenmann}, \citenamefont {Deacon}, \citenamefont {Klapwijk},
  \citenamefont {Platero}, \citenamefont {Molenkamp}, \citenamefont
  {Trauzettel},\ and\ \citenamefont {Hankiewicz}}]{Dominguez2017}%
  \BibitemOpen
  \bibfield  {author} {\bibinfo {author} {\bibfnamefont {F.}~\bibnamefont
  {Dom\'{\i}nguez}}, \bibinfo {author} {\bibfnamefont {O.}~\bibnamefont
  {Kashuba}}, \bibinfo {author} {\bibfnamefont {E.}~\bibnamefont {Bocquillon}},
  \bibinfo {author} {\bibfnamefont {J.}~\bibnamefont {Wiedenmann}}, \bibinfo
  {author} {\bibfnamefont {R.~S.}\ \bibnamefont {Deacon}}, \bibinfo {author}
  {\bibfnamefont {T.~M.}\ \bibnamefont {Klapwijk}}, \bibinfo {author}
  {\bibfnamefont {G.}~\bibnamefont {Platero}}, \bibinfo {author} {\bibfnamefont
  {L.~W.}\ \bibnamefont {Molenkamp}}, \bibinfo {author} {\bibfnamefont
  {B.}~\bibnamefont {Trauzettel}},\ and\ \bibinfo {author} {\bibfnamefont
  {E.~M.}\ \bibnamefont {Hankiewicz}},\ }\bibfield  {title} {\bibinfo {title}
  {{Josephson junction dynamics in the presence of $2\ensuremath{\pi}$- and
  $4\ensuremath{\pi}$-periodic supercurrents}},\ }\href
  {https://doi.org/10.1103/PhysRevB.95.195430} {\bibfield  {journal} {\bibinfo
  {journal} {Phys. Rev. B}\ }\textbf {\bibinfo {volume} {95}},\ \bibinfo
  {pages} {195430} (\bibinfo {year} {2017})}\BibitemShut {NoStop}%
\bibitem [{\citenamefont {Pic\'o-Cort\'es}\ \emph {et~al.}(2017)\citenamefont
  {Pic\'o-Cort\'es}, \citenamefont {Dom\'{\i}nguez},\ and\ \citenamefont
  {Platero}}]{Pico2017}%
  \BibitemOpen
  \bibfield  {author} {\bibinfo {author} {\bibfnamefont {J.}~\bibnamefont
  {Pic\'o-Cort\'es}}, \bibinfo {author} {\bibfnamefont {F.}~\bibnamefont
  {Dom\'{\i}nguez}},\ and\ \bibinfo {author} {\bibfnamefont {G.}~\bibnamefont
  {Platero}},\ }\bibfield  {title} {\bibinfo {title} {{Signatures of a
  $4\ensuremath{\pi}$-periodic supercurrent in the voltage response of
  capacitively shunted topological Josephson junctions}},\ }\href
  {https://doi.org/10.1103/PhysRevB.96.125438} {\bibfield  {journal} {\bibinfo
  {journal} {Phys. Rev. B}\ }\textbf {\bibinfo {volume} {96}},\ \bibinfo
  {pages} {125438} (\bibinfo {year} {2017})}\BibitemShut {NoStop}%
\bibitem [{\citenamefont {Rokhinson}\ \emph {et~al.}(2012)\citenamefont
  {Rokhinson}, \citenamefont {Liu},\ and\ \citenamefont
  {Furdyna}}]{Rokhinson2012}%
  \BibitemOpen
  \bibfield  {author} {\bibinfo {author} {\bibfnamefont {L.~P.}\ \bibnamefont
  {Rokhinson}}, \bibinfo {author} {\bibfnamefont {X.}~\bibnamefont {Liu}},\
  and\ \bibinfo {author} {\bibfnamefont {J.~K.}\ \bibnamefont {Furdyna}},\
  }\bibfield  {title} {\bibinfo {title} {{The fractional ac Josephson effect in
  a semiconductor--superconductor nanowire as a signature of Majorana
  particles}},\ }\href {https://doi.org/10.1038/nphys2429} {\bibfield
  {journal} {\bibinfo  {journal} {Nature Physics}\ }\textbf {\bibinfo {volume}
  {8}},\ \bibinfo {pages} {795} (\bibinfo {year} {2012})}\BibitemShut {NoStop}%
\bibitem [{\citenamefont {Le~Calvez}\ \emph {et~al.}(2019)\citenamefont
  {Le~Calvez}, \citenamefont {Veyrat}, \citenamefont {Gay}, \citenamefont
  {Plaindoux}, \citenamefont {Winkelmann}, \citenamefont {Courtois},\ and\
  \citenamefont {Sac{\'e}p{\'e}}}]{Calvez2019}%
  \BibitemOpen
  \bibfield  {author} {\bibinfo {author} {\bibfnamefont {K.}~\bibnamefont
  {Le~Calvez}}, \bibinfo {author} {\bibfnamefont {L.}~\bibnamefont {Veyrat}},
  \bibinfo {author} {\bibfnamefont {F.}~\bibnamefont {Gay}}, \bibinfo {author}
  {\bibfnamefont {P.}~\bibnamefont {Plaindoux}}, \bibinfo {author}
  {\bibfnamefont {C.~B.}\ \bibnamefont {Winkelmann}}, \bibinfo {author}
  {\bibfnamefont {H.}~\bibnamefont {Courtois}},\ and\ \bibinfo {author}
  {\bibfnamefont {B.}~\bibnamefont {Sac{\'e}p{\'e}}},\ }\bibfield  {title}
  {\bibinfo {title} {{Joule overheating poisons the fractional ac Josephson
  effect in topological Josephson junctions}},\ }\href
  {https://doi.org/10.1038/s42005-018-0100-x} {\bibfield  {journal} {\bibinfo
  {journal} {Communications Physics}\ }\textbf {\bibinfo {volume} {2}},\
  \bibinfo {pages} {4} (\bibinfo {year} {2019})}\BibitemShut {NoStop}%
\bibitem [{\citenamefont {Sch{\"u}ffelgen}\ \emph {et~al.}(2019)\citenamefont
  {Sch{\"u}ffelgen}, \citenamefont {Rosenbach}, \citenamefont {Li},
  \citenamefont {Schmitt}, \citenamefont {Schleenvoigt}, \citenamefont {Jalil},
  \citenamefont {Schmitt}, \citenamefont {K{\"o}lzer}, \citenamefont {Wang},
  \citenamefont {Bennemann} \emph {et~al.}}]{Schuffelgen2019}%
  \BibitemOpen
  \bibfield  {author} {\bibinfo {author} {\bibfnamefont {P.}~\bibnamefont
  {Sch{\"u}ffelgen}}, \bibinfo {author} {\bibfnamefont {D.}~\bibnamefont
  {Rosenbach}}, \bibinfo {author} {\bibfnamefont {C.}~\bibnamefont {Li}},
  \bibinfo {author} {\bibfnamefont {T.~W.}\ \bibnamefont {Schmitt}}, \bibinfo
  {author} {\bibfnamefont {M.}~\bibnamefont {Schleenvoigt}}, \bibinfo {author}
  {\bibfnamefont {A.~R.}\ \bibnamefont {Jalil}}, \bibinfo {author}
  {\bibfnamefont {S.}~\bibnamefont {Schmitt}}, \bibinfo {author} {\bibfnamefont
  {J.}~\bibnamefont {K{\"o}lzer}}, \bibinfo {author} {\bibfnamefont
  {M.}~\bibnamefont {Wang}}, \bibinfo {author} {\bibfnamefont {B.}~\bibnamefont
  {Bennemann}}, \emph {et~al.},\ }\bibfield  {title} {\bibinfo {title}
  {{Selective area growth and stencil lithography for in situ fabricated
  quantum devices}},\ }\href {https://doi.org/10.1038/s41565-019-0506-y}
  {\bibfield  {journal} {\bibinfo  {journal} {Nature nanotechnology}\ }\textbf
  {\bibinfo {volume} {14}},\ \bibinfo {pages} {825} (\bibinfo {year}
  {2019})}\BibitemShut {NoStop}%
\bibitem [{\citenamefont {Yano}\ \emph {et~al.}(2020)\citenamefont {Yano},
  \citenamefont {Koyanagi}, \citenamefont {Kashiwaya}, \citenamefont {Tsumura},
  \citenamefont {Hirose}, \citenamefont {Asano}, \citenamefont {Sasagawa},\
  and\ \citenamefont {Kashiwaya}}]{Yano2020}%
  \BibitemOpen
  \bibfield  {author} {\bibinfo {author} {\bibfnamefont {R.}~\bibnamefont
  {Yano}}, \bibinfo {author} {\bibfnamefont {M.}~\bibnamefont {Koyanagi}},
  \bibinfo {author} {\bibfnamefont {H.}~\bibnamefont {Kashiwaya}}, \bibinfo
  {author} {\bibfnamefont {K.}~\bibnamefont {Tsumura}}, \bibinfo {author}
  {\bibfnamefont {H.~T.}\ \bibnamefont {Hirose}}, \bibinfo {author}
  {\bibfnamefont {Y.}~\bibnamefont {Asano}}, \bibinfo {author} {\bibfnamefont
  {T.}~\bibnamefont {Sasagawa}},\ and\ \bibinfo {author} {\bibfnamefont
  {S.}~\bibnamefont {Kashiwaya}},\ }\bibfield  {title} {\bibinfo {title}
  {{Unusual Superconducting Proximity Effect in Magnetically Doped Topological
  Josephson Junctions}},\ }\href {https://doi.org/10.7566/JPSJ.89.034702}
  {\bibfield  {journal} {\bibinfo  {journal} {J. Phys. Soc. Jpn.}\ }\textbf
  {\bibinfo {volume} {89}},\ \bibinfo {pages} {034702} (\bibinfo {year}
  {2020})}\BibitemShut {NoStop}%
\bibitem [{\citenamefont {Wiedenmann}\ \emph {et~al.}(2016)\citenamefont
  {Wiedenmann}, \citenamefont {Bocquillon}, \citenamefont {Deacon},
  \citenamefont {Hartinger}, \citenamefont {Herrmann}, \citenamefont
  {Klapwijk}, \citenamefont {Maier}, \citenamefont {Ames}, \citenamefont
  {Br{\"{u}}ne}, \citenamefont {Gould}, \citenamefont {Oiwa}, \citenamefont
  {Ishibashi}, \citenamefont {Tarucha}, \citenamefont {Buhmann},\ and\
  \citenamefont {Molenkamp}}]{Wiedenmann2016a}%
  \BibitemOpen
  \bibfield  {author} {\bibinfo {author} {\bibfnamefont {J.}~\bibnamefont
  {Wiedenmann}}, \bibinfo {author} {\bibfnamefont {E.}~\bibnamefont
  {Bocquillon}}, \bibinfo {author} {\bibfnamefont {R.~S.}\ \bibnamefont
  {Deacon}}, \bibinfo {author} {\bibfnamefont {S.}~\bibnamefont {Hartinger}},
  \bibinfo {author} {\bibfnamefont {O.}~\bibnamefont {Herrmann}}, \bibinfo
  {author} {\bibfnamefont {T.~M.}\ \bibnamefont {Klapwijk}}, \bibinfo {author}
  {\bibfnamefont {L.}~\bibnamefont {Maier}}, \bibinfo {author} {\bibfnamefont
  {C.}~\bibnamefont {Ames}}, \bibinfo {author} {\bibfnamefont {C.}~\bibnamefont
  {Br{\"{u}}ne}}, \bibinfo {author} {\bibfnamefont {C.}~\bibnamefont {Gould}},
  \bibinfo {author} {\bibfnamefont {A.}~\bibnamefont {Oiwa}}, \bibinfo {author}
  {\bibfnamefont {K.}~\bibnamefont {Ishibashi}}, \bibinfo {author}
  {\bibfnamefont {S.}~\bibnamefont {Tarucha}}, \bibinfo {author} {\bibfnamefont
  {H.}~\bibnamefont {Buhmann}},\ and\ \bibinfo {author} {\bibfnamefont {L.~W.}\
  \bibnamefont {Molenkamp}},\ }\bibfield  {title} {\bibinfo {title}
  {{4$\pi$-periodic Josephson supercurrent in HgTe-based topological Josephson
  junctions}},\ }\href {https://doi.org/10.1038/ncomms10303} {\bibfield
  {journal} {\bibinfo  {journal} {Nat. Commun.}\ }\textbf {\bibinfo {volume}
  {7}},\ \bibinfo {pages} {10303} (\bibinfo {year} {2016})}\BibitemShut
  {NoStop}%
\bibitem [{\citenamefont {Bocquillon}\ \emph {et~al.}(2017)\citenamefont
  {Bocquillon}, \citenamefont {Deacon}, \citenamefont {Wiedenmann},
  \citenamefont {Leubner}, \citenamefont {Klapwijk}, \citenamefont {Br{\"u}ne},
  \citenamefont {Ishibashi}, \citenamefont {Buhmann},\ and\ \citenamefont
  {Molenkamp}}]{Bocquillon2017}%
  \BibitemOpen
  \bibfield  {author} {\bibinfo {author} {\bibfnamefont {E.}~\bibnamefont
  {Bocquillon}}, \bibinfo {author} {\bibfnamefont {R.~S.}\ \bibnamefont
  {Deacon}}, \bibinfo {author} {\bibfnamefont {J.}~\bibnamefont {Wiedenmann}},
  \bibinfo {author} {\bibfnamefont {P.}~\bibnamefont {Leubner}}, \bibinfo
  {author} {\bibfnamefont {T.~M.}\ \bibnamefont {Klapwijk}}, \bibinfo {author}
  {\bibfnamefont {C.}~\bibnamefont {Br{\"u}ne}}, \bibinfo {author}
  {\bibfnamefont {K.}~\bibnamefont {Ishibashi}}, \bibinfo {author}
  {\bibfnamefont {H.}~\bibnamefont {Buhmann}},\ and\ \bibinfo {author}
  {\bibfnamefont {L.~W.}\ \bibnamefont {Molenkamp}},\ }\bibfield  {title}
  {\bibinfo {title} {{Gapless Andreev bound states in the quantum spin Hall
  insulator HgTe}},\ }\href {https://doi.org/10.1038/nnano.2016.159} {\bibfield
   {journal} {\bibinfo  {journal} {Nature Nanotechnology}\ }\textbf {\bibinfo
  {volume} {12}},\ \bibinfo {pages} {137} (\bibinfo {year} {2017})}\BibitemShut
  {NoStop}%
\bibitem [{\citenamefont {Li}\ \emph {et~al.}(2018{\natexlab{b}})\citenamefont
  {Li}, \citenamefont {de~Boer}, \citenamefont {de~Ronde}, \citenamefont
  {Ramankutty}, \citenamefont {van Heumen}, \citenamefont {Huang},
  \citenamefont {de~Visser}, \citenamefont {Golubov}, \citenamefont {Golden},\
  and\ \citenamefont {Brinkman}}]{LiDS2018}%
  \BibitemOpen
  \bibfield  {author} {\bibinfo {author} {\bibfnamefont {C.}~\bibnamefont
  {Li}}, \bibinfo {author} {\bibfnamefont {J.~C.}\ \bibnamefont {de~Boer}},
  \bibinfo {author} {\bibfnamefont {B.}~\bibnamefont {de~Ronde}}, \bibinfo
  {author} {\bibfnamefont {S.~V.}\ \bibnamefont {Ramankutty}}, \bibinfo
  {author} {\bibfnamefont {E.}~\bibnamefont {van Heumen}}, \bibinfo {author}
  {\bibfnamefont {Y.}~\bibnamefont {Huang}}, \bibinfo {author} {\bibfnamefont
  {A.}~\bibnamefont {de~Visser}}, \bibinfo {author} {\bibfnamefont {A.~A.}\
  \bibnamefont {Golubov}}, \bibinfo {author} {\bibfnamefont {M.~S.}\
  \bibnamefont {Golden}},\ and\ \bibinfo {author} {\bibfnamefont
  {A.}~\bibnamefont {Brinkman}},\ }\bibfield  {title} {\bibinfo {title}
  {{4$\pi$-periodic Andreev bound states in a Dirac semimetal}},\ }\href
  {https://doi.org/10.1038/s41563-018-0158-6} {\bibfield  {journal} {\bibinfo
  {journal} {Nature materials}\ }\textbf {\bibinfo {volume} {17}},\ \bibinfo
  {pages} {875} (\bibinfo {year} {2018}{\natexlab{b}})}\BibitemShut {NoStop}%
\bibitem [{\citenamefont {Bai}\ \emph {et~al.}(2022)\citenamefont {Bai},
  \citenamefont {Wei}, \citenamefont {Feng}, \citenamefont {Luysberg},
  \citenamefont {Bliesener}, \citenamefont {Lippertz}, \citenamefont {Uday},
  \citenamefont {Taskin}, \citenamefont {Mayer},\ and\ \citenamefont
  {Ando}}]{Bai2022}%
  \BibitemOpen
  \bibfield  {author} {\bibinfo {author} {\bibfnamefont {M.}~\bibnamefont
  {Bai}}, \bibinfo {author} {\bibfnamefont {X.-K.}\ \bibnamefont {Wei}},
  \bibinfo {author} {\bibfnamefont {J.}~\bibnamefont {Feng}}, \bibinfo {author}
  {\bibfnamefont {M.}~\bibnamefont {Luysberg}}, \bibinfo {author}
  {\bibfnamefont {A.}~\bibnamefont {Bliesener}}, \bibinfo {author}
  {\bibfnamefont {G.}~\bibnamefont {Lippertz}}, \bibinfo {author}
  {\bibfnamefont {A.}~\bibnamefont {Uday}}, \bibinfo {author} {\bibfnamefont
  {A.~A.}\ \bibnamefont {Taskin}}, \bibinfo {author} {\bibfnamefont
  {J.}~\bibnamefont {Mayer}},\ and\ \bibinfo {author} {\bibfnamefont
  {Y.}~\bibnamefont {Ando}},\ }\bibfield  {title} {\bibinfo {title}
  {{Proximity-induced superconductivity in (Bi$_{1- x}$Sb$_x$)$_2$Te$_3$
  topological-insulator nanowires}},\ }\href
  {https://doi.org/10.1038/s43246-022-00242-6} {\bibfield  {journal} {\bibinfo
  {journal} {Communications Materials}\ }\textbf {\bibinfo {volume} {3}},\
  \bibinfo {pages} {20} (\bibinfo {year} {2022})}\BibitemShut {NoStop}%
\bibitem [{\citenamefont {Deacon}\ \emph {et~al.}(2017)\citenamefont {Deacon},
  \citenamefont {Wiedenmann}, \citenamefont {Bocquillon}, \citenamefont
  {Dom\'{\i}nguez}, \citenamefont {Klapwijk}, \citenamefont {Leubner},
  \citenamefont {Br\"une}, \citenamefont {Hankiewicz}, \citenamefont {Tarucha},
  \citenamefont {Ishibashi}, \citenamefont {Buhmann},\ and\ \citenamefont
  {Molenkamp}}]{Deaconb}%
  \BibitemOpen
  \bibfield  {author} {\bibinfo {author} {\bibfnamefont {R.~S.}\ \bibnamefont
  {Deacon}}, \bibinfo {author} {\bibfnamefont {J.}~\bibnamefont {Wiedenmann}},
  \bibinfo {author} {\bibfnamefont {E.}~\bibnamefont {Bocquillon}}, \bibinfo
  {author} {\bibfnamefont {F.}~\bibnamefont {Dom\'{\i}nguez}}, \bibinfo
  {author} {\bibfnamefont {T.~M.}\ \bibnamefont {Klapwijk}}, \bibinfo {author}
  {\bibfnamefont {P.}~\bibnamefont {Leubner}}, \bibinfo {author} {\bibfnamefont
  {C.}~\bibnamefont {Br\"une}}, \bibinfo {author} {\bibfnamefont {E.~M.}\
  \bibnamefont {Hankiewicz}}, \bibinfo {author} {\bibfnamefont
  {S.}~\bibnamefont {Tarucha}}, \bibinfo {author} {\bibfnamefont
  {K.}~\bibnamefont {Ishibashi}}, \bibinfo {author} {\bibfnamefont
  {H.}~\bibnamefont {Buhmann}},\ and\ \bibinfo {author} {\bibfnamefont {L.~W.}\
  \bibnamefont {Molenkamp}},\ }\bibfield  {title} {\bibinfo {title} {{Josephson
  Radiation from Gapless Andreev Bound States in HgTe-Based Topological
  Junctions}},\ }\href {https://doi.org/10.1103/PhysRevX.7.021011} {\bibfield
  {journal} {\bibinfo  {journal} {Phys. Rev. X}\ }\textbf {\bibinfo {volume}
  {7}},\ \bibinfo {pages} {021011} (\bibinfo {year} {2017})}\BibitemShut
  {NoStop}%
\bibitem [{\citenamefont {Laroche}\ \emph {et~al.}(2019)\citenamefont
  {Laroche}, \citenamefont {Bouman}, \citenamefont {van Woerkom}, \citenamefont
  {Proutski}, \citenamefont {Murthy}, \citenamefont {Pikulin}, \citenamefont
  {Nayak}, \citenamefont {van Gulik}, \citenamefont {Nyg{\aa}rd}, \citenamefont
  {Krogstrup}, \citenamefont {Kouwenhoven},\ and\ \citenamefont
  {Geresdi}}]{Laroche2019}%
  \BibitemOpen
  \bibfield  {author} {\bibinfo {author} {\bibfnamefont {D.}~\bibnamefont
  {Laroche}}, \bibinfo {author} {\bibfnamefont {D.}~\bibnamefont {Bouman}},
  \bibinfo {author} {\bibfnamefont {D.~J.}\ \bibnamefont {van Woerkom}},
  \bibinfo {author} {\bibfnamefont {A.}~\bibnamefont {Proutski}}, \bibinfo
  {author} {\bibfnamefont {C.}~\bibnamefont {Murthy}}, \bibinfo {author}
  {\bibfnamefont {D.~I.}\ \bibnamefont {Pikulin}}, \bibinfo {author}
  {\bibfnamefont {C.}~\bibnamefont {Nayak}}, \bibinfo {author} {\bibfnamefont
  {R.~J.~J.}\ \bibnamefont {van Gulik}}, \bibinfo {author} {\bibfnamefont
  {J.}~\bibnamefont {Nyg{\aa}rd}}, \bibinfo {author} {\bibfnamefont
  {P.}~\bibnamefont {Krogstrup}}, \bibinfo {author} {\bibfnamefont {L.~P.}\
  \bibnamefont {Kouwenhoven}},\ and\ \bibinfo {author} {\bibfnamefont
  {A.}~\bibnamefont {Geresdi}},\ }\bibfield  {title} {\bibinfo {title}
  {{Observation of the 4$\pi$-periodic Josephson effect in indium arsenide
  nanowires}},\ }\href {https://doi.org/10.1038/s41467-018-08161-2} {\bibfield
  {journal} {\bibinfo  {journal} {Nat. Commun.}\ }\textbf {\bibinfo {volume}
  {10}},\ \bibinfo {pages} {245} (\bibinfo {year} {2019})}\BibitemShut
  {NoStop}%
\bibitem [{\citenamefont {Kamata}\ \emph {et~al.}(2018)\citenamefont {Kamata},
  \citenamefont {Deacon}, \citenamefont {Matsuo}, \citenamefont {Li},
  \citenamefont {Jeppesen}, \citenamefont {Samuelson}, \citenamefont {Xu},
  \citenamefont {Ishibashi},\ and\ \citenamefont {Tarucha}}]{Kamata2018}%
  \BibitemOpen
  \bibfield  {author} {\bibinfo {author} {\bibfnamefont {H.}~\bibnamefont
  {Kamata}}, \bibinfo {author} {\bibfnamefont {R.~S.}\ \bibnamefont {Deacon}},
  \bibinfo {author} {\bibfnamefont {S.}~\bibnamefont {Matsuo}}, \bibinfo
  {author} {\bibfnamefont {K.}~\bibnamefont {Li}}, \bibinfo {author}
  {\bibfnamefont {S.}~\bibnamefont {Jeppesen}}, \bibinfo {author}
  {\bibfnamefont {L.}~\bibnamefont {Samuelson}}, \bibinfo {author}
  {\bibfnamefont {H.~Q.}\ \bibnamefont {Xu}}, \bibinfo {author} {\bibfnamefont
  {K.}~\bibnamefont {Ishibashi}},\ and\ \bibinfo {author} {\bibfnamefont
  {S.}~\bibnamefont {Tarucha}},\ }\bibfield  {title} {\bibinfo {title}
  {{Anomalous modulation of Josephson radiation in nanowire-based Josephson
  junctions}},\ }\href {https://doi.org/10.1103/PhysRevB.98.041302} {\bibfield
  {journal} {\bibinfo  {journal} {Phys. Rev. B}\ }\textbf {\bibinfo {volume}
  {98}},\ \bibinfo {pages} {041302(R)} (\bibinfo {year} {2018})}\BibitemShut
  {NoStop}%
\bibitem [{\citenamefont {Mudi}\ and\ \citenamefont {Frolov}(2022)}]{Mudi2021}%
  \BibitemOpen
  \bibfield  {author} {\bibinfo {author} {\bibfnamefont {S.~R.}\ \bibnamefont
  {Mudi}}\ and\ \bibinfo {author} {\bibfnamefont {S.~M.}\ \bibnamefont
  {Frolov}},\ }\href@noop {} {\bibinfo {title} {{Model for missing Shapiro
  steps due to bias-dependent resistance}}} (\bibinfo {year} {2022}),\ \Eprint
  {https://arxiv.org/abs/2106.00495} {arXiv:2106.00495 [cond-mat.mes-hall]}
  \BibitemShut {NoStop}%
\bibitem [{\citenamefont {Billangeon}\ \emph {et~al.}(2007)\citenamefont
  {Billangeon}, \citenamefont {Pierre}, \citenamefont {Bouchiat},\ and\
  \citenamefont {Deblock}}]{Billangeon2007}%
  \BibitemOpen
  \bibfield  {author} {\bibinfo {author} {\bibfnamefont {P.-M.}\ \bibnamefont
  {Billangeon}}, \bibinfo {author} {\bibfnamefont {F.}~\bibnamefont {Pierre}},
  \bibinfo {author} {\bibfnamefont {H.}~\bibnamefont {Bouchiat}},\ and\
  \bibinfo {author} {\bibfnamefont {R.}~\bibnamefont {Deblock}},\ }\bibfield
  {title} {\bibinfo {title} {{ac Josephson Effect and Resonant Cooper Pair
  Tunneling Emission of a Single Cooper Pair Transistor}},\ }\href
  {https://doi.org/10.1103/PhysRevLett.98.216802} {\bibfield  {journal}
  {\bibinfo  {journal} {Phys. Rev. Lett.}\ }\textbf {\bibinfo {volume} {98}},\
  \bibinfo {pages} {216802} (\bibinfo {year} {2007})}\BibitemShut {NoStop}%
\bibitem [{\citenamefont {Dartiailh}\ \emph {et~al.}(2021)\citenamefont
  {Dartiailh}, \citenamefont {Cuozzo}, \citenamefont {Elfeky}, \citenamefont
  {Mayer}, \citenamefont {Yuan}, \citenamefont {Wickramasinghe}, \citenamefont
  {Rossi},\ and\ \citenamefont {Shabani}}]{Dartiailh2021}%
  \BibitemOpen
  \bibfield  {author} {\bibinfo {author} {\bibfnamefont {M.~C.}\ \bibnamefont
  {Dartiailh}}, \bibinfo {author} {\bibfnamefont {J.~J.}\ \bibnamefont
  {Cuozzo}}, \bibinfo {author} {\bibfnamefont {B.~H.}\ \bibnamefont {Elfeky}},
  \bibinfo {author} {\bibfnamefont {W.}~\bibnamefont {Mayer}}, \bibinfo
  {author} {\bibfnamefont {J.}~\bibnamefont {Yuan}}, \bibinfo {author}
  {\bibfnamefont {K.~S.}\ \bibnamefont {Wickramasinghe}}, \bibinfo {author}
  {\bibfnamefont {E.}~\bibnamefont {Rossi}},\ and\ \bibinfo {author}
  {\bibfnamefont {J.}~\bibnamefont {Shabani}},\ }\bibfield  {title} {\bibinfo
  {title} {{Missing Shapiro steps in topologically trivial Josephson junction
  on InAs quantum well}},\ }\href {https://doi.org/10.1038/s41467-020-20382-y}
  {\bibfield  {journal} {\bibinfo  {journal} {Nature Communications}\ }\textbf
  {\bibinfo {volume} {12}},\ \bibinfo {pages} {78} (\bibinfo {year}
  {2021})}\BibitemShut {NoStop}%
\bibitem [{\citenamefont {Rosen}\ \emph {et~al.}(2021)\citenamefont {Rosen},
  \citenamefont {Trimble}, \citenamefont {Andersen}, \citenamefont {Mikheev},
  \citenamefont {Li}, \citenamefont {Liu}, \citenamefont {Tai}, \citenamefont
  {Zhang}, \citenamefont {Wang}, \citenamefont {Cui}, \citenamefont {Kastner},
  \citenamefont {Williams},\ and\ \citenamefont
  {Goldhaber-Gordon}}]{Rosen2021}%
  \BibitemOpen
  \bibfield  {author} {\bibinfo {author} {\bibfnamefont {I.~T.}\ \bibnamefont
  {Rosen}}, \bibinfo {author} {\bibfnamefont {C.~J.}\ \bibnamefont {Trimble}},
  \bibinfo {author} {\bibfnamefont {M.~P.}\ \bibnamefont {Andersen}}, \bibinfo
  {author} {\bibfnamefont {E.}~\bibnamefont {Mikheev}}, \bibinfo {author}
  {\bibfnamefont {Y.}~\bibnamefont {Li}}, \bibinfo {author} {\bibfnamefont
  {Y.}~\bibnamefont {Liu}}, \bibinfo {author} {\bibfnamefont {L.}~\bibnamefont
  {Tai}}, \bibinfo {author} {\bibfnamefont {P.}~\bibnamefont {Zhang}}, \bibinfo
  {author} {\bibfnamefont {K.~L.}\ \bibnamefont {Wang}}, \bibinfo {author}
  {\bibfnamefont {Y.}~\bibnamefont {Cui}}, \bibinfo {author} {\bibfnamefont
  {M.~A.}\ \bibnamefont {Kastner}}, \bibinfo {author} {\bibfnamefont {J.~R.}\
  \bibnamefont {Williams}},\ and\ \bibinfo {author} {\bibfnamefont
  {D.}~\bibnamefont {Goldhaber-Gordon}},\ }\href@noop {} {\bibinfo {title}
  {{Fractional AC Josephson effect in a topological insulator proximitized by a
  self-formed superconductor}}} (\bibinfo {year} {2021}),\ \Eprint
  {https://arxiv.org/abs/2110.01039} {arXiv:2110.01039 [cond-mat.mes-hall]}
  \BibitemShut {NoStop}%
\bibitem [{\citenamefont {Elfeky}\ \emph {et~al.}(2023)\citenamefont {Elfeky},
  \citenamefont {Cuozzo}, \citenamefont {Lotfizadeh}, \citenamefont {Schiela},
  \citenamefont {Farzaneh}, \citenamefont {Strickland}, \citenamefont
  {Langone}, \citenamefont {Rossi},\ and\ \citenamefont
  {Shabani}}]{Elfeky2023}%
  \BibitemOpen
  \bibfield  {author} {\bibinfo {author} {\bibfnamefont {B.~H.}\ \bibnamefont
  {Elfeky}}, \bibinfo {author} {\bibfnamefont {J.~J.}\ \bibnamefont {Cuozzo}},
  \bibinfo {author} {\bibfnamefont {N.}~\bibnamefont {Lotfizadeh}}, \bibinfo
  {author} {\bibfnamefont {W.~F.}\ \bibnamefont {Schiela}}, \bibinfo {author}
  {\bibfnamefont {S.~M.}\ \bibnamefont {Farzaneh}}, \bibinfo {author}
  {\bibfnamefont {W.~M.}\ \bibnamefont {Strickland}}, \bibinfo {author}
  {\bibfnamefont {D.}~\bibnamefont {Langone}}, \bibinfo {author} {\bibfnamefont
  {E.}~\bibnamefont {Rossi}},\ and\ \bibinfo {author} {\bibfnamefont
  {J.}~\bibnamefont {Shabani}},\ }\bibfield  {title} {\bibinfo {title}
  {{Evolution of 4$\pi$-Periodic Supercurrent in the Presence of an In-Plane
  Magnetic Field}},\ }\href {https://doi.org/10.1021/acsnano.2c10880}
  {\bibfield  {journal} {\bibinfo  {journal} {ACS Nano}\ }\textbf {\bibinfo
  {volume} {17}},\ \bibinfo {pages} {4650} (\bibinfo {year}
  {2023})}\BibitemShut {NoStop}%
\bibitem [{\citenamefont {Houzet}\ \emph {et~al.}(2013)\citenamefont {Houzet},
  \citenamefont {Meyer}, \citenamefont {Badiane},\ and\ \citenamefont
  {Glazman}}]{Houzet2013}%
  \BibitemOpen
  \bibfield  {author} {\bibinfo {author} {\bibfnamefont {M.}~\bibnamefont
  {Houzet}}, \bibinfo {author} {\bibfnamefont {J.~S.}\ \bibnamefont {Meyer}},
  \bibinfo {author} {\bibfnamefont {D.~M.}\ \bibnamefont {Badiane}},\ and\
  \bibinfo {author} {\bibfnamefont {L.~I.}\ \bibnamefont {Glazman}},\
  }\bibfield  {title} {\bibinfo {title} {{Dynamics of Majorana States in a
  Topological Josephson Junction}},\ }\href
  {https://doi.org/10.1103/PhysRevLett.111.046401} {\bibfield  {journal}
  {\bibinfo  {journal} {Phys. Rev. Lett.}\ }\textbf {\bibinfo {volume} {111}},\
  \bibinfo {pages} {046401} (\bibinfo {year} {2013})}\BibitemShut {NoStop}%
\bibitem [{\citenamefont {Park}\ \emph {et~al.}(2021)\citenamefont {Park},
  \citenamefont {Choi}, \citenamefont {Lee},\ and\ \citenamefont
  {Lee}}]{Park2021}%
  \BibitemOpen
  \bibfield  {author} {\bibinfo {author} {\bibfnamefont {J.}~\bibnamefont
  {Park}}, \bibinfo {author} {\bibfnamefont {Y.-B.}\ \bibnamefont {Choi}},
  \bibinfo {author} {\bibfnamefont {G.-H.}\ \bibnamefont {Lee}},\ and\ \bibinfo
  {author} {\bibfnamefont {H.-J.}\ \bibnamefont {Lee}},\ }\bibfield  {title}
  {\bibinfo {title} {{Characterization of Shapiro steps in the presence of a
  4\ensuremath{\pi}-periodic Josephson current}},\ }\href
  {https://doi.org/10.1103/PhysRevB.103.235428} {\bibfield  {journal} {\bibinfo
   {journal} {Phys. Rev. B}\ }\textbf {\bibinfo {volume} {103}},\ \bibinfo
  {pages} {235428} (\bibinfo {year} {2021})}\BibitemShut {NoStop}%
\bibitem [{\citenamefont {Jang}\ and\ \citenamefont {Doh}(2021)}]{Jang2021}%
  \BibitemOpen
  \bibfield  {author} {\bibinfo {author} {\bibfnamefont {Y.}~\bibnamefont
  {Jang}}\ and\ \bibinfo {author} {\bibfnamefont {Y.-J.}\ \bibnamefont {Doh}},\
  }\bibfield  {title} {\bibinfo {title} {{Optimal conditions for observing
  fractional Josephson effect in topological Josephson junctions}},\ }\href
  {https://doi.org/10.1007/s40042-020-00035-5} {\bibfield  {journal} {\bibinfo
  {journal} {J. Korean Phys. Soc.}\ }\textbf {\bibinfo {volume} {78}},\
  \bibinfo {pages} {58} (\bibinfo {year} {2021})}\BibitemShut {NoStop}%
\bibitem [{\citenamefont {San-Jose}\ \emph {et~al.}(2012)\citenamefont
  {San-Jose}, \citenamefont {Prada},\ and\ \citenamefont
  {Aguado}}]{San-Jose2012}%
  \BibitemOpen
  \bibfield  {author} {\bibinfo {author} {\bibfnamefont {P.}~\bibnamefont
  {San-Jose}}, \bibinfo {author} {\bibfnamefont {E.}~\bibnamefont {Prada}},\
  and\ \bibinfo {author} {\bibfnamefont {R.}~\bibnamefont {Aguado}},\
  }\bibfield  {title} {\bibinfo {title} {{ac Josephson Effect in Finite-Length
  Nanowire Junctions with Majorana Modes}},\ }\href
  {https://doi.org/10.1103/PhysRevLett.108.257001} {\bibfield  {journal}
  {\bibinfo  {journal} {Phys. Rev. Lett.}\ }\textbf {\bibinfo {volume} {108}},\
  \bibinfo {pages} {257001} (\bibinfo {year} {2012})}\BibitemShut {NoStop}%
\bibitem [{\citenamefont {Takeshige}\ \emph {et~al.}(2020)\citenamefont
  {Takeshige}, \citenamefont {Matsuo}, \citenamefont {Deacon}, \citenamefont
  {Ueda}, \citenamefont {Sato}, \citenamefont {Zhao}, \citenamefont {Zhou},
  \citenamefont {Chang}, \citenamefont {Ishibashi},\ and\ \citenamefont
  {Tarucha}}]{Takeshige2020}%
  \BibitemOpen
  \bibfield  {author} {\bibinfo {author} {\bibfnamefont {Y.}~\bibnamefont
  {Takeshige}}, \bibinfo {author} {\bibfnamefont {S.}~\bibnamefont {Matsuo}},
  \bibinfo {author} {\bibfnamefont {R.~S.}\ \bibnamefont {Deacon}}, \bibinfo
  {author} {\bibfnamefont {K.}~\bibnamefont {Ueda}}, \bibinfo {author}
  {\bibfnamefont {Y.}~\bibnamefont {Sato}}, \bibinfo {author} {\bibfnamefont
  {Y.~F.}\ \bibnamefont {Zhao}}, \bibinfo {author} {\bibfnamefont
  {L.}~\bibnamefont {Zhou}}, \bibinfo {author} {\bibfnamefont {C.~Z.}\
  \bibnamefont {Chang}}, \bibinfo {author} {\bibfnamefont {K.}~\bibnamefont
  {Ishibashi}},\ and\ \bibinfo {author} {\bibfnamefont {S.}~\bibnamefont
  {Tarucha}},\ }\bibfield  {title} {\bibinfo {title} {{Experimental study of ac
  Josephson effect in gate-tunable (Bi$_{1-x}$Sb$_x$)$_2$Te$_3$ thin-film
  Josephson junctions}},\ }\href {https://doi.org/10.1103/PhysRevB.101.115410}
  {\bibfield  {journal} {\bibinfo  {journal} {Phys. Rev. B}\ }\textbf {\bibinfo
  {volume} {101}},\ \bibinfo {pages} {115410} (\bibinfo {year}
  {2020})}\BibitemShut {NoStop}%
\bibitem [{\citenamefont {Shapiro}(1963)}]{Shapiro1963}%
  \BibitemOpen
  \bibfield  {author} {\bibinfo {author} {\bibfnamefont {S.}~\bibnamefont
  {Shapiro}},\ }\bibfield  {title} {\bibinfo {title} {{Josephson Currents in
  Superconducting Tunneling: The Effect of Microwaves and Other
  Observations}},\ }\href {https://doi.org/10.1103/PhysRevLett.11.80}
  {\bibfield  {journal} {\bibinfo  {journal} {Phys. Rev. Lett.}\ }\textbf
  {\bibinfo {volume} {11}},\ \bibinfo {pages} {80} (\bibinfo {year}
  {1963})}\BibitemShut {NoStop}%
\bibitem [{Note52()}]{Note52}%
  \BibitemOpen
  \bibinfo {note} {See Supplemental Material for details about the RCSJ
  model\protect \tmspace +\thinmuskip {.1667em}(\ref {sec:modeling}), the
  device fabrication\protect \tmspace +\thinmuskip {.1667em}(\ref
  {sec:methodes}), the measurement setup\protect \tmspace +\thinmuskip
  {.1667em}(\ref {sec:setup}); valuations of the Josephson emission
  power\protect \tmspace +\thinmuskip {.1667em}(\ref {sec:transfer}), the
  $I_{\protect \mathrm {c}}R_N$ product\protect \tmspace +\thinmuskip
  {.1667em}(\ref {sec:icrn}); the evaluation of Shapiro patterns\protect
  \tmspace +\thinmuskip {.1667em}(\ref {sec:shapiro_eva}).}\BibitemShut {Stop}%
\bibitem [{\citenamefont {Heikkil\"a}\ \emph {et~al.}(2002)\citenamefont
  {Heikkil\"a}, \citenamefont {S\"arkk\"a},\ and\ \citenamefont
  {Wilhelm}}]{Heikkila}%
  \BibitemOpen
  \bibfield  {author} {\bibinfo {author} {\bibfnamefont {T.~T.}\ \bibnamefont
  {Heikkil\"a}}, \bibinfo {author} {\bibfnamefont {J.}~\bibnamefont
  {S\"arkk\"a}},\ and\ \bibinfo {author} {\bibfnamefont {F.~K.}\ \bibnamefont
  {Wilhelm}},\ }\bibfield  {title} {\bibinfo {title} {{Supercurrent-carrying
  density of states in diffusive mesoscopic Josephson weak links}},\ }\href
  {https://doi.org/10.1103/PhysRevB.66.184513} {\bibfield  {journal} {\bibinfo
  {journal} {Phys. Rev. B}\ }\textbf {\bibinfo {volume} {66}},\ \bibinfo
  {pages} {184513} (\bibinfo {year} {2002})}\BibitemShut {NoStop}%
\bibitem [{\citenamefont {Chauvin}\ \emph {et~al.}(2006)\citenamefont
  {Chauvin}, \citenamefont {vom Stein}, \citenamefont {Pothier}, \citenamefont
  {Joyez}, \citenamefont {Huber}, \citenamefont {Esteve},\ and\ \citenamefont
  {Urbina}}]{Chauvin2006}%
  \BibitemOpen
  \bibfield  {author} {\bibinfo {author} {\bibfnamefont {M.}~\bibnamefont
  {Chauvin}}, \bibinfo {author} {\bibfnamefont {P.}~\bibnamefont {vom Stein}},
  \bibinfo {author} {\bibfnamefont {H.}~\bibnamefont {Pothier}}, \bibinfo
  {author} {\bibfnamefont {P.}~\bibnamefont {Joyez}}, \bibinfo {author}
  {\bibfnamefont {M.~E.}\ \bibnamefont {Huber}}, \bibinfo {author}
  {\bibfnamefont {D.}~\bibnamefont {Esteve}},\ and\ \bibinfo {author}
  {\bibfnamefont {C.}~\bibnamefont {Urbina}},\ }\bibfield  {title} {\bibinfo
  {title} {{Superconducting Atomic Contacts under Microwave Irradiation}},\
  }\href {https://doi.org/10.1103/PhysRevLett.97.067006} {\bibfield  {journal}
  {\bibinfo  {journal} {Phys. Rev. Lett.}\ }\textbf {\bibinfo {volume} {97}},\
  \bibinfo {pages} {067006} (\bibinfo {year} {2006})}\BibitemShut {NoStop}%
\bibitem [{\citenamefont {Dayem}\ and\ \citenamefont
  {Grimes}(1966)}]{Dayem1966}%
  \BibitemOpen
  \bibfield  {author} {\bibinfo {author} {\bibfnamefont {A.~H.}\ \bibnamefont
  {Dayem}}\ and\ \bibinfo {author} {\bibfnamefont {C.~C.}\ \bibnamefont
  {Grimes}},\ }\bibfield  {title} {\bibinfo {title} {{Microwave emission from
  superconducting point-contacts}},\ }\href {https://doi.org/10.1063/1.1754595}
  {\bibfield  {journal} {\bibinfo  {journal} {Appl. Phys. Lett.}\ }\textbf
  {\bibinfo {volume} {9}},\ \bibinfo {pages} {47} (\bibinfo {year}
  {1966})}\BibitemShut {NoStop}%
\bibitem [{\citenamefont {Bretheau}(2013)}]{Bretheau2013}%
  \BibitemOpen
  \bibfield  {author} {\bibinfo {author} {\bibfnamefont {L.}~\bibnamefont
  {Bretheau}},\ }\emph {\bibinfo {title} {{Localized excitations in
  superconducting atomic contacts: Probing the Andreev doublet}}},\ \href
  {https://pastel.archives-ouvertes.fr/pastel-00862029} {\bibinfo {type}
  {Theses}},\ \bibinfo  {school} {{Ecole Polytechnique X}} (\bibinfo {year}
  {2013})\BibitemShut {NoStop}%
\bibitem [{\citenamefont {Basset}\ \emph {et~al.}(2019)\citenamefont {Basset},
  \citenamefont {Kuzmanovi\ifmmode~\acute{c}\else \'{c}\fi{}}, \citenamefont
  {Virtanen}, \citenamefont {Heikkil\"a}, \citenamefont {Est\`eve},
  \citenamefont {Gabelli}, \citenamefont {Strunk},\ and\ \citenamefont
  {Aprili}}]{Basset2019}%
  \BibitemOpen
  \bibfield  {author} {\bibinfo {author} {\bibfnamefont {J.}~\bibnamefont
  {Basset}}, \bibinfo {author} {\bibfnamefont {M.}~\bibnamefont
  {Kuzmanovi\ifmmode~\acute{c}\else \'{c}\fi{}}}, \bibinfo {author}
  {\bibfnamefont {P.}~\bibnamefont {Virtanen}}, \bibinfo {author}
  {\bibfnamefont {T.~T.}\ \bibnamefont {Heikkil\"a}}, \bibinfo {author}
  {\bibfnamefont {J.}~\bibnamefont {Est\`eve}}, \bibinfo {author}
  {\bibfnamefont {J.}~\bibnamefont {Gabelli}}, \bibinfo {author} {\bibfnamefont
  {C.}~\bibnamefont {Strunk}},\ and\ \bibinfo {author} {\bibfnamefont
  {M.}~\bibnamefont {Aprili}},\ }\bibfield  {title} {\bibinfo {title}
  {{Nonadiabatic dynamics in strongly driven diffusive Josephson junctions}},\
  }\href {https://doi.org/10.1103/PhysRevResearch.1.032009} {\bibfield
  {journal} {\bibinfo  {journal} {Phys. Rev. Res.}\ }\textbf {\bibinfo {volume}
  {1}},\ \bibinfo {pages} {032009(R)} (\bibinfo {year} {2019})}\BibitemShut
  {NoStop}%
\bibitem [{\citenamefont {Ridderbos}\ \emph {et~al.}(2019)\citenamefont
  {Ridderbos}, \citenamefont {Brauns}, \citenamefont {Li}, \citenamefont
  {Bakkers}, \citenamefont {Brinkman}, \citenamefont {van~der Wiel},\ and\
  \citenamefont {Zwanenburg}}]{Ridderbos2019}%
  \BibitemOpen
  \bibfield  {author} {\bibinfo {author} {\bibfnamefont {J.}~\bibnamefont
  {Ridderbos}}, \bibinfo {author} {\bibfnamefont {M.}~\bibnamefont {Brauns}},
  \bibinfo {author} {\bibfnamefont {A.}~\bibnamefont {Li}}, \bibinfo {author}
  {\bibfnamefont {E.~P. A.~M.}\ \bibnamefont {Bakkers}}, \bibinfo {author}
  {\bibfnamefont {A.}~\bibnamefont {Brinkman}}, \bibinfo {author}
  {\bibfnamefont {W.~G.}\ \bibnamefont {van~der Wiel}},\ and\ \bibinfo {author}
  {\bibfnamefont {F.~A.}\ \bibnamefont {Zwanenburg}},\ }\bibfield  {title}
  {\bibinfo {title} {{Multiple Andreev reflections and Shapiro steps in a Ge-Si
  nanowire Josephson junction}},\ }\href
  {https://doi.org/10.1103/PhysRevMaterials.3.084803} {\bibfield  {journal}
  {\bibinfo  {journal} {Phys. Rev. Mater.}\ }\textbf {\bibinfo {volume} {3}},\
  \bibinfo {pages} {084803} (\bibinfo {year} {2019})}\BibitemShut {NoStop}%
\bibitem [{\citenamefont {Kautz}\ and\ \citenamefont {Martinis}(1990)}]{Kautz}%
  \BibitemOpen
  \bibfield  {author} {\bibinfo {author} {\bibfnamefont {R.~L.}\ \bibnamefont
  {Kautz}}\ and\ \bibinfo {author} {\bibfnamefont {J.~M.}\ \bibnamefont
  {Martinis}},\ }\bibfield  {title} {\bibinfo {title} {{Noise-affected $I$-$V$
  curves in small hysteretic Josephson junctions}},\ }\href
  {https://doi.org/10.1103/PhysRevB.42.9903} {\bibfield  {journal} {\bibinfo
  {journal} {Phys. Rev. B}\ }\textbf {\bibinfo {volume} {42}},\ \bibinfo
  {pages} {9903} (\bibinfo {year} {1990})}\BibitemShut {NoStop}%
\bibitem [{Note60()}]{Note60}%
  \BibitemOpen
  \bibinfo {note} {Python code of the RCSJ model is available on\\\protect
  \href
  {https://github.com/donfuge/rcsj_sde}{https://github.com/donfuge/rcsj\protect
  \_sde}}\BibitemShut {NoStop}%
\bibitem [{\citenamefont {Stewart}(1968)}]{Stewart}%
  \BibitemOpen
  \bibfield  {author} {\bibinfo {author} {\bibfnamefont {W.~C.}\ \bibnamefont
  {Stewart}},\ }\bibfield  {title} {\bibinfo {title} {{Current‐voltage
  characteristics of Josephson junctions}},\ }\href
  {https://doi.org/10.1063/1.1651991} {\bibfield  {journal} {\bibinfo
  {journal} {Appl. Phys. Lett.}\ }\textbf {\bibinfo {volume} {12}},\ \bibinfo
  {pages} {277} (\bibinfo {year} {1968})}\BibitemShut {NoStop}%
\end{thebibliography}

\begin{thebibliography}{19}%
\makeatletter
\providecommand \@ifxundefined [1]{%
 \@ifx{#1\undefined}
}%
\providecommand \@ifnum [1]{%
 \ifnum #1\expandafter \@firstoftwo
 \else \expandafter \@secondoftwo
 \fi
}%
\providecommand \@ifx [1]{%
 \ifx #1\expandafter \@firstoftwo
 \else \expandafter \@secondoftwo
 \fi
}%
\providecommand \natexlab [1]{#1}%
\providecommand \enquote  [1]{``#1''}%
\providecommand \bibnamefont  [1]{#1}%
\providecommand \bibfnamefont [1]{#1}%
\providecommand \citenamefont [1]{#1}%
\providecommand \href@noop [0]{\@secondoftwo}%
\providecommand \href [0]{\begingroup \@sanitize@url \@href}%
\providecommand \@href[1]{\@@startlink{#1}\@@href}%
\providecommand \@@href[1]{\endgroup#1\@@endlink}%
\providecommand \@sanitize@url [0]{\catcode `\\12\catcode `\$12\catcode
  `\&12\catcode `\#12\catcode `\^12\catcode `\_12\catcode `\%12\relax}%
\providecommand \@@startlink[1]{}%
\providecommand \@@endlink[0]{}%
\providecommand \url  [0]{\begingroup\@sanitize@url \@url }%
\providecommand \@url [1]{\endgroup\@href {#1}{\urlprefix }}%
\providecommand \urlprefix  [0]{URL }%
\providecommand \Eprint [0]{\href }%
\providecommand \doibase [0]{https://doi.org/}%
\providecommand \selectlanguage [0]{\@gobble}%
\providecommand \bibinfo  [0]{\@secondoftwo}%
\providecommand \bibfield  [0]{\@secondoftwo}%
\providecommand \translation [1]{[#1]}%
\providecommand \BibitemOpen [0]{}%
\providecommand \bibitemStop [0]{}%
\providecommand \bibitemNoStop [0]{.\EOS\space}%
\providecommand \EOS [0]{\spacefactor3000\relax}%
\providecommand \BibitemShut  [1]{\csname bibitem#1\endcsname}%
\let\auto@bib@innerbib\@empty
%</preamble>
\bibitem [{Note1()}]{Note1}%
  \BibitemOpen
  \bibinfo {note} {Python code of the RCSJ model is available on\\\protect
  \href
  {https://github.com/donfuge/rcsj_sde}{https://github.com/donfuge/rcsj\protect
  \_sde}}\BibitemShut {NoStop}%
\bibitem [{\citenamefont {Kautz}\ and\ \citenamefont
  {Martinis}(1990)}]{S_Kautz}%
  \BibitemOpen
  \bibfield  {author} {\bibinfo {author} {\bibfnamefont {R.~L.}\ \bibnamefont
  {Kautz}}\ and\ \bibinfo {author} {\bibfnamefont {J.~M.}\ \bibnamefont
  {Martinis}},\ }\bibfield  {title} {\bibinfo {title} {{Noise-affected $I$-$V$
  curves in small hysteretic Josephson junctions}},\ }\href
  {https://doi.org/10.1103/PhysRevB.42.9903} {\bibfield  {journal} {\bibinfo
  {journal} {Phys. Rev. B}\ }\textbf {\bibinfo {volume} {42}},\ \bibinfo
  {pages} {9903} (\bibinfo {year} {1990})}\BibitemShut {NoStop}%
\bibitem [{\citenamefont {Gross}\ \emph {et~al.}(2016)\citenamefont {Gross},
  \citenamefont {Marx},\ and\ \citenamefont {Deppe}}]{S_AppliedS}%
  \BibitemOpen
  \bibfield  {author} {\bibinfo {author} {\bibfnamefont {R.}~\bibnamefont
  {Gross}}, \bibinfo {author} {\bibfnamefont {A.}~\bibnamefont {Marx}},\ and\
  \bibinfo {author} {\bibfnamefont {F.}~\bibnamefont {Deppe}},\ }\href@noop {}
  {\emph {\bibinfo {title} {{Applied superconductivity: Josephson effect and
  superconducting electronics}}}}\ (\bibinfo  {publisher} {De Gruyter},\
  \bibinfo {year} {2016})\BibitemShut {NoStop}%
\bibitem [{\citenamefont {Rümelin}(1982)}]{S_Rumelin}%
  \BibitemOpen
  \bibfield  {author} {\bibinfo {author} {\bibfnamefont {W.}~\bibnamefont
  {Rümelin}},\ }\bibfield  {title} {\bibinfo {title} {{Numerical Treatment of
  Stochastic Differential Equations}},\ }\href
  {http://www.jstor.org/stable/2156972} {\bibfield  {journal} {\bibinfo
  {journal} {SIAM Journal on Numerical Analysis}\ }\textbf {\bibinfo {volume}
  {19}},\ \bibinfo {pages} {604} (\bibinfo {year} {1982})}\BibitemShut
  {NoStop}%
\bibitem [{\citenamefont {Welch}(1967)}]{S_1161901}%
  \BibitemOpen
  \bibfield  {author} {\bibinfo {author} {\bibfnamefont {P.}~\bibnamefont
  {Welch}},\ }\bibfield  {title} {\bibinfo {title} {{The use of fast Fourier
  transform for the estimation of power spectra: A method based on time
  averaging over short, modified periodograms}},\ }\href
  {https://doi.org/10.1109/TAU.1967.1161901} {\bibfield  {journal} {\bibinfo
  {journal} {IEEE Transactions on Audio and Electroacoustics}\ }\textbf
  {\bibinfo {volume} {15}},\ \bibinfo {pages} {70} (\bibinfo {year}
  {1967})}\BibitemShut {NoStop}%
\bibitem [{\citenamefont {Jung}\ \emph {et~al.}(2018)\citenamefont {Jung},
  \citenamefont {Yoshida}, \citenamefont {Park}, \citenamefont {Zhang},
  \citenamefont {Yesilyurt}, \citenamefont {Siu}, \citenamefont {Jalil},
  \citenamefont {Park}, \citenamefont {Park}, \citenamefont {Nagaosa},
  \citenamefont {Seo},\ and\ \citenamefont {Hirakawa}}]{S_Minkyung2018}%
  \BibitemOpen
  \bibfield  {author} {\bibinfo {author} {\bibfnamefont {M.}~\bibnamefont
  {Jung}}, \bibinfo {author} {\bibfnamefont {K.}~\bibnamefont {Yoshida}},
  \bibinfo {author} {\bibfnamefont {K.}~\bibnamefont {Park}}, \bibinfo {author}
  {\bibfnamefont {X.-X.}\ \bibnamefont {Zhang}}, \bibinfo {author}
  {\bibfnamefont {C.}~\bibnamefont {Yesilyurt}}, \bibinfo {author}
  {\bibfnamefont {Z.~B.}\ \bibnamefont {Siu}}, \bibinfo {author} {\bibfnamefont
  {M.~B.~A.}\ \bibnamefont {Jalil}}, \bibinfo {author} {\bibfnamefont
  {J.}~\bibnamefont {Park}}, \bibinfo {author} {\bibfnamefont {J.}~\bibnamefont
  {Park}}, \bibinfo {author} {\bibfnamefont {N.}~\bibnamefont {Nagaosa}},
  \bibinfo {author} {\bibfnamefont {J.}~\bibnamefont {Seo}},\ and\ \bibinfo
  {author} {\bibfnamefont {K.}~\bibnamefont {Hirakawa}},\ }\bibfield  {title}
  {\bibinfo {title} {{Quantum Dots Formed in Three-dimensional Dirac Semimetal
  Cd$_3$As$_2$ Nanowires}},\ }\href
  {https://doi.org/10.1021/acs.nanolett.7b05165} {\bibfield  {journal}
  {\bibinfo  {journal} {Nano Letters}\ }\textbf {\bibinfo {volume} {18}},\
  \bibinfo {pages} {1863} (\bibinfo {year} {2018})}\BibitemShut {NoStop}%
\bibitem [{\citenamefont {Jeon}\ \emph {et~al.}(2014)\citenamefont {Jeon},
  \citenamefont {Zhou}, \citenamefont {Gyenis}, \citenamefont {Feldman},
  \citenamefont {Kimchi}, \citenamefont {Potter}, \citenamefont {Gibson},
  \citenamefont {Cava}, \citenamefont {Vishwanath},\ and\ \citenamefont
  {Yazdani}}]{S_Jeon2014a}%
  \BibitemOpen
  \bibfield  {author} {\bibinfo {author} {\bibfnamefont {S.}~\bibnamefont
  {Jeon}}, \bibinfo {author} {\bibfnamefont {B.~B.}\ \bibnamefont {Zhou}},
  \bibinfo {author} {\bibfnamefont {A.}~\bibnamefont {Gyenis}}, \bibinfo
  {author} {\bibfnamefont {B.~E.}\ \bibnamefont {Feldman}}, \bibinfo {author}
  {\bibfnamefont {I.}~\bibnamefont {Kimchi}}, \bibinfo {author} {\bibfnamefont
  {A.~C.}\ \bibnamefont {Potter}}, \bibinfo {author} {\bibfnamefont {Q.~D.}\
  \bibnamefont {Gibson}}, \bibinfo {author} {\bibfnamefont {R.~J.}\
  \bibnamefont {Cava}}, \bibinfo {author} {\bibfnamefont {A.}~\bibnamefont
  {Vishwanath}},\ and\ \bibinfo {author} {\bibfnamefont {A.}~\bibnamefont
  {Yazdani}},\ }\bibfield  {title} {\bibinfo {title} {{Landau quantization and
  quasiparticle interference in the three-dimensional Dirac semimetal
  Cd$_3$As$_2$}},\ }\href {https://doi.org/10.1038/nmat4023} {\bibfield
  {journal} {\bibinfo  {journal} {Nat. Mater.}\ }\textbf {\bibinfo {volume}
  {13}},\ \bibinfo {pages} {851} (\bibinfo {year} {2014})}\BibitemShut
  {NoStop}%
\bibitem [{\citenamefont {Park}\ \emph {et~al.}(2020)\citenamefont {Park},
  \citenamefont {Jung}, \citenamefont {Kim}, \citenamefont {Bayogan},
  \citenamefont {Lee}, \citenamefont {An}, \citenamefont {Seo}, \citenamefont
  {Seo}, \citenamefont {Ahn},\ and\ \citenamefont {Park}}]{S_Kidong2020}%
  \BibitemOpen
  \bibfield  {author} {\bibinfo {author} {\bibfnamefont {K.}~\bibnamefont
  {Park}}, \bibinfo {author} {\bibfnamefont {M.}~\bibnamefont {Jung}}, \bibinfo
  {author} {\bibfnamefont {D.}~\bibnamefont {Kim}}, \bibinfo {author}
  {\bibfnamefont {J.~R.}\ \bibnamefont {Bayogan}}, \bibinfo {author}
  {\bibfnamefont {J.~H.}\ \bibnamefont {Lee}}, \bibinfo {author} {\bibfnamefont
  {S.~J.}\ \bibnamefont {An}}, \bibinfo {author} {\bibfnamefont
  {J.}~\bibnamefont {Seo}}, \bibinfo {author} {\bibfnamefont {J.}~\bibnamefont
  {Seo}}, \bibinfo {author} {\bibfnamefont {J.-P.}\ \bibnamefont {Ahn}},\ and\
  \bibinfo {author} {\bibfnamefont {J.}~\bibnamefont {Park}},\ }\bibfield
  {title} {\bibinfo {title} {{Phase Controlled Growth of Cd$_3$As$_2$ Nanowires
  and Their Negative Photoconductivity}},\ }\href
  {https://doi.org/10.1021/acs.nanolett.0c01010} {\bibfield  {journal}
  {\bibinfo  {journal} {Nano Letters}\ }\textbf {\bibinfo {volume} {20}},\
  \bibinfo {pages} {4939} (\bibinfo {year} {2020})}\BibitemShut {NoStop}%
\bibitem [{\citenamefont {Li}\ \emph {et~al.}(2018)\citenamefont {Li},
  \citenamefont {Li}, \citenamefont {Wang}, \citenamefont {Wang}, \citenamefont
  {Liao}, \citenamefont {Brinkman},\ and\ \citenamefont {Yu}}]{S_Li2018b}%
  \BibitemOpen
  \bibfield  {author} {\bibinfo {author} {\bibfnamefont {C.~Z.}\ \bibnamefont
  {Li}}, \bibinfo {author} {\bibfnamefont {C.}~\bibnamefont {Li}}, \bibinfo
  {author} {\bibfnamefont {L.~X.}\ \bibnamefont {Wang}}, \bibinfo {author}
  {\bibfnamefont {S.}~\bibnamefont {Wang}}, \bibinfo {author} {\bibfnamefont
  {Z.~M.}\ \bibnamefont {Liao}}, \bibinfo {author} {\bibfnamefont
  {A.}~\bibnamefont {Brinkman}},\ and\ \bibinfo {author} {\bibfnamefont
  {D.~P.}\ \bibnamefont {Yu}},\ }\bibfield  {title} {\bibinfo {title} {{Bulk
  and surface states carried supercurrent in ballistic Nb-Dirac semimetal
  Cd$_3$As$_2$ nanowire-Nb junctions}},\ }\href
  {https://doi.org/10.1103/PhysRevB.97.115446} {\bibfield  {journal} {\bibinfo
  {journal} {Phys. Rev. B}\ }\textbf {\bibinfo {volume} {97}},\ \bibinfo
  {pages} {115446} (\bibinfo {year} {2018})}\BibitemShut {NoStop}%
\bibitem [{\citenamefont {Li}\ \emph {et~al.}(2015)\citenamefont {Li},
  \citenamefont {Wang}, \citenamefont {Liu}, \citenamefont {Wang},
  \citenamefont {Liao},\ and\ \citenamefont {Yu}}]{S_Li2015giant}%
  \BibitemOpen
  \bibfield  {author} {\bibinfo {author} {\bibfnamefont {C.-Z.}\ \bibnamefont
  {Li}}, \bibinfo {author} {\bibfnamefont {L.-X.}\ \bibnamefont {Wang}},
  \bibinfo {author} {\bibfnamefont {H.}~\bibnamefont {Liu}}, \bibinfo {author}
  {\bibfnamefont {J.}~\bibnamefont {Wang}}, \bibinfo {author} {\bibfnamefont
  {Z.-M.}\ \bibnamefont {Liao}},\ and\ \bibinfo {author} {\bibfnamefont
  {D.-P.}\ \bibnamefont {Yu}},\ }\bibfield  {title} {\bibinfo {title} {{Giant
  negative magnetoresistance induced by the chiral anomaly in individual
  Cd$_3$As$_2$ nanowires}},\ }\href
  {https://doi.org/https://doi.org/10.1038/ncomms10137} {\bibfield  {journal}
  {\bibinfo  {journal} {Nature communications}\ }\textbf {\bibinfo {volume}
  {6}},\ \bibinfo {pages} {10137} (\bibinfo {year} {2015})}\BibitemShut
  {NoStop}%
\bibitem [{\citenamefont {Zhang}\ \emph {et~al.}(2015)\citenamefont {Zhang},
  \citenamefont {Liu}, \citenamefont {Wang}, \citenamefont {Zhang},
  \citenamefont {Zhou}, \citenamefont {Chen}, \citenamefont {Zou},\ and\
  \citenamefont {Xiu}}]{S_Zhang2015}%
  \BibitemOpen
  \bibfield  {author} {\bibinfo {author} {\bibfnamefont {E.}~\bibnamefont
  {Zhang}}, \bibinfo {author} {\bibfnamefont {Y.}~\bibnamefont {Liu}}, \bibinfo
  {author} {\bibfnamefont {W.}~\bibnamefont {Wang}}, \bibinfo {author}
  {\bibfnamefont {C.}~\bibnamefont {Zhang}}, \bibinfo {author} {\bibfnamefont
  {P.}~\bibnamefont {Zhou}}, \bibinfo {author} {\bibfnamefont {Z.-G.}\
  \bibnamefont {Chen}}, \bibinfo {author} {\bibfnamefont {J.}~\bibnamefont
  {Zou}},\ and\ \bibinfo {author} {\bibfnamefont {F.}~\bibnamefont {Xiu}},\
  }\bibfield  {title} {\bibinfo {title} {{Magnetotransport Properties of
  Cd$_3$As$_2$ Nanostructures}},\ }\href
  {https://doi.org/10.1021/acsnano.5b02243} {\bibfield  {journal} {\bibinfo
  {journal} {ACS Nano}\ }\textbf {\bibinfo {volume} {9}},\ \bibinfo {pages}
  {8843} (\bibinfo {year} {2015})}\BibitemShut {NoStop}%
\bibitem [{\citenamefont {Deacon}\ \emph {et~al.}(2017)\citenamefont {Deacon},
  \citenamefont {Wiedenmann}, \citenamefont {Bocquillon}, \citenamefont
  {Dom\'{\i}nguez}, \citenamefont {Klapwijk}, \citenamefont {Leubner},
  \citenamefont {Br\"une}, \citenamefont {Hankiewicz}, \citenamefont {Tarucha},
  \citenamefont {Ishibashi}, \citenamefont {Buhmann},\ and\ \citenamefont
  {Molenkamp}}]{S_Deaconb}%
  \BibitemOpen
  \bibfield  {author} {\bibinfo {author} {\bibfnamefont {R.~S.}\ \bibnamefont
  {Deacon}}, \bibinfo {author} {\bibfnamefont {J.}~\bibnamefont {Wiedenmann}},
  \bibinfo {author} {\bibfnamefont {E.}~\bibnamefont {Bocquillon}}, \bibinfo
  {author} {\bibfnamefont {F.}~\bibnamefont {Dom\'{\i}nguez}}, \bibinfo
  {author} {\bibfnamefont {T.~M.}\ \bibnamefont {Klapwijk}}, \bibinfo {author}
  {\bibfnamefont {P.}~\bibnamefont {Leubner}}, \bibinfo {author} {\bibfnamefont
  {C.}~\bibnamefont {Br\"une}}, \bibinfo {author} {\bibfnamefont {E.~M.}\
  \bibnamefont {Hankiewicz}}, \bibinfo {author} {\bibfnamefont
  {S.}~\bibnamefont {Tarucha}}, \bibinfo {author} {\bibfnamefont
  {K.}~\bibnamefont {Ishibashi}}, \bibinfo {author} {\bibfnamefont
  {H.}~\bibnamefont {Buhmann}},\ and\ \bibinfo {author} {\bibfnamefont {L.~W.}\
  \bibnamefont {Molenkamp}},\ }\bibfield  {title} {\bibinfo {title} {{Josephson
  Radiation from Gapless Andreev Bound States in HgTe-Based Topological
  Junctions}},\ }\href {https://doi.org/10.1103/PhysRevX.7.021011} {\bibfield
  {journal} {\bibinfo  {journal} {Phys. Rev. X}\ }\textbf {\bibinfo {volume}
  {7}},\ \bibinfo {pages} {021011} (\bibinfo {year} {2017})}\BibitemShut
  {NoStop}%
\bibitem [{\citenamefont {Hasler}(2016)}]{S_hasler2016microwave}%
  \BibitemOpen
  \bibfield  {author} {\bibinfo {author} {\bibfnamefont {T.}~\bibnamefont
  {Hasler}},\ }\emph {\bibinfo {title} {{Microwave noise detection of a quantum
  dot with stub impedance matching}}},\ \href
  {https://doi.org/10.5451/unibas-006624242} {Ph.D. thesis},\ \bibinfo
  {school} {University of Basel} (\bibinfo {year} {2016})\BibitemShut {NoStop}%
\bibitem [{\citenamefont {Scheller}\ \emph {et~al.}(2014)\citenamefont
  {Scheller}, \citenamefont {Heizmann}, \citenamefont {Bedner}, \citenamefont
  {Giss}, \citenamefont {Meschke}, \citenamefont {Zumb{\"u}hl}, \citenamefont
  {Zimmerman},\ and\ \citenamefont {Gossard}}]{S_scheller2014silver}%
  \BibitemOpen
  \bibfield  {author} {\bibinfo {author} {\bibfnamefont {C.~P.}\ \bibnamefont
  {Scheller}}, \bibinfo {author} {\bibfnamefont {S.}~\bibnamefont {Heizmann}},
  \bibinfo {author} {\bibfnamefont {K.}~\bibnamefont {Bedner}}, \bibinfo
  {author} {\bibfnamefont {D.}~\bibnamefont {Giss}}, \bibinfo {author}
  {\bibfnamefont {M.}~\bibnamefont {Meschke}}, \bibinfo {author} {\bibfnamefont
  {D.~M.}\ \bibnamefont {Zumb{\"u}hl}}, \bibinfo {author} {\bibfnamefont
  {J.~D.}\ \bibnamefont {Zimmerman}},\ and\ \bibinfo {author} {\bibfnamefont
  {A.~C.}\ \bibnamefont {Gossard}},\ }\bibfield  {title} {\bibinfo {title}
  {{Silver-epoxy microwave filters and thermalizers for millikelvin
  experiments}},\ }\href {https://doi.org/10.1063/1.4880099} {\bibfield
  {journal} {\bibinfo  {journal} {Applied physics letters}\ }\textbf {\bibinfo
  {volume} {104}},\ \bibinfo {pages} {211106} (\bibinfo {year}
  {2014})}\BibitemShut {NoStop}%
\bibitem [{\citenamefont {Haller}(2021)}]{S_haller2021probing}%
  \BibitemOpen
  \bibfield  {author} {\bibinfo {author} {\bibfnamefont {R.}~\bibnamefont
  {Haller}},\ }\emph {\bibinfo {title} {{Probing the microwave response of
  novel Josephson elements}}},\ \href {https://doi.org/10.5451/unibas-ep91822}
  {Ph.D. thesis},\ \bibinfo  {school} {University of Basel} (\bibinfo {year}
  {2021})\BibitemShut {NoStop}%
\bibitem [{\citenamefont {Park}\ \emph {et~al.}(2021)\citenamefont {Park},
  \citenamefont {Choi}, \citenamefont {Lee},\ and\ \citenamefont
  {Lee}}]{S_Park2021}%
  \BibitemOpen
  \bibfield  {author} {\bibinfo {author} {\bibfnamefont {J.}~\bibnamefont
  {Park}}, \bibinfo {author} {\bibfnamefont {Y.-B.}\ \bibnamefont {Choi}},
  \bibinfo {author} {\bibfnamefont {G.-H.}\ \bibnamefont {Lee}},\ and\ \bibinfo
  {author} {\bibfnamefont {H.-J.}\ \bibnamefont {Lee}},\ }\bibfield  {title}
  {\bibinfo {title} {{Characterization of Shapiro steps in the presence of a
  4\ensuremath{\pi}-periodic Josephson current}},\ }\href
  {https://doi.org/10.1103/PhysRevB.103.235428} {\bibfield  {journal} {\bibinfo
   {journal} {Phys. Rev. B}\ }\textbf {\bibinfo {volume} {103}},\ \bibinfo
  {pages} {235428} (\bibinfo {year} {2021})}\BibitemShut {NoStop}%
\bibitem [{\citenamefont {Jang}\ and\ \citenamefont {Doh}(2021)}]{S_Jang2021}%
  \BibitemOpen
  \bibfield  {author} {\bibinfo {author} {\bibfnamefont {Y.}~\bibnamefont
  {Jang}}\ and\ \bibinfo {author} {\bibfnamefont {Y.-J.}\ \bibnamefont {Doh}},\
  }\bibfield  {title} {\bibinfo {title} {{Optimal conditions for observing
  fractional Josephson effect in topological Josephson junctions}},\ }\href
  {https://doi.org/10.1007/s40042-020-00035-5} {\bibfield  {journal} {\bibinfo
  {journal} {J. Korean Phys. Soc.}\ }\textbf {\bibinfo {volume} {78}},\
  \bibinfo {pages} {58} (\bibinfo {year} {2021})}\BibitemShut {NoStop}%
\bibitem [{\citenamefont {Wang}\ \emph {et~al.}(2018)\citenamefont {Wang},
  \citenamefont {Li}, \citenamefont {Li}, \citenamefont {Liao}, \citenamefont
  {Brinkman},\ and\ \citenamefont {Yu}}]{S_Wang2018c}%
  \BibitemOpen
  \bibfield  {author} {\bibinfo {author} {\bibfnamefont {A.-Q.}\ \bibnamefont
  {Wang}}, \bibinfo {author} {\bibfnamefont {C.-Z.}\ \bibnamefont {Li}},
  \bibinfo {author} {\bibfnamefont {C.}~\bibnamefont {Li}}, \bibinfo {author}
  {\bibfnamefont {Z.-M.}\ \bibnamefont {Liao}}, \bibinfo {author}
  {\bibfnamefont {A.}~\bibnamefont {Brinkman}},\ and\ \bibinfo {author}
  {\bibfnamefont {D.-P.}\ \bibnamefont {Yu}},\ }\bibfield  {title} {\bibinfo
  {title} {{$4\ensuremath{\pi}$-Periodic Supercurrent from Surface States in
  ${\mathrm{Cd}}_{3}{\mathrm{As}}_{2}$ Nanowire-Based Josephson Junctions}},\
  }\href {https://doi.org/10.1103/PhysRevLett.121.237701} {\bibfield  {journal}
  {\bibinfo  {journal} {Phys. Rev. Lett.}\ }\textbf {\bibinfo {volume} {121}},\
  \bibinfo {pages} {237701} (\bibinfo {year} {2018})}\BibitemShut {NoStop}%
\bibitem [{Note2()}]{Note2}%
  \BibitemOpen
  \bibinfo {note} {See the Python script in folder '\protect \texttt
  {shapiro\protect \_evaluation}' in the Zenodo repository~\protect \href
  {https://doi.org/10.5281/zenodo.7961884}{10.5281/zenodo.7961884} to extract
  the current plateau widths from Shapiro maps.}\BibitemShut {Stop}%
\end{thebibliography}
